\documentclass[pdflatex,sn-mathphys-num]{sn-jnl}
\usepackage{graphicx}%
\usepackage{multirow}%
\usepackage{amsmath,amssymb,amsfonts}%
\usepackage{amsthm}%
\usepackage{mathrsfs}%
\usepackage[title]{appendix}%
\usepackage{xcolor}%
\usepackage{textcomp}%
\usepackage{manyfoot}%
\usepackage{booktabs}%
\usepackage{booktabs}%
\usepackage[utf8]{inputenc}
\usepackage{algorithm}%
\usepackage{algorithmicx}%
\usepackage{algpseudocode}%
\usepackage{listings}%
\usepackage{float}
\usepackage{graphicx}
\usepackage{amsmath}
\usepackage[version=4]{mhchem}
\usepackage{siunitx}
\usepackage{longtable,tabularx}
\usepackage{epstopdf} 
\usepackage{subcaption}
\usepackage{float} 
\usepackage{xcolor}
\usepackage{placeins}
\usepackage{booktabs}
\DeclareGraphicsExtensions{.eps,.pdf,.png,.jpg}
\usepackage{titlesec}
\usepackage{doi}
\usepackage{algorithm}
\usepackage{algpseudocode}
\usepackage[numbers,sort&compress]{natbib}
\usepackage{hyperref}
\usepackage{chngcntr}
\usepackage{fullpage}
\usepackage{xcolor}

\theoremstyle{thmstyleone}%
\theoremstyle{thmstyletwo}%

\theoremstyle{thmstylethree}%
\counterwithout{figure}{section}
\counterwithout{table}{section}
\counterwithout{equation}{section}

\begin{document}

\title[Detonation propagation in weakly confined gases]{Detonation propagation in weakly confined gases}


\author*[1]{\fnm{Youssef K.} \sur{Wahba}}\email{youssef.wahba@mail.mcgill.ca}

\author[2,3]{\fnm{XiaoCheng} \sur{Mi}}\email{x.c.mi@tue.nl}

\author[4]{\fnm{Charles B.} \sur{Kiyanda}}\email{charles.kiyanda@concordia.ca}

\author[1]{\fnm{Andrew J.} \sur{Higgins}}\email{andrew.higgins@mcgill.ca}

\affil*[1]{\orgdiv{Department of Mechanical Engineering}, \orgname{McGill University}, \orgaddress{\street{817 Sherbrooke St W}, \city{Montreal}, \postcode{H3A 0C3}, \state{Quebec}, \country{Canada}}}

\affil[2]{\orgdiv{Department of Mechanical Engineering}, \orgname{Eindhoven University of Technology}, \city{Eindhoven}, \postcode{5600 MB},  \country{The Netherlands}}

\affil[3]{\orgdiv{Eindhoven Institute of Renewable Energy Systems}, \orgname{Eindhoven University of Technology}, \city{Eindhoven}, \postcode{5600 MB},  \country{The Netherlands}}

\affil[4]{\orgdiv{Department of Mechanical, Industrial, and Aerospace Engineering}, \orgname{Concordia University}, \orgaddress{\street{1515 Ste-Catherine St W}, \city{Montreal}, \postcode{H3G 2W1}, \state{Quebec}, \country{Canada}}}


\abstract{This study investigates the propagation of detonations along a layered configuration where a reactive gas is weakly confined by a hotter inert layer. Computational Fluid Dynamics (CFD) simulations are performed using a single-step, non-Arrhenius reaction model designed to intentionally suppress cellular instabilities, thereby enabling the formulation of a theoretical framework that can be directly compared with the simulation results. The simulations reach a quasi-steady state, revealing distinct flowfield regimes that depend on the acoustic-impedance ratio and relative layer thicknesses, with some detonations exhibiting velocity deficits while others propagate above the ideal Chapman–Jouguet (CJ) speed. Analytical models are developed to interpret these regimes. When a precursor shock is observed in the inert layer, the detonation is overdriven; this phenomenon is modeled using shock-polar analysis and velocity estimates based on the approach of Mitrofanov (Acta Astronaut. \textbf{3}:995-1004, 1976). An analytical criterion for the onset of the precursor shock is proposed. In underdriven scenarios, the detonation front exhibits positive curvature, which is analyzed using a geometric construction approach, wherein the relationship between wave speed and front curvature is evaluated \textit{a priori}. In underdriven cases, a simplified characteristic-based model is introduced to capture the decay of the shock wave in the inert layer, after which a~shock-polar~analysis is applied to determine the resulting wave interaction. The predictions from these analytical models are assembled into a phase map that delineates regions of overdriven and underdriven behavior, along with the corresponding shock interactions, in the space of acoustic impedance and area ratios. This map is then compared directly with the CFD results. The combined numerical--theoretical framework presented in this paper clarifies transition mechanisms that govern layered detonations and provides insights into detonation dynamics relevant to rotating detonation engines in which the detonation is bounded by the hotter combustion products formed from a previous cycle.}




\maketitle

\section{Introduction}\label{sec1}
Layered detonation problems, where an explosive layer is laterally bounded by a layer of inert material, have long been of interest to the detonation research community, initially in high-explosive and mining applications involving confined explosives. With high explosives, a layer of explosive bounded by a layer of air or other gas can generate a precursor shock that is driven ahead of the detonation, potentially modifying the explosive prior to arrival of the detonation \cite{Tanguay2004a, Tanguay2004b, Vu2009}. More recently, these problems have received attention in gaseous detonation research owing to their relevance to rotating detonation engines (RDEs). An RDE \cite{WOLANSKI2013125} is a pressure-gain combustion engine in which a detonation wave propagates continuously in an annular chamber. Fresh fuel and oxidizer are injected into the annulus, displacing the combustion products that expand to a high velocity towards the exhaust. Consequently, the detonation is bounded on one side by high-temperature combustion products from the previous cycle, analogous to a detonation weakly confined by a low-impedance gas. 

Experimental studies have investigated gaseous detonations bounded by inert gases \cite{Sommers_Morrison, DABORA1965817, ADAMS}. Initially motivated as an analog to condensed-phase explosives, these studies explored interactions between gas-phase detonations and confining gaseous media. Sommers and Morrison examined detonations in reactive gases confined by non-reactive gases (air or helium) \cite{Sommers_Morrison}. When confined by air, an oblique shock trailed the detonation, whereas helium confinement resulted in a precursor shock moving ahead of the detonation. Their theoretical framework assumed the Chapman--Jouguet (CJ) plane coincided with the shock front, implying a detonation traveling at the CJ velocity, with a Prandtl--Meyer fan expanding the combustion products to match deflection and pressure conditions behind the oblique shock. Dabora et al. \cite{DABORA1965817} later refined the wave velocity of Sommers and Morrison's model by accounting for lateral gas expansion into the confining medium, reducing the detonation velocity below CJ. In early experiments \cite{DABORA1965817, ADAMS}, a membrane was used to separate the explosive layer from the bounding gas. As discussed by Liu et al. \cite{Liu_1987}, the thickness of these membranes plays a critical role: if too thin, diffusion between the mixtures may occur, whereas if too thick, the membrane can interfere with the overall wave interaction being studied.

\textcolor{black}{Lieberman and Shepherd \cite{Lieberman2007-yc} investigated detonation interaction with a diffuse composition interface formed by a gravity current, eliminating the need for a plastic film separator. They showed that spatial variations in local Chapman–Jouguet speed caused the detonation to curve, and that increasing dilution led to decoupling of the reaction zone from the leading shock, producing a transmitted shock and turbulent mixing zone.} Metrow et al. \cite{METROW20213565} used a gravity current to separate argon from a stoichiometric hydrogen-oxygen mixture, allowing the heavy inert gas to settle in the lower half of the linear channel. In the reactive layer, a curved detonation was observed, which decoupled at the interface into an oblique shock and a trailing contact surface, with the shock curving and reflecting off the bottom channel wall. Cheevers et al. \cite{CHEEVERS20233095} ignited a flame along the top wall of a channel to replicate the combustion products found in RDEs. Schlieren imaging of the detonation interacting with the inert gases, represented by the combustion products, revealed a shock wave consistently propagating ahead of the detonation with a greater velocity. This discrepancy in speeds led to the quenching of the detonation.  These findings underscore the role of acoustic impedance on the overall wave structure, although a thorough analysis of the onset of the precursor shock was not performed. 

\textcolor{black}{Menezes et al.~\cite{MENEZES20241223} investigated detonation propagation in a diffuse reactive layer and demonstrated that increasing inert dilution induces front curvature, promotes shock–reaction decoupling, and ultimately results in detonation failure once at sufficiently high dilution levels.} Li et al.~\cite{LiMiHiggins} conducted computational fluid dynamics (CFD) simulations using a pressure-dependent reaction rate to investigate how varying inert gas impedances affect detonation propagation subject to losses. They found that when confined by a high impedance inert gas, the velocity deficit results were consistent with a Newtonian model \cite{Tsuge_et_al}. In contrast, with low impedance confinement, the detonation velocities appear to conform more closely with the front-curvature model proposed by Wood and Kirkwood \cite{Wood_Kirkwood}. No cases involved sufficiently low impedance ratios to observe a precursor shock. It was observed that a sonic locus stems from the reactive--inert interface when the density of the inert medium is equal to or less than that of the reactive layer, corresponding to weak confinement. This phenomenon is widely reported in the literature \cite{aslam2002numerical, Bdzil_Short_Chiquete_2020}. Houim and Fievisohn \cite{HOUIM2017185} conducted CFD simulations across a broad range of acoustic impedance ratios in a two-dimensional channel-flow configuration, confirming that at low enough impedance ratios between the inert and the reactive gases the wave structure transitions from attached oblique shocks in the inert layer to detached shocks; their study employed polar analysis using Mach numbers from CFD simulations to characterize wave interactions at the reactive--inert gas interface. Shepherd and Kasahara \cite{shepherd_2023_f4bv2-11306} theoretically validated this transition, showing through polar analysis that a detached shock forms when the expansion polar at the CJ plane fails to intersect the oblique shock polar in the inert layer. This approach is similar to Dabora and Desbordes \cite{DaboraDesbordes}, who analyzed reactive gas layers and predicted that for an oblique detonation to stabilize in a confining reactive medium, the expansion polar must intersect the weak overdriven detonation branch. \textcolor{black}{Moreover, Reynaud et al.~\cite{Reynaud2020-rx} investigated detonations with lateral losses and demonstrated that the velocity deficit scales with both the mean front curvature and the hydrodynamic thickness of the reaction zone. In their study, detached shocks were observed when the temperature of the inert gas layer was increased, yet far from the reactive--inert interface, the shock in the inert layer retained its obliquity.}

Polar analysis has also been extensively applied to condensed-phase explosives \cite{ShortQuirk, Chiquete_Short_2019, Short_Voelkel_Chiquete_2020}, demonstrating that for low-impedance confiners, shocks in the confiner may propagate ahead of the detonation. Liu et al. \cite{QiuyueLiu} investigated how the height ratio between an air layer, which mimics an inert gas, and a reactive layer can affect wave propagation. Their analysis involved using shock polars informed by their CFD simulation results to examine the resulting wave structures in the inert layer where both regular and Mach reflections were observed. Since no precursor shock was seen, the results demonstrated that the detonation was traveling at less than the CJ speed. Li et al.~\cite{Li_Yang_Wang_2025} examined a configuration in which the reactive layer is bounded above and below by inert gas, varying the inert-gas molecular weight and temperature. Increasing the inert temperature and decreasing its molecular weight, both of which lower the acoustic impedance of the inert layer, promoted the formation of a precursor shock that overdrove the detonation above the CJ speed and produced a concave detonation front. \textcolor{black}{Liu et al. \cite{Liu2026} demonstrated that, in addition to acoustic impedance ratio, the acoustic velocity ratio can act as an independent governing parameter. When the impedance ratio was held constant and the acoustic velocity ratio was increased, the detonation–shock complex transitioned to more complicated regimes, including the emergence of a precursor shock ahead of the detonation front that overdrove it.} A similar precursor-driven overdrive was observed by the present authors \cite{Wahba_et_al} when the impedance of the inert gas was set to be less than half that of the reactive gas. The shock structure of the cases examined in \cite{Wahba_et_al} were also resolved using shock polars. These findings emphasize the importance of polar analysis as a valuable tool for understanding the interactions between reactive media and their confinement. They also highlight the significance of confinement properties in influencing the overall detonation dynamics, with some cases leading to underdriven detonations and others to overdriven detonations. However, there are currently no criteria to characterize comprehensively the combined effects of the area ratio and acoustic impedance ratio between the inert and reactive layers on detonation behavior.

In this paper, we examine a channel flow configuration in which a reactive layer is \textcolor{black}{weakly} confined by an inert gas, \textcolor{black}{which is itself} bounded by a rigid wall. The problem is studied through CFD simulations and theoretical analysis. The computational simulations elucidate the underlying physics, thereby enabling the construction of an analytical solution. Section~\ref{Problem_Statement} provides an overview of the problem and introduces the relevant non-dimensional parameters. In Section~\ref{Numerical_Methodology}, the numerical methods used to simulate the problem computationally are presented. Section~\ref{sample_results} provides sample computational results and identifies the different types of flow structures. Guided by the computational results, Section~\ref{Theory} develops analytical treatments of the flowfield elements, demonstrating how they can be assembled to reconstruct the entire flowfield. Comparisons between the analytical constructions and the computational simulations are made in Section~\ref{results}, and a phase map is presented, indicating the conditions under which the different possible solutions arise. Section~\ref{Discussion_Interpretattion} examines the performance and limitations of the analytical models, as well as the relevance of the analyzed flowfields to engineering applications. 

\section{Problem Statement}
\label{Problem_Statement}
\begin{figure}[H]
\centering
\includegraphics[width=0.9\textwidth]{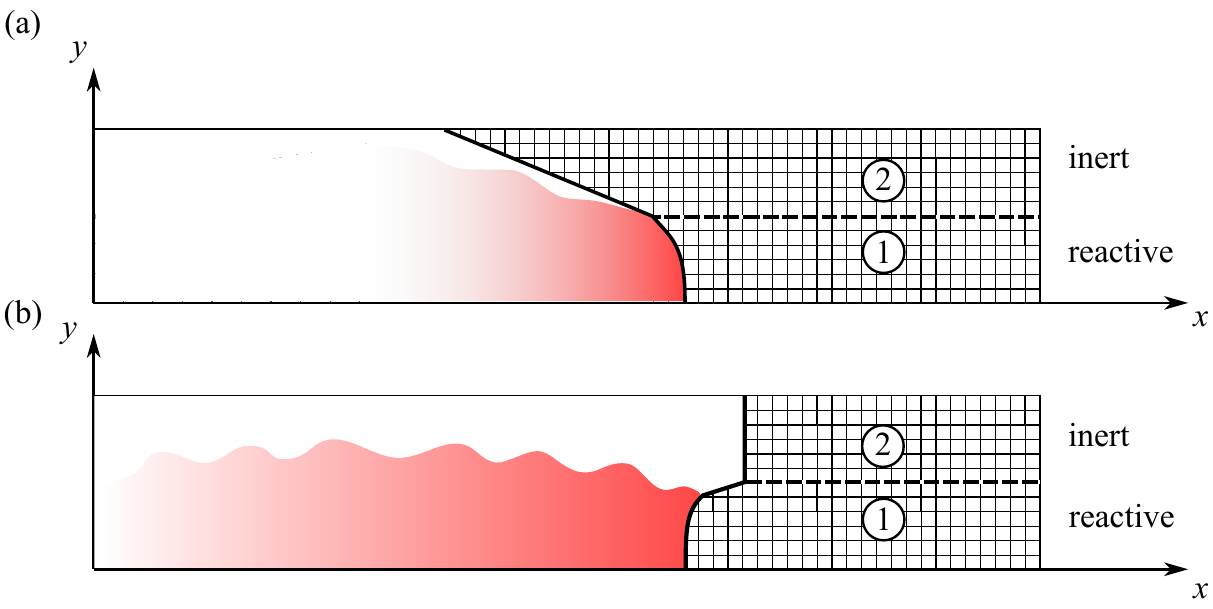}
\caption{Schematic representation of the problem, showing terminal shock structures relative to the detonation front: (a) attached shock (behind the front); (b) precursor shock (ahead of the front).}
\label{fig_statement}
\end{figure}

We consider a 2D,  two-layer, uniform initial pressure configuration with a reactive layer at the bottom and an inert layer above it, confined between horizontal walls at the top and bottom. The properties of the reactive and inert layers are denoted through subscripts 1 and 2, respectively, with both layers composed of perfect gases, as shown in Fig.~\ref{fig_statement}. The uniform pressure assumption mimics conditions relevant to an RDE combustion chamber, in which fresh reactants and products are separated by a contact surface, or relevant to a weakly bounded, high-density explosive. The initial temperature of the reactive layer is fixed, while the initial temperature of the inert layer is varied from one case to another. The inert layer density varies to accommodate the varying initial temperature. The reactive gas properties are chosen to mimic those of a stoichiometric hydrogen–oxygen mixture, with an adiabatic index $\gamma = 1.333$ and a nondimensional heat release of ${\Delta q}/{\left(R_1 T_1\right)} = 24$. For simplicity of analysis, a uniform property model was adopted such that $\gamma_2 = \gamma_1 = \gamma$, and $R_2 = R_1 = R$.

In this 2D configuration, the heights of the inert and reactive layers correspond to the unit cross-sectional area of each layer. For conciseness, we use the two terms interchangeably and denote them both by $A$. The total height of the computational domain was largely held constant while the ratio of the inert layer area, $A_2$, to the reactive layer area, $A_1$, was varied across a simulation matrix detailed in the Supplementary Information. Ratios in the range $0.25 \leq A_2/A_1 \leq 15$ were considered.

As the detonation interacts with the inert layer, a resulting wave structure appears, where the shock wave in the inert layer can travel either ahead of or behind the detonation, depending on the acoustic impedance ratio between both layers and the ratio of their respective heights.

The acoustic impedance, $\Upsilon$, is defined as 
\begin{equation}
\Upsilon = \rho c.
\end{equation}
The ratio of the acoustic impedance between the inert and the reactive layer, $Z$, is then related to the layers' thermodynamic properties according to

\begin{equation}
Z = \frac{\Upsilon_{\text{2}}}{\Upsilon_{\text{1}}} = \frac{\rho_{\text{2}} \sqrt{\gamma R_2 T_{\text{2}}}}{\rho_{\text{1}} \sqrt{\gamma R_1 T_{\text{1}}}} = \frac{\left(\frac{p_{\text{2}}}{R_2 T_{\text{2}}}\right)}{\left(\frac{p_{\text{1}}}{R_1 T_{\text{1}}}\right)} \sqrt{\frac{T_{\text{2}}}{T_{\text{1}}}} = \sqrt{\frac{T_{\text{1}}}{T_{\text{2}}}} .
\label{eq:impedance_ratio}
\end{equation}
For this study, only configurations where the acoustic impedance of the inert gas is less than that of the reactive gas, $\textit{Z} < 1$, were considered. Equation~\eqref{eq:impedance_ratio} demonstrates that, in our study, $Z$ depends solely on the initial temperatures of the two layers. It can also be shown that the acoustic velocity ratio between the inert and reactive medium, $\chi$, is the reciprocal of the acoustic impedance ratio
\begin{equation}
\chi = \frac{c_2}{c_1} = \frac{\sqrt{\gamma_2 R_2 T_2}}{\sqrt{\gamma_1 R_1 T_1}}
\end{equation}
\begin{equation}
\chi = \frac{1}{Z} \cdot \frac{\gamma_2}{\gamma_1} = \frac{1}{Z}
\end{equation}
for $\gamma_1 = \gamma_2$.

The reaction rate model consists of single-step, irreversible chemistry, where reactant \textbf{A} is converted into product \textbf{B}. The reaction progress variable, $\lambda$, ranges from 0 to 1, corresponding to unburnt and fully burnt states, respectively. The reaction rate is described by 
\begin{equation}
\label{lambdadot}
\dot{\lambda} = k (1 - \lambda) .
\end{equation}
and is chosen to be linear in $\lambda$, with no exponential dependence. 
This assumption keeps the simulations free of cellular structure, thereby facilitating analytical comparison focused on $Z$ and $A_2/A_1$. The rate constant, \textit{k}, is chosen such that the half-reaction zone length of the ideal Chapman--Jouguet (CJ) detonation is normalized to unity, giving \textit{k} = 1.05. To prevent premature reaction, it was necessary to introduce a switching function to the reaction rate, consisting of a Heaviside function of pressure, with the switching value between the initial pressure and the CJ pressure. This switch was set such that the weak oblique shock transmitted into the reactive layer (in the case where a precursor was present) would not initiate reaction. 

All variables are nondimensionalized with respect to the unburnt mixture state ahead of the detonation and defined as 
\vspace{-10pt}
\begin{equation}
p = \frac{\tilde{p}}{\tilde{p_1}}, \quad \rho = \frac{\tilde{\rho}}{\tilde{\rho_1}}, \quad T = \frac{\tilde{T}}{\tilde{T_1}}, \quad Q = \frac{\tilde{Q}}{\tilde{R T_1}}, \quad {x} = \frac{\tilde{x}}{\tilde{l}_{1/2}}, \quad t = \frac{\tilde{t} \sqrt{\tilde{R} \tilde{T_1}}}{\tilde{l}_{1/2}}, \quad k = \frac{\tilde{k} \tilde{l}_{1/2}}{\sqrt{\tilde{R} \tilde{T_1}}} .
\end{equation}

\noindent Table~\ref{table:thermochemical} summarizes the characteristics of the mixture, and its calculated thermodynamic variables.
\begin{table}[h]
    \centering
    \begin{tabular}{|c|c!{\vrule width 1pt}c|c|}
        \hline
        \multicolumn{2}{|c!{\vrule width 1pt}}{\textbf{Inputs}} & \multicolumn{2}{c|}{\textbf{Outputs}} \\
        \hline
        \textbf{Nondimensional} & \textbf{Value} & \textbf{Nondimensional} & \textbf{Value} \\
        \hline
        $\gamma$ & 1.333 & $M_{\text{CJ}}$ & 5.47 \\
        \textit{R} & 1 & $p_{\text{vN}}$ & 34.07 \\
        $T_{1}$ & 1 & $p_{\text{CJ}}$ & 17.54 \\
        $p_{1}$ & 1 & $T_{\text{vN}}$ & 5.84 \\
        $p_{2}$ & 1 & $T_{\text{CJ}}$ & 10.27 \\
        \textit{Q} & 24 & -- & -- \\
        \hline
    \end{tabular}
    \caption{Thermochemical parameters: input values and corresponding computed detonation properties (all nondimensionalized).}
    \label{table:thermochemical}
\end{table}

The paper aims to establish the criteria, depending on $Z$ and $A_2/A_1$, that govern the transition from cases without a precursor shock to those with a precursor shock. When no precursor is present and an attached shock forms, as in Fig.~\ref{fig_statement}(a), the focus is on what detonation velocity is obtained and how the shock interactions develop within the \emph{inert} layer. When a precursor is present, as in Fig.~\ref{fig_statement}(b), the analysis examines the corresponding detonation velocity and the resulting shock interactions in the \emph{reactive} layer.

\section{Numerical Methodology}
\label{Numerical_Methodology}
\begin{figure}[H]
\centering
\includegraphics[width=1\textwidth]{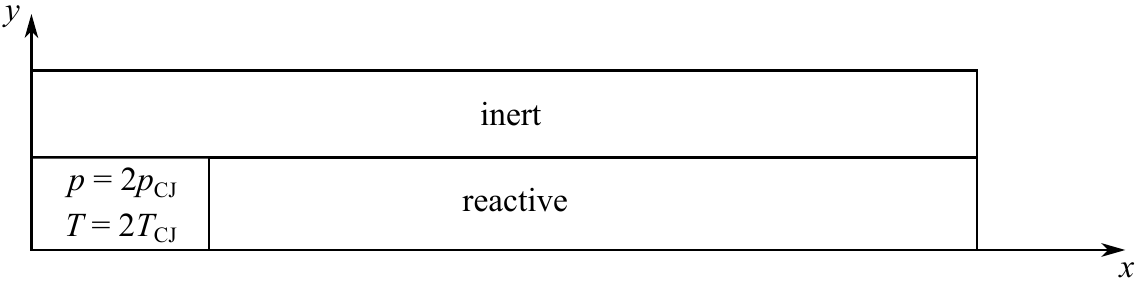}
\caption{Initial conditions showing abrupt initiation via a high-pressure, high-temperature region.}
\label{fig_initiation}
\end{figure}

The two-dimensional, reactive, Euler equations were solved using a MUSCL-Hancock scheme coupled with a Harten–Lax–van Leer–Contact (HLLC) approximate Riemann solver and a van Leer slope limiter and satisfy the governing equation
\begin{equation}
\frac{\partial \boldsymbol{\mathrm{U}}}{\partial t} + \frac{\partial \boldsymbol{\mathrm{F}}(\boldsymbol{\mathrm{U}})}{\partial x} + \frac{\partial \boldsymbol{\mathrm{G}}(\boldsymbol{\mathrm{U}})}{\partial y} = \boldsymbol{\mathrm{S}}(\boldsymbol{\mathrm{U}})
\end{equation}
where the vector of conserved variables, \textbf{U}, the flux terms \textbf{F} and \textbf{G}, in \textit{x} and \textit{y} directions, respectively, and the chemical source term \textbf{S}, are defined as 
\begin{equation}
\mathbf{U} = 
\begin{pmatrix}
\rho \\
\rho u \\
\rho v \\
\rho e \\
\rho \lambda
\end{pmatrix},
\quad
\mathbf{F} = 
\begin{pmatrix}
\rho u \\
\rho u^2 + p \\
\rho u v \\
(\rho e + p)u \\
\rho \lambda u
\end{pmatrix},
\quad
\mathbf{G} = 
\begin{pmatrix}
\rho v \\
\rho u v \\
\rho v^2 + p \\
(\rho e + p)v \\
\rho \lambda v
\end{pmatrix},
\quad
\mathbf{S} = 
\begin{pmatrix}
0 \\
0 \\
0 \\
0 \\
\rho \dot{\lambda}
\end{pmatrix}
.
\end{equation}
The specific internal energy is given by
\begin{equation}
e = \frac{p}{(\gamma - 1)\rho} + \frac{(u^2 + v^2)}{2} - \lambda Q
\end{equation}
and the reaction rate is given by Eq.~(\ref{lambdadot}). The simulations were initiated through a \textcolor{black}{reactive ($\lambda = 0$)} hotspot region that spans the height of the reactive layer as demonstrated in Fig.~\ref{fig_initiation}.

A Courant-Friedrichs-Lewy (CFL) number of 0.8 was used in all simulations. Further details of the numerical solver used can be found in \cite{GMorgan, Kiyanda2015-cf, Mi2018-wm, mi_effect_2018}. A numerical grid resolution of three points per half-reaction zone length of the ideal CJ detonation was used. \textcolor{black}{The supplementary Information accompanying this paper reports a numerical resolution study that shows doubling or halving the resolution from this value does not change the simulation outcome in terms of the observed wave structures.} The chemical source term is integrated through a Strang-splitting approach, preserving second-order accuracy. The code was implemented in NVIDIA's CUDA programming language and was executed on NVIDIA V100, V100l, and A100 GPUs. The total computational cost of the simulations amounted to approximately 3.5 core-years.

The horizontal boundaries are slip walls, mimicking solid surfaces, while the vertical boundaries \textcolor{black}{are treated using transmissive (outflow-type) boundary conditions, where ghost-cell values are obtained by zero-order extrapolation from the neighboring interior cells. To prevent boundary effects from influencing the solution, the computational window is periodically truncated and the simulation is reinitialized with a new left boundary, such that the detonation--shock complex does not reach the right boundary while the left boundary remains downstream of the sonic locus.} The simulations are run for a sufficiently long time to achieve a steady-state configuration and reach a terminal velocity. This is indicated by a shock-tracking feature along the upper and lower boundaries, which records the location of the shock in the inert layer and the detonation in the reactive layer. Therefore, the instantaneous velocity can be calculated at various positions through numerical differentiation and compared. Further details regarding the implementation are provided in Appendix~\ref{Appendi_steady_state}. To ensure that the final wave structure and velocity in the quasi-steady state are not affected by the choice of initiation, Appendix~\ref{Initial_conditions_test} includes a validation using different initial conditions. Additionally, since part of the domain is clipped once the wave approaches the right boundary, care is taken to ensure that this does not remove any subsonic regions, either within the reaction zone or elsewhere, that could influence the propagation of the detonation wave or shock in the inert layer. 
\newpage
\section{Sample Computational Results}
\label{sample_results}

Following the CFD methodology described previously, representative cases from the simulations are summarized. In cases without a precursor shock, an attached shock trails the detonation in the inert layer; the front acquires positive curvature ($\kappa>0$), indicating an underdriven detonation with associated losses in the reaction zone. The front is therefore \emph{convex} with respect to the unreacted mixture. By contrast, when a precursor shock forms ahead of the detonation in the inert layer, the front attains negative curvature ($\kappa<0$) and the detonation is overdriven. In this regime, the detonation front is \emph{concave} with respect to the unreacted mixture.

Based on the simulations, five cases are observed:  
\begin{itemize}
  \item[---] \textbf{Case A:} The detonation is underdriven and the trailing shock in the inert layer results in a \textbf{regular reflection (RR)}.
    \item[---] \textbf{Case B:} The detonation is underdriven and the trailing shock in the inert layer results in a \textbf{Mach reflection (MR)}.
    \item[---] \textbf{Case C:} A precursor shock appears in the inert layer, driving a \textbf{Mach~reflection~(MR)} in the reactive layer.
    \item[---] \textbf{Case D:} A precursor shock appears in the inert layer, driving a \textbf{regular~reflection~(RR)} in the reactive layer.
    \item[---] \textbf{Case E:} The shock in the inert layer protrudes ahead of the detonation, forming a \textbf{detached-shock} configuration. The shock interaction in the inert layer may either form a regular or Mach reflection.
\end{itemize}

\begin{figure}[H]
\centering
\includegraphics[width=0.905\textwidth]{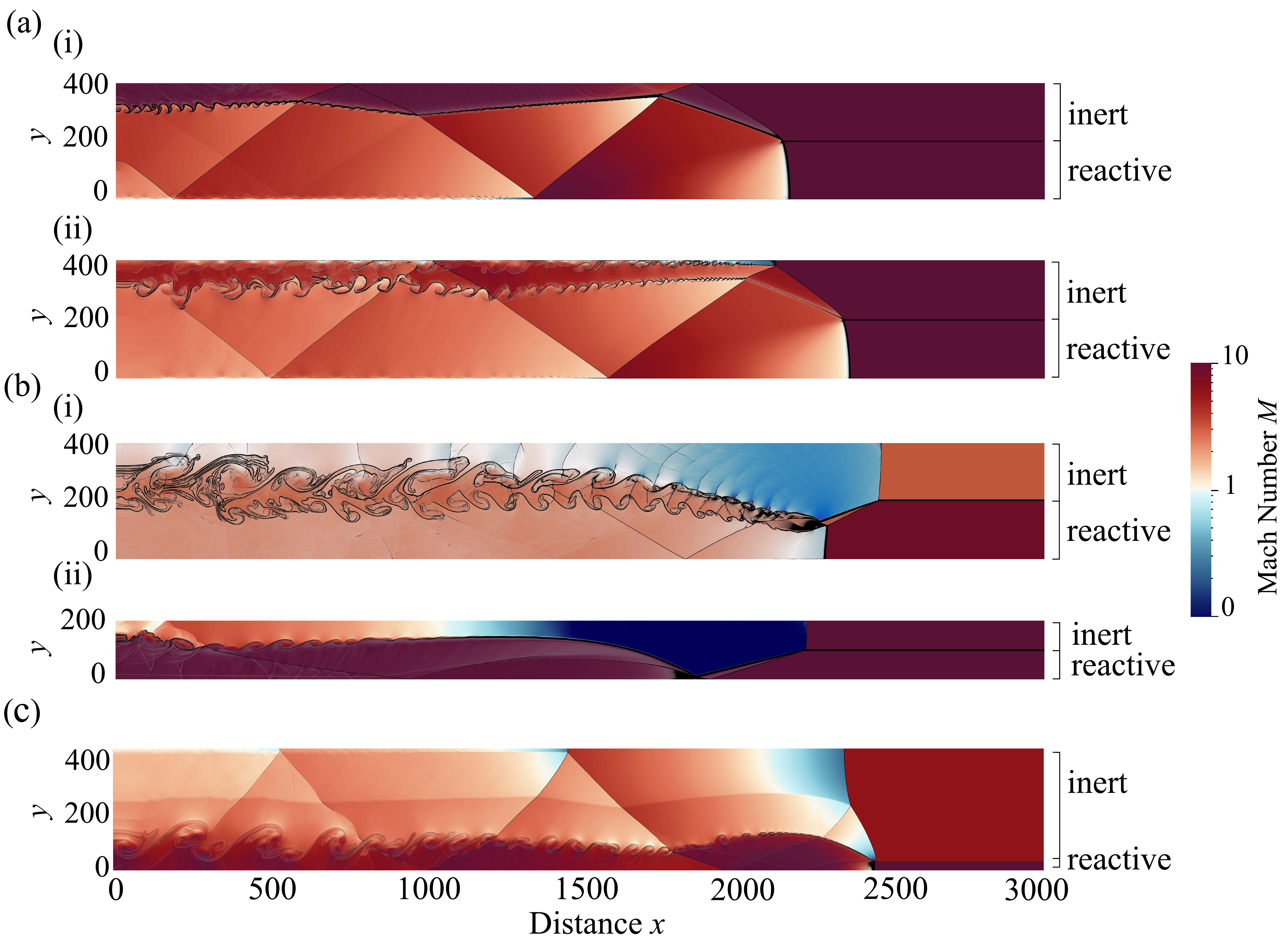}
\caption{Representative flowfields with wave-fixed Mach number overlaid on numerical schlieren for $A_2/A_1 = 1$.
(a) Underdriven cases, $\kappa > 0$: i) $Z = 0.80$ (Case~A), ii) $Z = 0.70$ (Case~B).
(b) Precursor shock cases, $\kappa < 0$: i) $Z = 0.45$ (Case~C), ii) $Z = 0.30$ (Case~D). (c) Detached shock, $A_2/A_1 = 11$, $Z = 0.40$ (Case E).}
\label{fig:mach_schlieren_A2A1_1}
\end{figure}

\begin{figure}[H]
\centering
\includegraphics[width=0.838\textwidth]{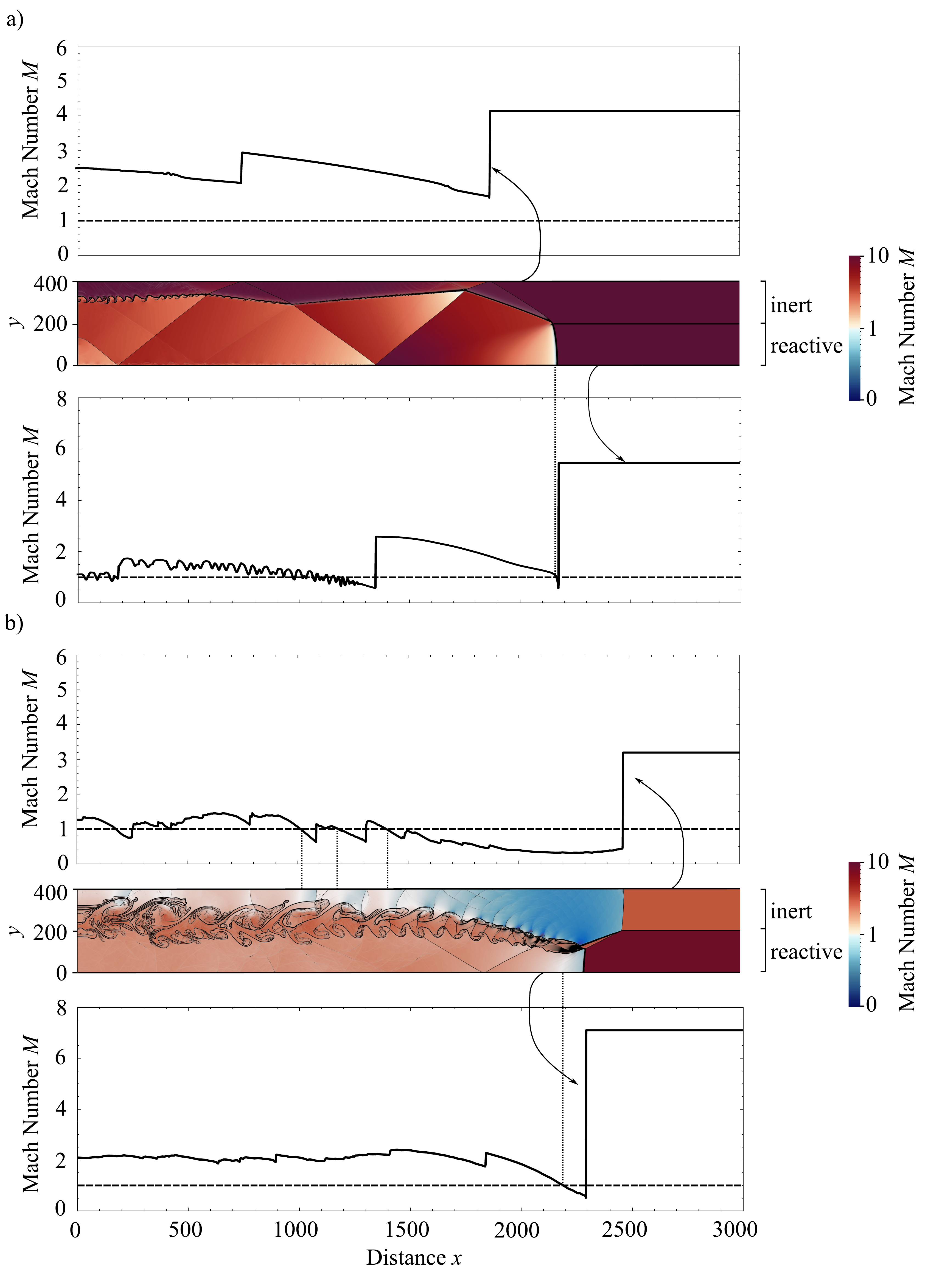}
\vspace{-5pt}
\caption{Example flowfield of wave-fixed Mach number overlay with schlieren for (a) $\kappa > 0$ for $A_2/A_1 = 1$ and $Z = 0.80$ (Case A from Fig.~\ref{fig:mach_schlieren_A2A1_1}), and (b) $\kappa < 0$ for $A_2/A_1 = 1$ and $Z = 0.45$ (Case C from Fig.~\ref{fig:mach_schlieren_A2A1_1}). Wave-fixed Mach number profiles are extracted along the top and bottom boundaries are shown above and below the flowfield. Dashed lines indicate $M = 1$ on the Mach number plots, while dotted lines project from the plots onto the flowfield to mark approximate locations where flow becomes sonic.}
\label{fig:mach_overlay}
\end{figure}

Sample flowfields showing the wave-fixed Mach number are presented in Fig.~\ref{fig:mach_schlieren_A2A1_1} which describe Cases A-E explained above. These examples are shown for a specific set of conditions, but a broader sweep over the parameters $A_2/A_1$ and $Z$ was carried out to identify a wide range of such cases. The blue-shaded and red-shaded regions highlight the wave-fixed subsonic and supersonic regions observed in each case. 
The wave-fixed Mach numbers along the top and bottom boundaries for Cases~A and~C 
(Fig.~\ref{fig:mach_schlieren_A2A1_1}(a)(i) and Fig.~\ref{fig:mach_schlieren_A2A1_1}(b)(i))
are extracted and shown above and below the flowfield overlays in Fig.~\ref{fig:mach_overlay}. 
The locations where the boundary wave-fixed Mach number reaches unity are identified in the plots and 
projected back onto the flowfields. As shown in the CFD flowfield in Fig.~\ref{fig:mach_overlay}(b), a sonic region appears in the inert layer when a precursor shock propagates ahead of the detonation, separating the subsonic and supersonic regions. This condition can be interpreted as \emph{choking} of the inert layer. 
\vspace{-20pt}

\begin{figure}[H]
\centering
\includegraphics[width=0.95\textwidth]{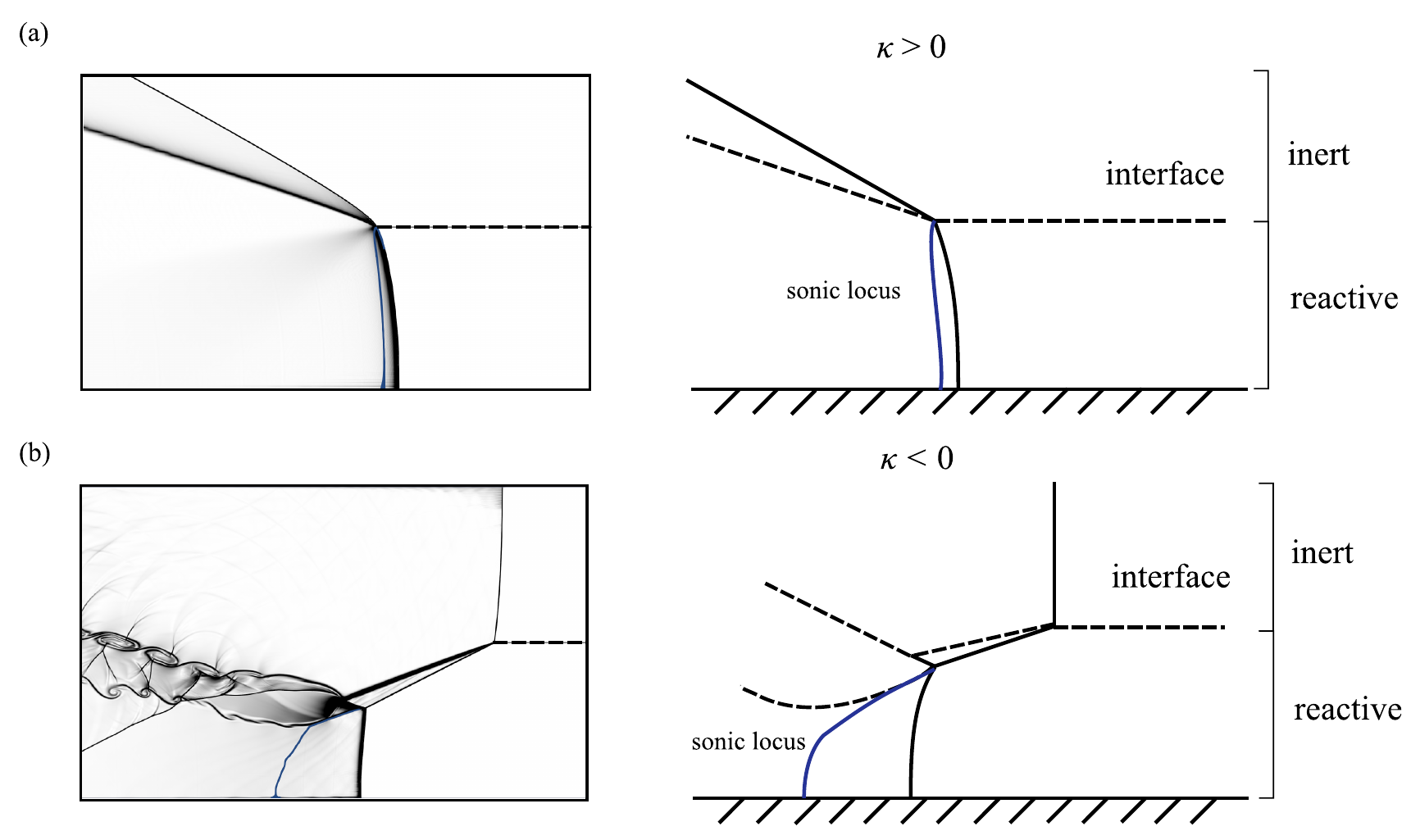}
\caption{Closeup of the schlieren image of the detonation structure in Fig.~\ref{fig:mach_overlay}, showing (a) the underdriven case ($\kappa > 0$) and (b) the overdriven case ($\kappa < 0$), with the sonic locus overlay and corresponding simplified schematics on the right.}
\label{fig:detonation_structures}
\end{figure}

\begin{figure}[H]
\centering
\includegraphics[width=1\textwidth]{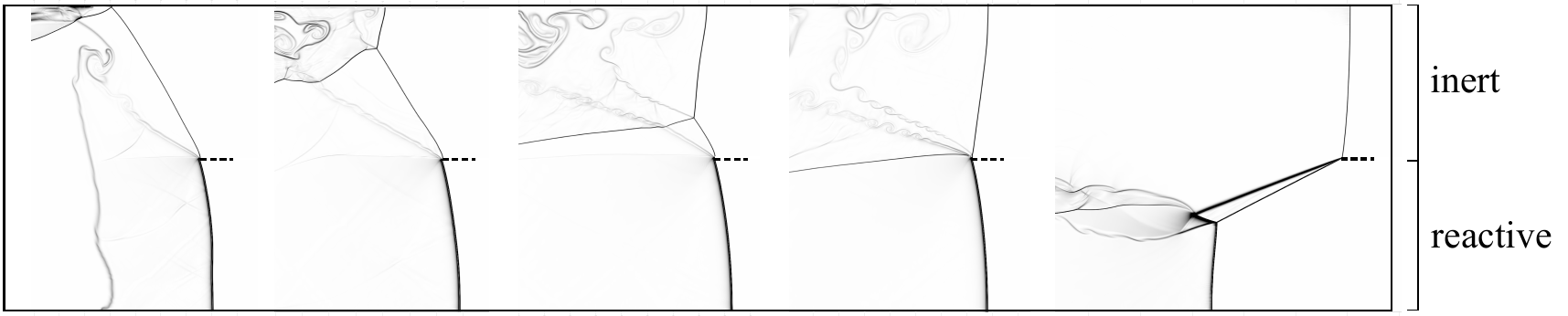}
\caption{Collage of CFD schlieren images showing precursor shock formation for $A_2/A_1 = 1$, $Z = 0.45$}
\label{CFD_snapshots}
\end{figure}

Figure~\ref{fig:detonation_structures} provides both a CFD snapshot and schematic illustrating the wave structures observed in these underdriven and overdriven regimes. In the underdriven configuration, the post-shock sonic locus originates from the reactive–inert interface, whereas in the overdriven case, it emerges from the triple point in the reactive layer. Figure~\ref{CFD_snapshots} shows several snapshots from the CFD simulations for the case where a precursor shock develops at $A_2/A_1 = 1$ and $Z = 0.45$. These results demonstrate that, in this scenario, the Mach stem in the inert layer eventually overtakes the detonation front and forms a precursor shock. 

\vspace{-15pt}
\begin{figure}[H]
\centering
\includegraphics[width=0.92\textwidth]{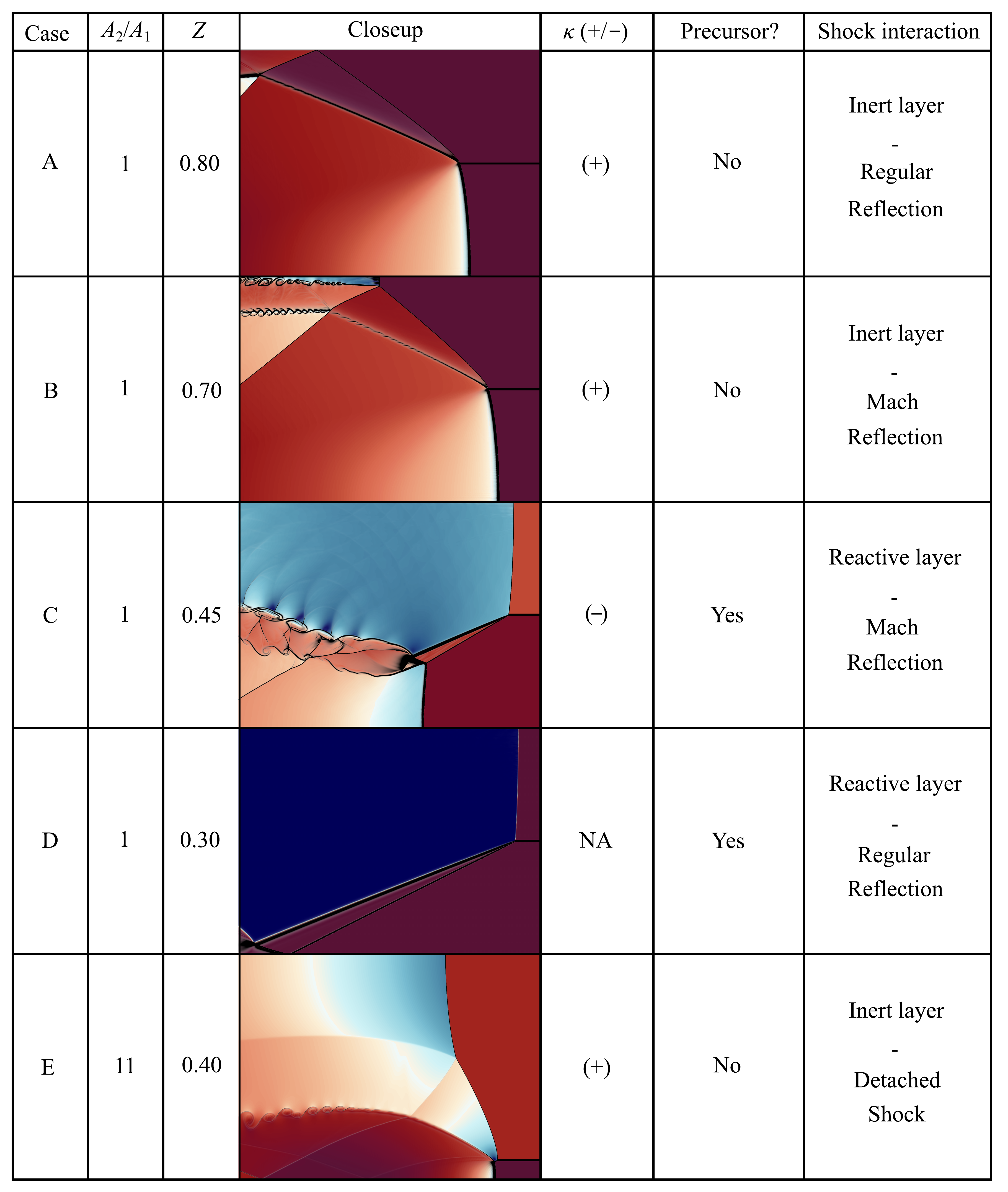}
\captionsetup{type=table}
\caption{Overview of detonation structures occurring across different $Z$ and $A_2/A_1$ values, with close-up views of representative simulations (from Fig.~\ref{fig:mach_schlieren_A2A1_1}) corresponding to Cases A–E and highlighting the corresponding detonation properties and associated shock behavior.}
\label{tab:detonation_structures}
\end{figure}

\vspace{-15pt}
A summary of the detonation behaviors identified from Fig.~\ref{fig:mach_schlieren_A2A1_1}, along with the corresponding wave structures and parameters, is provided in Table~\ref{tab:detonation_structures} for Cases A-E. For the remainder of the main body of the paper, the focus is devoted entirely to Cases A-D, while the emergence of a detached shock (Case E) is discussed in Appendix~\ref{Detached_Precursor_Comparison}.

\section{Theory}
\label{Theory}
Building on the sample CFD results from the previous section, an analytical formulation 
is developed to evaluate the influence of the parameters $Z$ and $A_2/A_1$ on the overall 
solution and to classify the corresponding wave-structure case. As summarized in Table~\ref{tab:detonation_structures}, the emergence of a precursor shock governs the detonation behavior: Cases~A and~B correspond to underdriven detonations, whereas Cases~C and~D are overdriven. Accordingly, Section~\ref{Precusor_Onset} introduces two alternative criteria for predicting the onset of a precursor shock, based on a quasi-one-dimensional analysis of the detonation 
products expanding into the inert layer. The underdriven detonation behavior is then analyzed in Section~\ref{Underdriven_Cases} using a ZND-based model and the Eyring construction method~\cite{Eyring1949-vl}, with polar analysis employed to predict the shock interaction in the inert layer. Section~\ref{Overdriven_Cases} examines the overdriven detonation behavior based on Mitrofanov's two-layer approach~\cite{MITROFANOV1976995}, with polar 
analysis similarly applied to predict the shock–detonation interaction in the reactive layer.

\subsection{Onset of precursor shock}
\label{Precusor_Onset}
Two criteria for the onset of a precursor shock in the inert layer are investigated. In both cases, a Chapman--Jouguet (CJ) detonation is assumed to propagate in the reactive layer. The first criterion considers a configuration where a shock is present directly above the detonation front in the inert layer, resulting in stagnation pressure losses and requiring a full accounting of shock-induced total pressure drop/entropy generation. This will be referred to as the \emph{Kantrowitz-type criterion}. The second criterion neglects such a shock and assumes isentropic compression in the inert layer, implying that the stagnation pressure is conserved in the inert layer. Although both criteria are introduced for completeness, subsequent comparison with CFD shows that the isentropic assumption does not reproduce the observed behavior. Therefore, only the Kantrowitz-type criterion is pursued in the analysis that follows. 

\begin{figure}[H]
\centering
\includegraphics[width=0.75\textwidth]{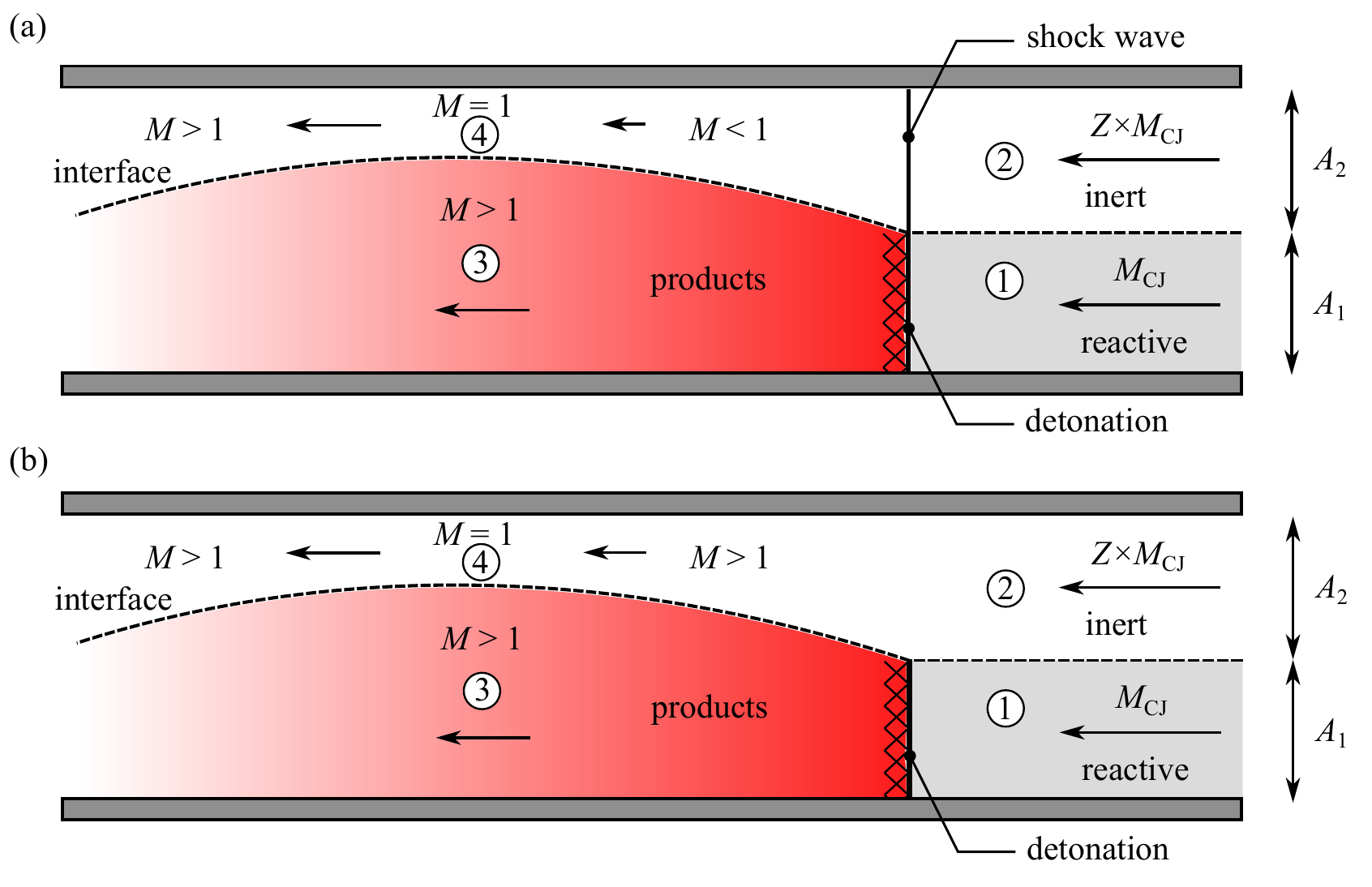}
\vspace{-5pt}
\caption{Schematic of the proposed theoretical models predicting choking of the inert layer by the expanding detonation products, resulting in a precursor shock: (a)~Kantrowitz criterion with a shock wave positioned above the detonation, and (b)~Isentropic criterion assuming no stagnation pressure losses in the inert layer. Both models are presented in a wave-fixed framework, with subsonic and supersonic regions indicated. The detonation is assumed to be travelling at the CJ speed.}
\label{precursor_schematics}
\end{figure}

The phenomenon here is analogous to the \textcolor{black}{classical} problem of a high-speed train moving through a tunnel. \textcolor{black}{As the train advances, the available flow area between the train body and the tunnel wall decreases, potentially choking the flow and generating a compression wave that can steepen into a shock propagating ahead of the train, even when the train itself is subsonic \cite{LANG2024106202}}. In the present problem, the expansion of the
detonation products plays a role \textcolor{black}{analogous} to the advancing train, \textcolor{black}{reducing the flow area available to the inert layer and potentially inducing choking.} \textcolor{black}{This behavior is closely related to the} Kantrowitz limit \cite{kantrowitz1945supersonic}, originally formulated to determine the critical area ratio required for a supersonic diffuser to spontaneously \emph{start}, i.e., establish supersonic flow through the inlet. At the Kantrowitz criterion, a normal shock is located at the entrance plane of the inlet, and either increasing Mach number or decreasing the compression ratio will result in the normal shock being swallowed, establishing supersonic flow through the inlet. Similarly, in our model, the shock in the inert layer imposes a constraint on the allowable area ratio between the inert and reactive layers. The flow must adjust its expansion and area ratio at the choking point in the inert layer such that the flow becomes choked, while maintaining a constant total height of the configuration. A schematic explaining the two criteria examined for choking of the inert layer is provided in Fig.~\ref{precursor_schematics}

\vspace{-10pt}

Solving the conservation of mass, momentum and energy for the control volumes in the inert and reactive layer, while imposing that at the choking point in the inert layer $A_4 + A_3 = A_2 + A_1$, the equations become quite similar to \cite{VUetal}, \textcolor{black}{who developed a quasi-one-dimensional choking model for precursor shock formation in explosive channel detonations. In their formulation, stagnation pressure losses in the inert layer arise from a train of reflected oblique shocks, whereas in the present model they are represented by a single normal shock. Despite this difference in shock structure, both approaches enforce area conservation and pressure matching at the critical (choked) section of the inert layer.}

\begin{equation}
M_2 = M_{\text{CJ}} \cdot Z
\label{eq:M_2}
\end{equation}
\begin{equation}
\frac{p_{04}}{p_{02}}\footnotemark \footnotetext{If the isentropic assumption were applied, the stagnation pressure ratio, $p_{04}/p_{02}$, would be $1$. This case is shown in Fig.~\ref{precursor_schematics}(b) for completeness but was found to be inconsistent with CFD results and is not pursued further.} =
\left( \frac{\gamma + 1}{2\gamma M_2^2 \ - (\gamma - 1)} \right)^{\frac{1}{\gamma - 1}}
\cdot
\left( \frac{(\gamma + 1) M_2^2 }{2 + (\gamma - 1) M_2^2 } \right)^{\frac{\gamma}{\gamma - 1}}
\label{eq:p_o_ratio}
\end{equation}

\begin{equation}
M_3 =
\sqrt{
\frac{\gamma + 1}{\gamma - 1} \cdot
\left[
\left( \frac{p_{04}}{p_{02}} \cdot \frac{1}{p_\mathrm{CJ}} \right)
\cdot
\left( \frac{2 + (\gamma - 1) M_2^2}{\gamma + 1} \right)^{\frac{\gamma}{\gamma - 1}}
\right]^{\frac{1 - \gamma}{\gamma}} - \frac{2}{\gamma - 1}
}
\label{eq:M_3}
\end{equation}

\begin{equation}
\left( \frac{A_2}{A_1} \right)_{\text{critical}} =
\frac{
1 - \left( \frac{1}{M_3} \left( \frac{2 + (\gamma - 1) M_3^2}{\gamma + 1} \right)^{\frac{\gamma + 1}{2(\gamma - 1)}} \right)
}{
\left( \frac{M_2}{p_{04}/p_{02}} \left( \frac{\gamma + 1}{2 + (\gamma - 1) M_2^2} \right)^{\frac{\gamma + 1}{2(\gamma - 1)}} \right) - 1
}
\label{eq:A_2_A_1}
\end{equation}

\begin{figure}[H]
\centering
\includegraphics[width=0.45\textwidth]{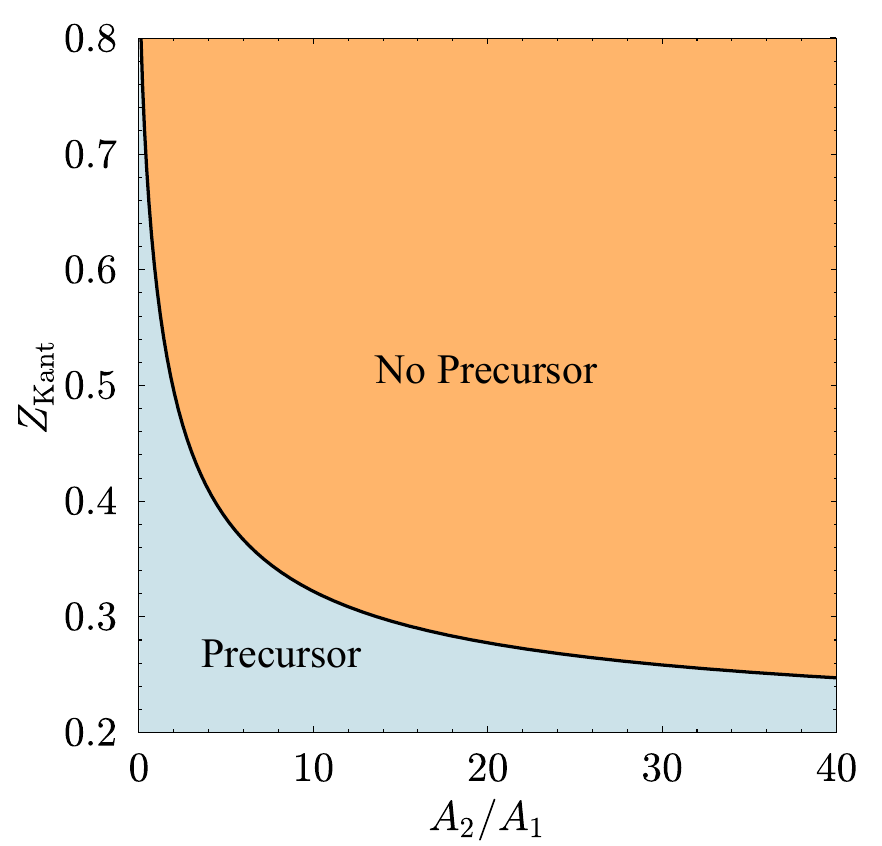}
\vspace{-5pt}
\caption{Critical acoustic impedance ratio, $Z_\mathrm{Kant}$, as a function of the area ratio $A_2/A_1$, delineating the boundary between configurations with an attached inert-layer shock and those exhibiting a precursor shock.}
\label{precursor_noprecursor}
\end{figure}

Using Eqs.~\eqref{eq:M_2}--\eqref{eq:M_3}, it is evident that $Z$ enters implicitly in Eq.~\eqref{eq:A_2_A_1}. Thus, a relationship can be derived between the area ratio $A_2/A_1$ and the critical acoustic impedance ratio, $Z_\mathrm{Kant}$, corresponding to choked flow in the inert layer. The critical value, $Z_\mathrm{Kant}$, corresponds to the configuration where the shock lies directly above the detonation, as shown in Fig.~\ref{precursor_schematics}(a). For $Z > Z_\mathrm{Kant}$, the shock in the inert layer remains attached behind the detonation; the theoretical analysis of this case is presented in Section~\ref{Underdriven_Cases}. In contrast, for $Z < Z_\mathrm{Kant}$, the shock propagates ahead of the detonation and forms a precursor shock that overdrives it; the corresponding theoretical analysis is given in Section~\ref{Overdriven_Cases}. \textcolor{black}{The distinction between these regimes is summarized in Fig.~\ref{precursor_noprecursor}.}

\subsection{Underdriven detonations (\textit{Z} $> Z_\mathrm{Kant}$, $\kappa > 0$ cases)}
\label{Underdriven_Cases}
\subsubsection{Modeling of detonation front curvature and velocity}
\label{velocity_deficit}
In the case of an underdriven detonation with weak confinement ($Z < 1$), a sonic locus originates from the interface between the reactive and inert layers, as shown in Fig.~\ref{fig:detonation_structures}(a). The detonation behavior in this regime can be modeled by analyzing the relationship between the normal detonation velocity, $D_n$, and the front curvature, $\kappa$, in conjunction with a shock polar analysis to determine the shock angle at the interface that results in post-shock sonic flow. \textcolor{black}{The front curvature $\kappa$ is defined geometrically in a two-dimensional slab configuration as $\kappa = 1/R$, where $R$ denotes the local radius of curvature of the detonation front.} \textcolor{black}{To construct the curved detonation front consistent with these conditions, the classical Eyring geometric construction for curved detonations is employed \cite{Eyring1949-vl}. This method relates the local front curvature to the normal detonation velocity through a compatibility condition along the interface. Algorithm~\ref{algorithmic_1} outlines the procedure used to determine the detonation-front curvature}, with further details provided in \cite{LiMiHiggins} and summarized in Appendix~\ref{Eyring_Appendix}.

\begin{figure}[H]
\centering
\includegraphics[width=0.85\textwidth]{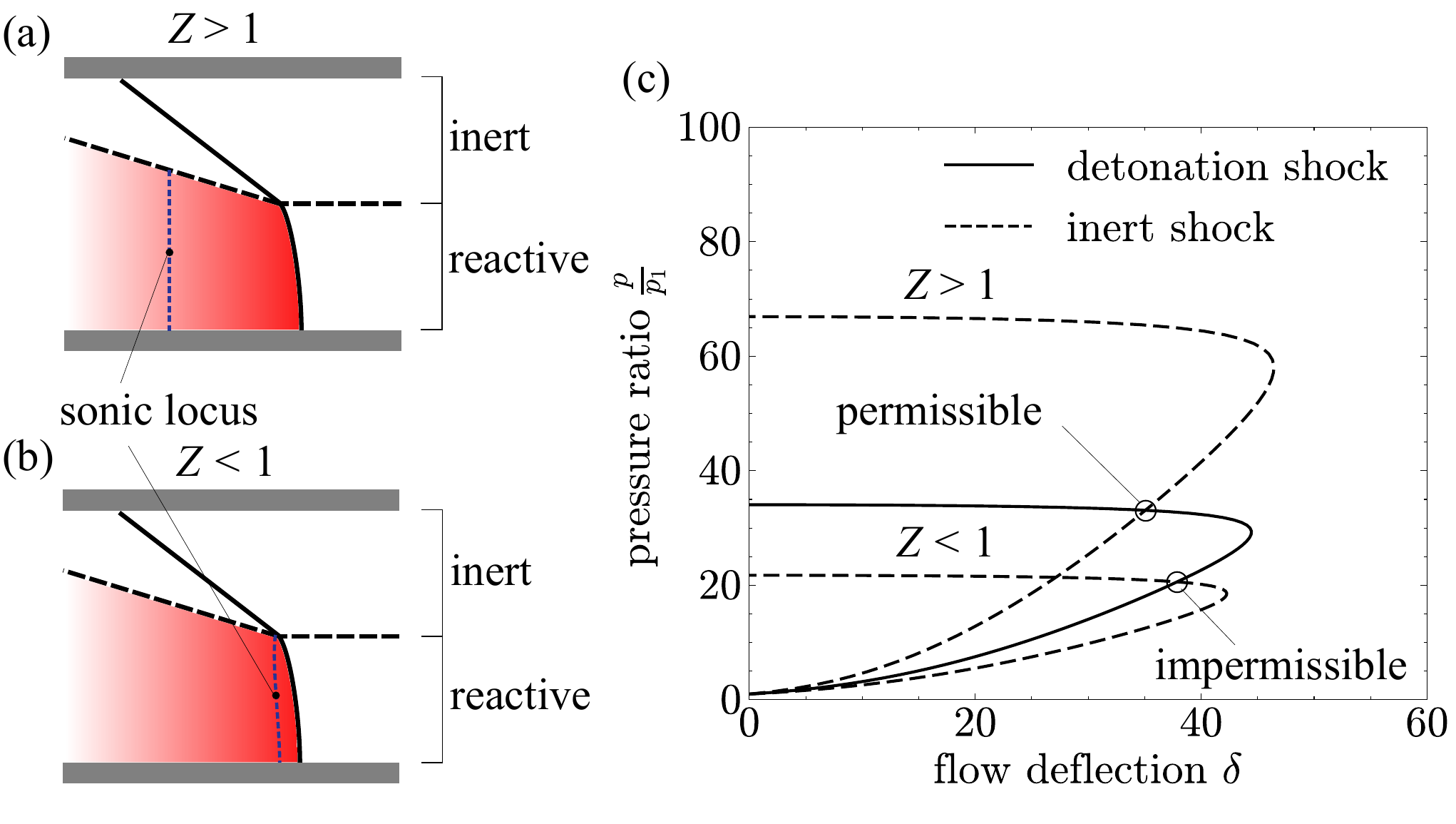}
\caption{(a) High-impedance confinement ($Z>1$): the sonic locus is detached from the detonation shock front.
(b) Low-impedance confinement ($Z<1$): the sonic locus attaches to the detonation shock front.
(c) Inert shock polars showing weak-branch intersection for $Z > 1$ and strong-branch intersection for $Z < 1$.}
\label{permissible_inpermissible}
\vspace{-30pt}
\end{figure}

\textcolor{black}{Implementation of the Eyring construction requires determination of the detonation shock angle at the reactive--inert interface, which depends on whether the detonation is strongly ($Z>1$) or weakly ($Z<1$) confined. The confinement further determines the location of the sonic locus. Schematics illustrating representative strong and weak confinement configurations are shown in Fig.~\ref{permissible_inpermissible}(a) and Fig.~\ref{permissible_inpermissible}(b), respectively.} For strong confinement, the shock angle at the interface is obtained from the intersection of the shock polar in the inert medium with that of the detonation shock. In this case, the solution corresponds to the weak branch of the inert shock polar, indicating that the post-shock flow in the inert layer is supersonic. For weak confinement, a similar intersection would lie on the strong branch of the inert shock polar, \textcolor{black}{as seen in Fig.~\ref{permissible_inpermissible}(c)}, leading to subsonic post-shock flow in the inert layer. This outcome is not physical, particularly when more complex wave structures, such as regular or Mach reflections, form in the inert layer. Consequently, for weak confinement, the detonation shock angle at the interface is instead taken to correspond to post-shock sonic flow, which agrees with CFD results showing a sonic locus emerging from the reactive–inert interface.

As such, since the shock angle at the reactive–inert interface for weak confinement cases is independent of the acoustic impedance ratio, the extent of the velocity deficit becomes dominated by the height of the reactive layer, with lower heights leading to greater velocity deficits. \textcolor{black}{A comparison between the detonation front curvature predicted by the Eyring construction and the corresponding detonation front extracted from the CFD simulations is provided in Appendix~\ref{Eyring_Construction_CFD}.}

\begin{algorithm}[H]
\caption{Eyring Construction for Detonation Velocity}
\begin{algorithmic}
\State \textbf{Input:} $D_n–\kappa$ relation, charge thickness $A_1$, shock-polar data, tolerance $\varepsilon$
\State \textbf{Initialize:} Choose initial $D_n^{(0)}$, find $\kappa^{(0)}$ from $D_{n}$–$\kappa$ relation
\For{$i = 0, 1, 2, \dots$}
    \State Integrate flow equations from $y = 0$ to $y = h_1$ with $\kappa^{(i)}$
    \State Compute slope $\left. \frac{\mathrm{d}y}{\mathrm{d}x} \right|_{y = h_1}$
    \State Compute $\cot\phi_\mathrm{s}$ from shock polar
    \If{$\left| \left. \frac{\mathrm{d}y}{\mathrm{d}x} \right|_{y = h_1} - \cot\phi_\mathrm{s} \right| < \varepsilon$}
        \State \textbf{Return:} $D_{n}^{(i)}$, $\kappa^{(i)}$ as solution
        \State \textbf{Break}
    \Else
        \State Update $D_{n}^{(i+1)}$ 
        \State Update $\kappa^{(i+1)}$ from $D_{n}$–$\kappa$ relation
    \EndIf
\EndFor
\end{algorithmic}
\label{algorithmic_1}
\end{algorithm}
\vspace{1em}
\vspace{-30pt}
\subsubsection{Case $A_2/A_1 \leq 1$: Straight incident shock}
\label{Straight_Incident_Shock}
\begin{figure}[H]
\centering
\includegraphics[width=0.63\textwidth]{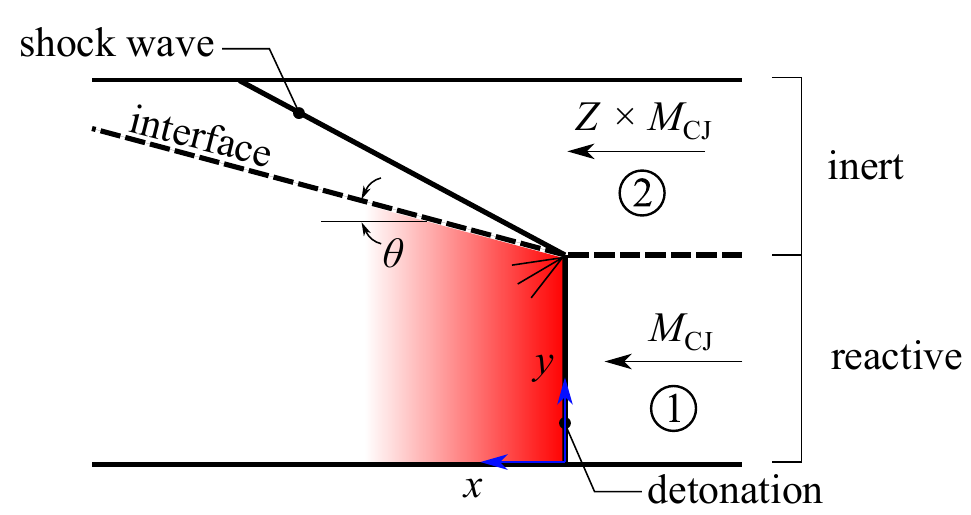}
\caption{\textcolor{black}{Schematic of a CJ detonation propagating adjacent to an inert layer, with a straight incident shock transmitted into the inert medium and a Prandtl–Meyer expansion fan adjusting the detonation products to match the interface angle.}}
\label{straight_shock}
\end{figure}

For cases where the reactive layer is at least as thick as the inert layer ($A_2/A_1 \leq 1$), the incident shock in the inert layer behind the detonation is assumed to be straight, \textcolor{black}{as shown in Fig.~\ref{straight_shock}}. \textcolor{black}{The incident shock angle} is determined from a shock-polar analysis of a CJ detonation interacting with a Prandtl–Meyer expansion fan at the CJ plane, which expands the detonation products to a \textcolor{black}{pressure equal to that behind the incident shock in the inert layer}; this analysis was also used in \cite{Sommers_Morrison}.

\begin{figure}[H]
\centering
\includegraphics[width=0.62\textwidth]{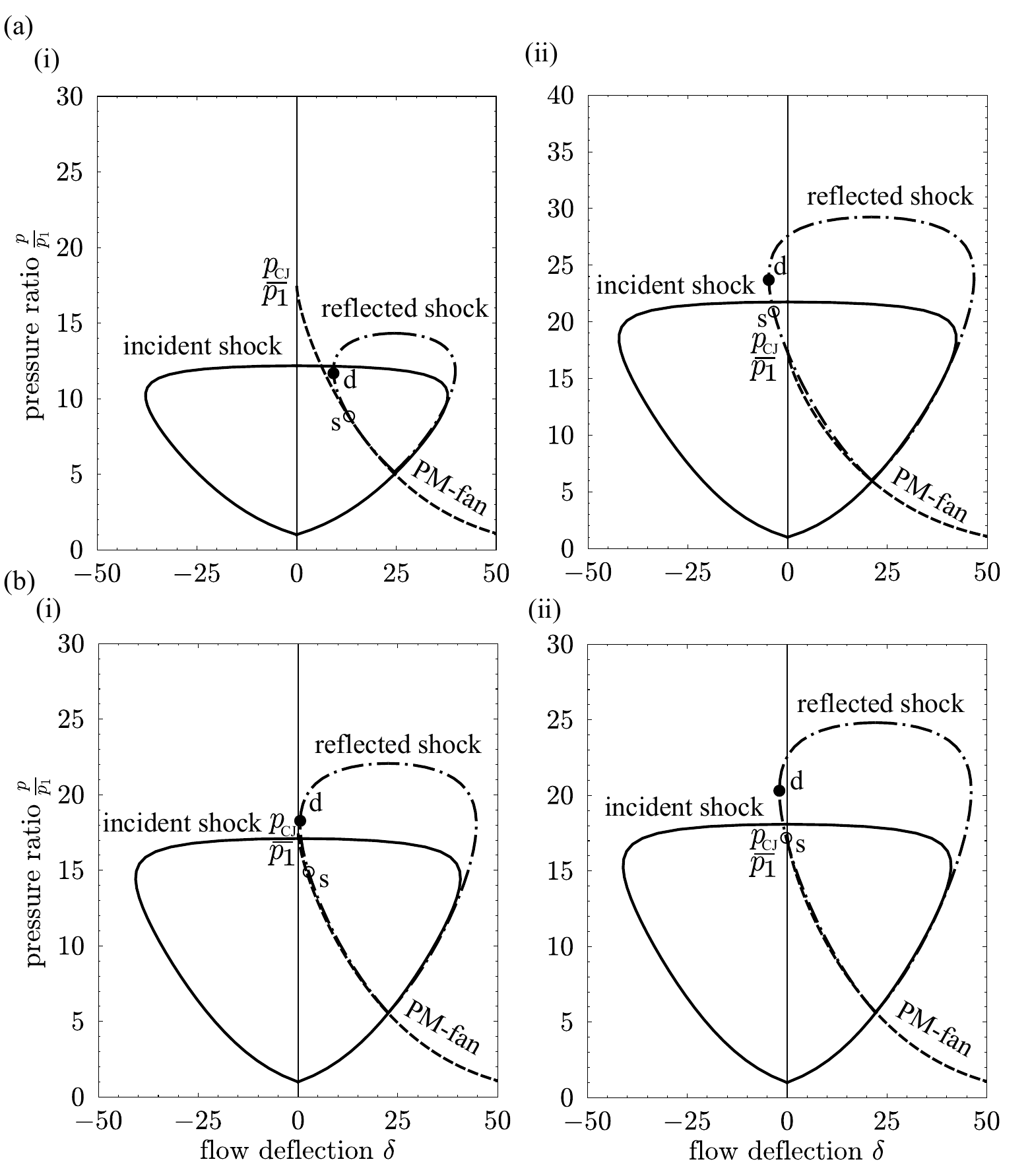}
\caption{Shock-polar analysis of the RR-MR transition in the inert layer for $A_2/A_1 \leq 1$. (a) (i) Mach reflection in inert layer (Case B) at $Z = 0.60$ (ii) regular reflection in inert layer (Case A) at $Z = 0.80$. The detonation is assumed to be travelling at the CJ speed with the expansion fan polar being extended from the point corresponding to $p_\mathrm{CJ}$ on $\delta =0$ axis. Transition depending on the detachment and sonic criteria (b) (i) $Z = 0.71$, (ii) $Z = 0.73$, which is denoted as $Z_\mathrm{detach (inert)}$ and $Z_\mathrm{sonic (inert)}$.}
\label{fig:detonation_transition_inert}
\end{figure}

Knowing both the angle of the incident shock and the wave-fixed Mach number in the inert layer ($M_2 = M_\mathrm{CJ} \cdot Z$), the reflected shock polar can be constructed. From this construction, the type of interaction, i.e., Mach reflection or regular reflection, can be identified, as demonstrated in Fig.~\ref{fig:detonation_transition_inert}. The type of reflection depends on $Z$, since different values of $Z$ correspond to different $M_2$ values. For example, Fig.~\ref{fig:detonation_transition_inert}(a)(i) shows that at $Z = 0.60$, the reflected shock polar fails to intersect the $\delta = 0$ axis and thus a Mach reflection results (Case B). In contrast, Fig.~\ref{fig:detonation_transition_inert}(a)(ii) shows that when $Z$ is increased to 0.80, the reflected shock polar does intersect the $\delta = 0$ axis and thereby forms a regular reflection (Case A).

The transition between these two cases can be evaluated using either the detachment criterion or the sonic criterion \textcolor{black}{\cite{BenDor}} as shown in Fig.~\ref{fig:detonation_transition_inert}(b)(i) and Fig.~\ref{fig:detonation_transition_inert}(b)(i), with the points d and s denoting the maximum-deflection and sonic points on the shock polar, respectively. The $Z$ values at which this transition occurs are denoted as $Z_\mathrm{detach(inert)}$ and $Z_\mathrm{sonic(inert)}$, respectively.

\label{Underdriven}
\vspace{-30pt}
\subsubsection{Case $A_2/A_1 > 1$: Decaying incident shock}
\label{Decaying_Incident_Shock}
\begin{figure}[H]
\centering
\includegraphics[width=0.55\textwidth]{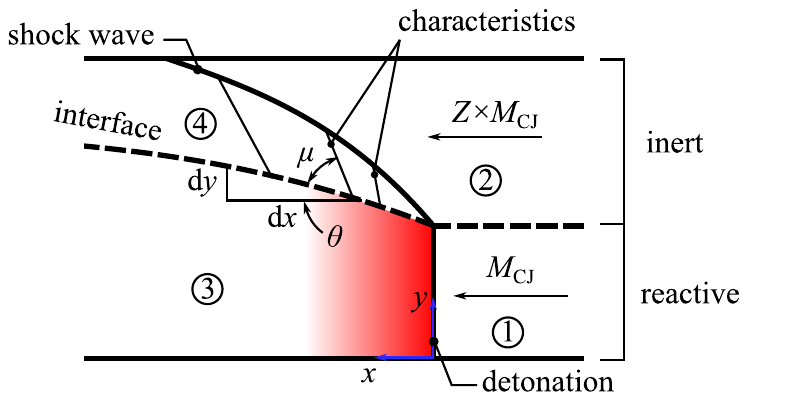}
\caption{\textcolor{black}{Schematic of the shock--expansion method applied to a decaying attached shock trailing the detonation front, with straight-line characteristics emanating from the reactive–inert interface.}}
\label{fig:detonation_attached_characteristics}
\end{figure}

\textcolor{black}{For cases where the reactive layer is thinner than the inert layer ($A_2/A_1 > 1$), the incident shock in the inert layer decays as it approaches the upper boundary. This decay is modeled using the shock--expansion method, which determines the evolution of the attached shock and its angle at the upper wall. Once this angle is obtained, a shock–polar analysis can be performed. A schematic of the shock–expansion method applied to the decaying attached shock is shown in Fig.~\ref{fig:detonation_attached_characteristics}, and a corresponding quasi-one-dimensional model is developed to determine the detonation–inert interface angle and inert-layer Mach number as the shock weakens.
}
\vspace{-20pt}
The deflection angle of the combustion products, denoted by $\theta$, inclines the interface between the reactive and inert layers. The slope of this inclined interface can therefore be written as
\begin{equation}
\left(\frac{\mathrm{d}y}{\mathrm{d}x}\right)_{\mathrm{interface}} = \tan\theta
\label{eq:dy_dx}
\end{equation}
which can also be expressed in terms of the local area variation as
\begin{equation}
\frac{1}{A}\,\frac{\mathrm{d}A}{\mathrm{d}x} = \frac{1}{y}\,\tan\theta
\label{eq:1_A_dA_dx} .
\end{equation}

By assuming the presence of a local Prandtl--Meyer fan in region~4 issuing from the inclined interface, an analogy to the shock--expansion method can be drawn, whereby a decaying shock is related to a varying interface angle. Under this assumption, the variation of $\theta$ along the $x$-direction can be described as follows:

\begin{equation}
\frac{\mathrm{d}\theta}{\mathrm{d}x}
= -\,\frac{\sqrt{M_{4}(x)^{2}-1}}{1+\frac{\gamma-1}{2}\,M_{4}(x)^{2}}
\frac{1}{M_{4}(x)}\,\frac{\mathrm{d}M_{4}(x)}{\mathrm{d}x} .
\label{eq:dtheta_dx}
\end{equation}

\noindent Similarly, the pressure variation along the interface in region~4 can be described using an isentropic relation, expressed in terms of the local Mach number in that region.

\begin{equation}
\frac{\mathrm{d}p}{\mathrm{d}x}
= -\,\frac{\gamma\,p\,M_{4}(x)}{1+\frac{\gamma-1}{2}\,M_{4}(x)^{2}}\,
\frac{\mathrm{d}M_{4}(x)}{\mathrm{d}x}
\label{eq:dp_dx_inert}
\end{equation}

In region~3, the pressure of the expanding detonation products can also be evaluated by rearranging the differential form of the momentum equation, which yields
 
\begin{equation}
\frac{\mathrm{d}p}{\mathrm{d}x} 
= -\,\rho\,u\,\frac{\mathrm{d}u}{\mathrm{d}x} 
= -\,\rho_\mathrm{vN}\,u_\mathrm{vN}\frac{y_1}{y}\,\frac{\mathrm{d}u}{\mathrm{d}x}
\label{eq:dp_dx_reactive}
\end{equation}
\noindent
where $u_\mathrm{vN}$ and $\rho_\mathrm{vN}$ denote the wave-fixed velocity and density at the von~Neumann state. Equating Eqs.~\eqref{eq:dp_dx_inert} and~\eqref{eq:dp_dx_reactive} then yields an implicit expression for the variation of the Mach number $M_4(x)$ along the interface.

The reaction rate and the variation of the local velocity of the reacting gas are given by

\begin{equation}
\frac{\mathrm{d}\lambda}{\mathrm{d}x} = \frac{k\left(1 - \lambda\right)}{u}
\label{eq:dlambda_dx}
\end{equation}

\begin{equation}
\frac{\mathrm{d}u}{\mathrm{d}x} = \frac{\dfrac{\Delta q \dot{\lambda}}{c_p T} - u \left(\dfrac{1}{A} \dfrac{\mathrm{d}A}{\mathrm{d}x} \right)}{1 - M^2}
\label{eq:du_dx}
\end{equation}
\noindent
where $c_p$ denotes the specific heat capacity of the gas. Equations~\eqref{eq:dy_dx}--\eqref{eq:du_dx} form a coupled set of ODEs that govern the divergence at the detonation–inert interface. It is evident that Eq.~\eqref{eq:du_dx} can exhibit a singularity at the sonic condition when the denominator vanishes. To obtain a continuous solution through this point, an eigenvalue iteration is required to pass smoothly into the supersonic regime.

An alternative to eliminate this sonic-point singularity is by setting the divergence term, $\left(\tfrac{1}{A}\right)\left(\tfrac{\mathrm{d}A}{\mathrm{d}x}\right)$, in the reaction zone ($\lambda < 1$) to zero. This approximation is adopted throughout the main body, since it retains the equilibrium CJ solution while restricting divergence effects to the isentropic expansion of the detonation products outside the reaction zone.

With this treatment, the decaying shock can be traced along straight-line characteristics emanating from the interface. These characteristics carry information of $\theta(x)$, and since $M_4(x)$ is known, the shock angle at the top wall can be determined. With this angle, a polar analysis can be performed to establish whether the interaction results in a regular or Mach reflection. \textcolor{black}{A detailed formulation of the shock–expansion method, including a representative solution for $A_2/A_1 = 6.5$ and $Z = 0.60$, is provided in Appendix~\ref{Shock_Expansion}.}

The coupled ODEs are initialized at the sonic point of the inert-layer shock polar, with $M_4(x)=1$ and $\theta(x)$ equal to the corresponding sonic deflection. Since this deflection lies below the maximum allowable deflection angle on the inert-layer shock polar, the analysis is naturally restricted to attached-shock cases, thereby excluding detached bow shocks (see Appendix~\ref{Detached_Precursor_Comparison}).

\subsection{Overdriven detonations ($Z < Z_\mathrm{Kant}$, $\kappa < 0$ cases)}
\label{Overdriven_Cases}

\begin{figure}[H]
\centering
\includegraphics[width=0.71\textwidth]{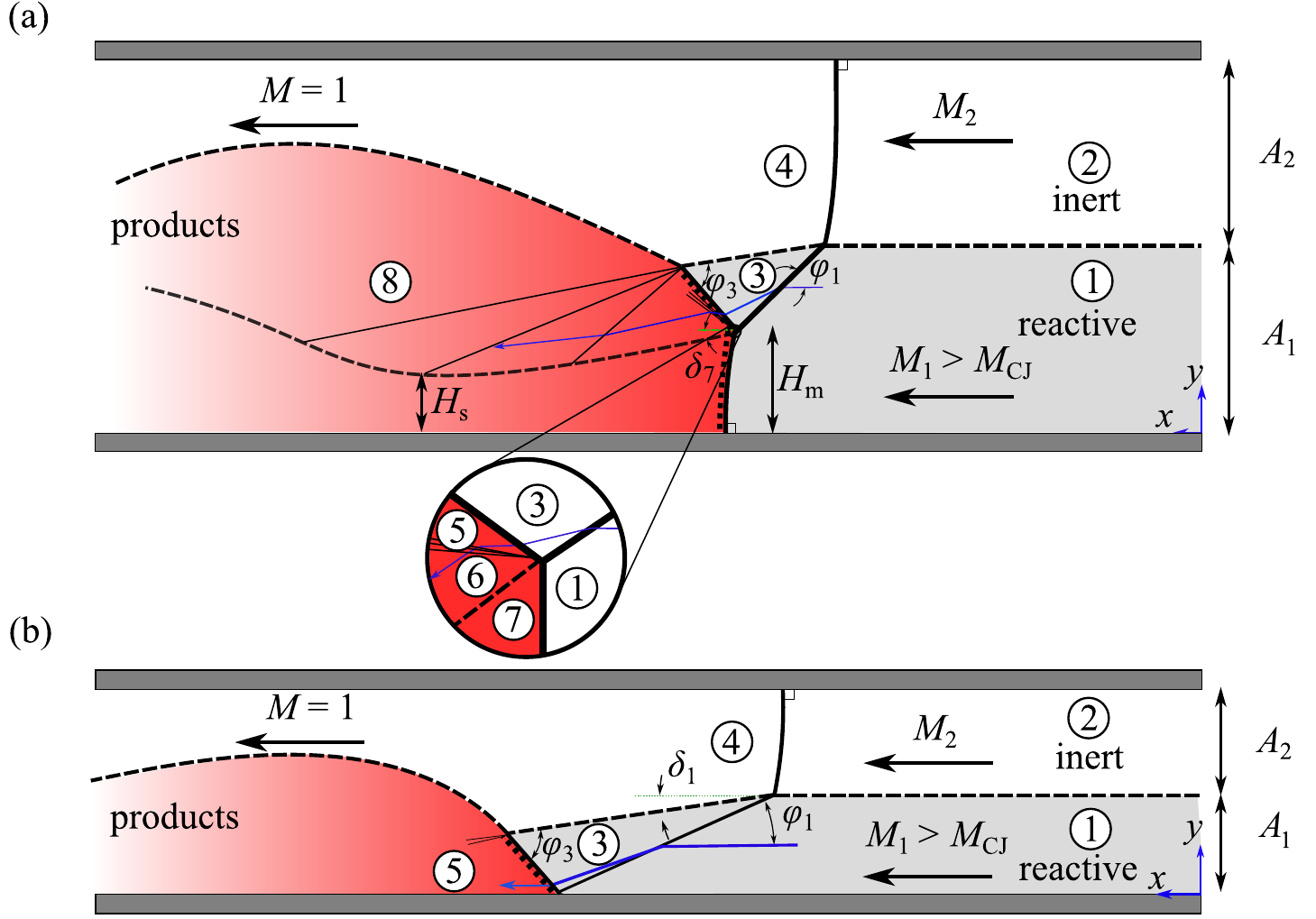}
\caption{Schematics of the observed flowfields when a precursor shock is present. Red-shaded regions past the black lines indicate reacted gas, while gray-shaded regions correspond to unreacted gas. (a) Reactive Mach reflection (Case C), in which an inert shock, an oblique detonation, and a detonative Mach stem intersect to form a triple point in the reactive layer. (b) Reactive regular reflection (Case D), in which an incident oblique shock reflects from the bottom wall as an oblique detonation within the reactive layer.}
\label{Overdriven}
\end{figure}

A precursor shock forms when the acoustic impedance ratio $Z$ is less than $Z_\mathrm{Kant}$ for a given area ratio, $A_2/A_1$. In this regime, the resulting wave structure in the reactive layer may take the form of either a Mach reflection or a regular reflection, which were introduced in Section~\ref{sample_results} as cases C and D, respectively. A regular reflection can be obtained by decreasing $Z$ or by simultaneously reducing $A_2/A_1$. A schematic of the observed wave structures is provided in Fig.~\ref{Overdriven}.  

In all cases, choking occurs in the inert layer. The primary distinction lies in whether a reactive Mach stem develops within the detonable layer. When it does, the Mach stem is overdriven and eventually curves such that the local curvature becomes negative ($\kappa < 0$).

\begin{figure}[H]
\centering
\includegraphics[width=0.63\textwidth]{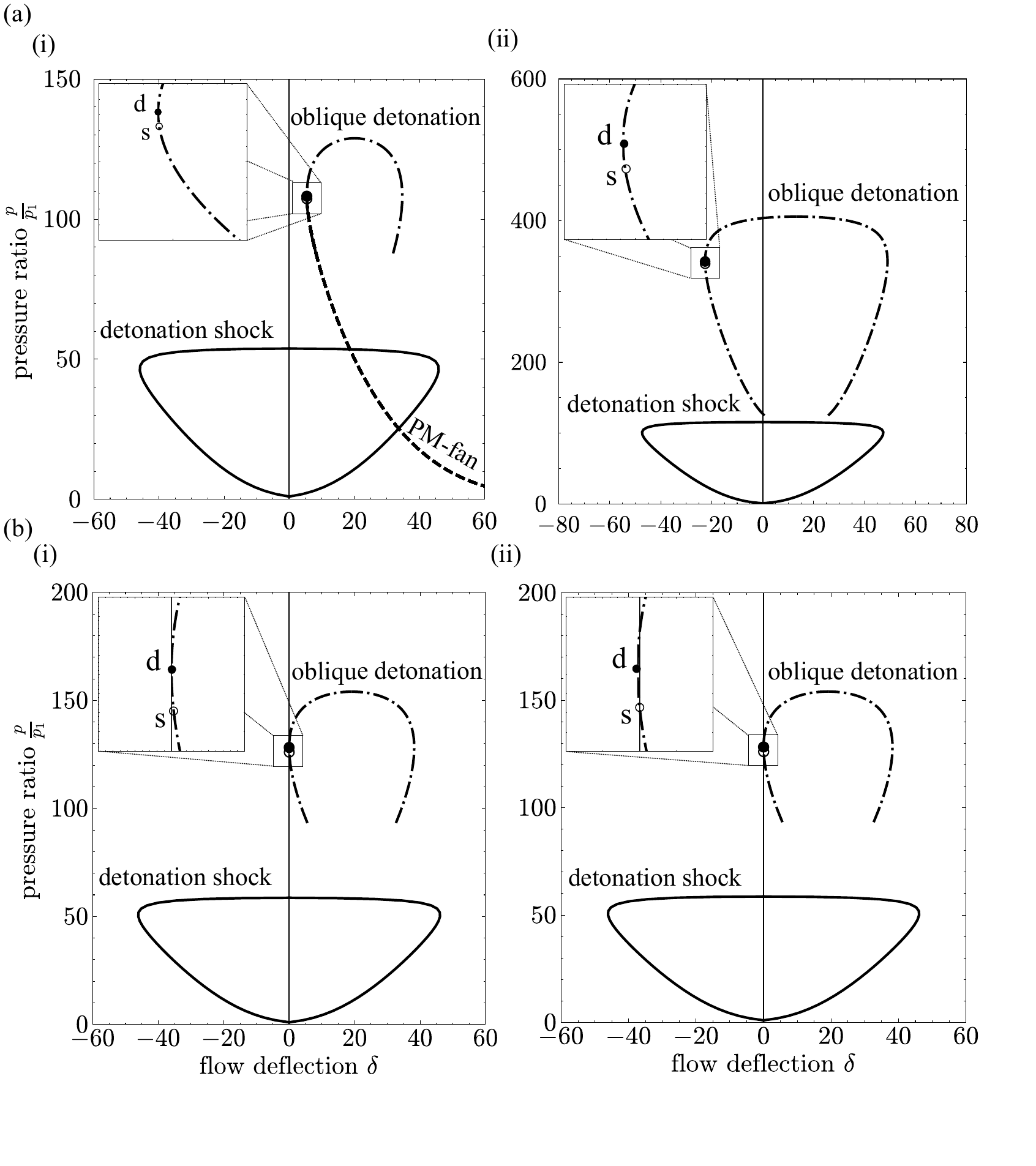}
\vspace{-20pt}
\caption{Shock-polar analysis of the RR–MR transition in the reactive layer for $A_2/A_1 = 1$. \textcolor{black}{A polar for the leading shock front of the overdriven detonation is drawn and labeled as ``detonation shock''.}  (a) (i) Reactive Mach reflection (Case C) at $Z = 0.45$; the oblique detonation polar does not intersect the $\delta = 0$ axis. \textcolor{black}{In this case, a Prandtl-Meyer fan originates from the triple point of the Mach reflection in order to expand the post-oblique detonation flow and match pressure and deflection with the detonation shock polar.} (ii) Reactive regular reflection (Case D) at $Z = 0.30$; \textcolor{black}{the oblique detonation polar intersects the $\delta = 0$ axis, resulting in a regular reflection structure in the reactive layer.} Transition between a reactive Mach reflection and a regular reflection depending on the detachment and sonic criteria (b) (i) $Z_\mathrm{detach(react)} = 0.4295$, (ii) $Z_\mathrm{sonic(react)} = 0.4294$.}
\label{RR_RMR_Transition}
\end{figure}

\vspace{-15pt}
The velocity associated with the configurations shown in Fig.~\ref{Overdriven} can be evaluated using the model of Mitrofanov~\cite{MITROFANOV1976995}. \textcolor{black}{This model treats each layer with independent quasi-one-dimensional conservation equations for mass, momentum, and energy, coupled through pressure equilibrium at a critical cross-section. The total cross-sectional area is conserved, but the reactive layer is free to expand at the expense of the inert layer. A sonic condition is imposed on the inert layer at the critical section, serving the same role as the Chapman–Jouguet condition in classical detonation theory. The model neglects heat and mass transfer between layers, so the coupling is entirely through pressure forces at the interface. Beyond thermodynamic properties, the required inputs are the acoustic impedance ratio between the inert and reactive layers and the ratio of their heights.} Although Mitrofanov’s model does not predict the terminal wave structure, this limitation can be overcome by applying a shock-polar analysis at the triple point in the reactive layer. The results of this polar analysis are presented in Fig.~\ref{RR_RMR_Transition}. Depending on the solution, the reactive layer may exhibit either a Mach reflection or a regular reflection. 
\vspace{-20pt}
In Fig.~\ref{RR_RMR_Transition}(a)(i), the oblique detonation polar fails to intersect the $\delta = 0$ axis, and a Mach reflection is obtained. The oblique detonation polar is constructed from knowledge of the wave-fixed Mach number $M_1$ and the incident shock angle $\phi_1$, where $\phi_1$ is determined from the relation $\sin \phi_1 = Z$. This relation arises by equating the pressures behind the precursor shock in the inert layer and the incident non-reactive shock in the reactive layer, together with the condition $M_2 = Z \cdot M_1$. The full derivation is provided in Appendix~\ref{Detonation_Geometric}.

Conversely, if the oblique detonation polar intersects the $\delta = 0$ axis, as shown in Fig.~\ref{RR_RMR_Transition}(a)(ii), a regular reflection occurs. When the intersection does not occur, the flow is instead closed by an expansion fan behind the weakly overdriven oblique detonation, which supplies the additional deflection and reduction in pressure needed to match the detonation shock.

Finally, the transition between Mach and regular reflection in the reactive layer is illustrated in Fig.~\ref{RR_RMR_Transition}(b). The detachment and sonic criteria specify the conditions under which the transition occurs, with the corresponding acoustic impedance ratios denoted as $Z_\mathrm{detach(react)}$ and $Z_\mathrm{sonic(react)}$. 
\newpage
\subsection{Flowchart of theoretical framework}
\begin{figure}[H]
\centering
\includegraphics[width=1.0\textwidth]{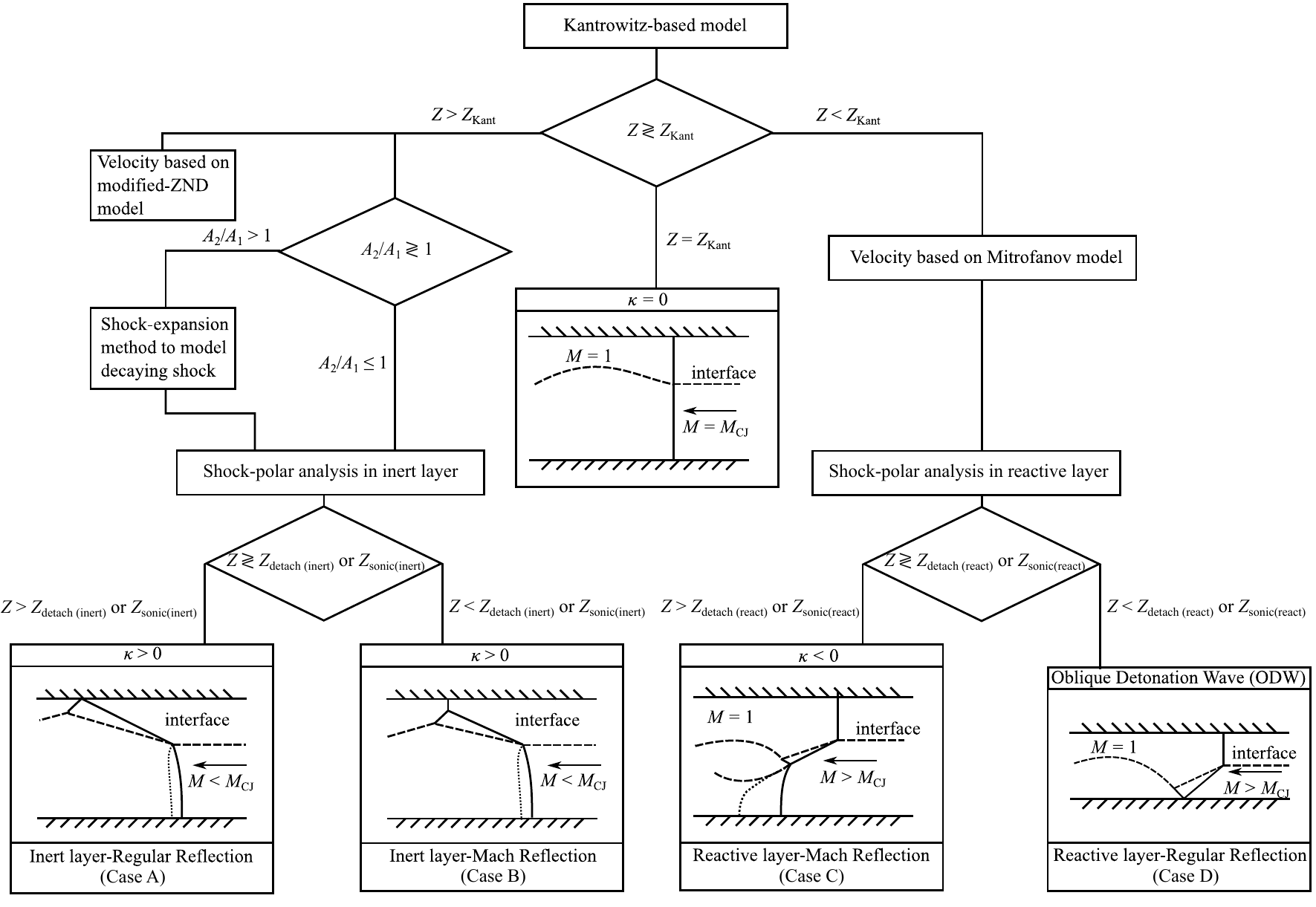}
\caption{Flowchart summarizing the criteria and analytical models used to determine the steady-state detonation structure. For the analysis presented, the area ratio $A_2/A_1$ is fixed, making the outcome dependent only on $Z$. The middle branch follows Section~\ref{Precusor_Onset} to establish the critical value $Z_\mathrm{Kant}$ for the onset of a precursor shock. Detonations with $Z > Z_\mathrm{Kant}$ are classified as underdriven (left branch), while those with $Z < Z_\mathrm{Kant}$ are overdriven (right branch). In the underdriven case, a shock-polar analysis is applied at the top wall, with the incident shock modeled as straight ($A_2/A_1 \leq 1$) or decaying ($A_2/A_1 > 1$). In the overdriven case, the wave structure is resolved by combining Mitrofanov’s model with polar analysis (Section~\ref{Overdriven_Cases}).}
\label{Flowchart}
\end{figure}

An overview of the theoretical modeling is provided in Fig.~\ref{Flowchart}. The diagram organizes the different detonation modes and indicates the sequence of analytical tools used to determine the overall steady-state wave structure as was described in this section.

\newpage
\section{Results}
\label{results}

\begin{figure}[H]
\centering
\includegraphics[width=1.0\textwidth]{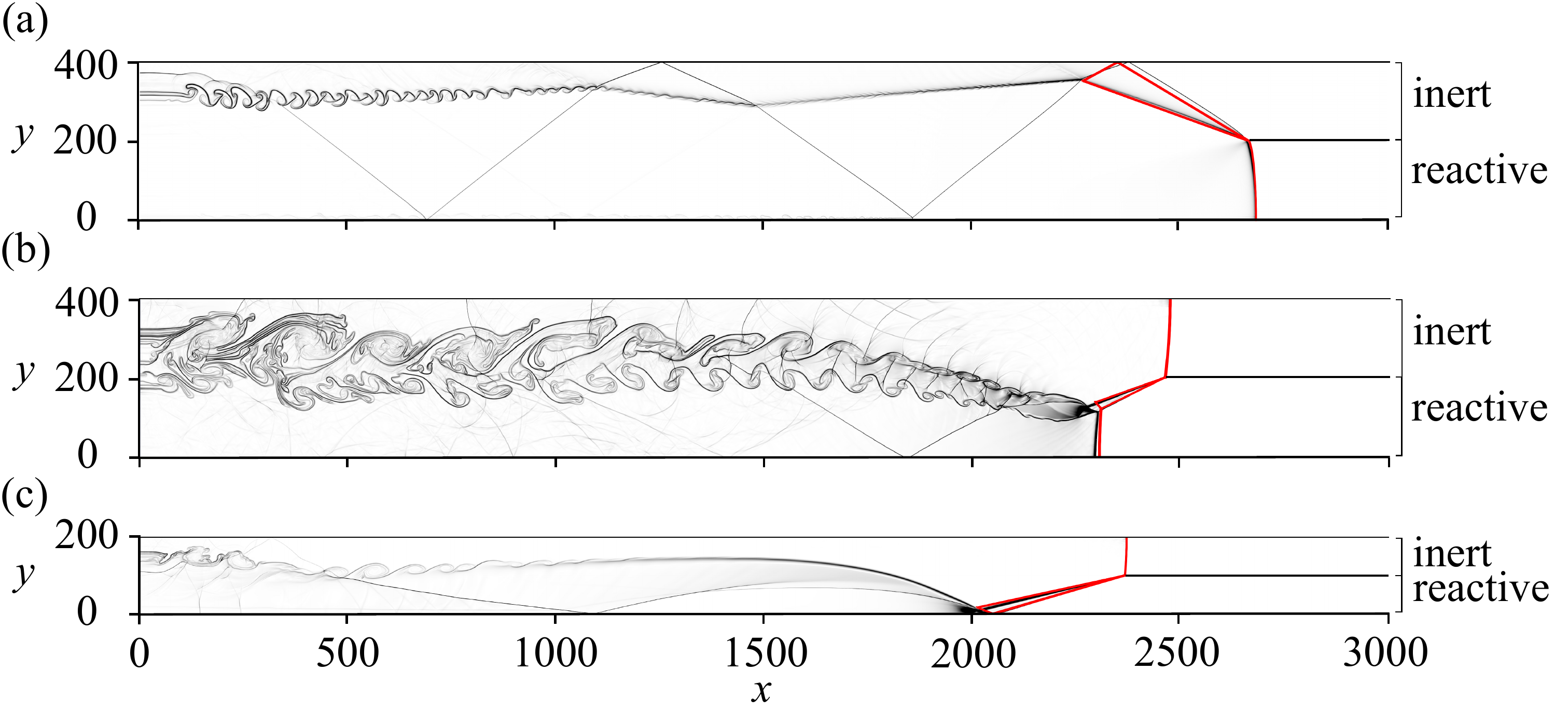}
\caption{Overlay comparisons, with red lines indicating the analytical solution, for representative Cases~A, C, and D (top to bottom). (a) $A_2/A_1=1$, $Z=0.80$; (b) $A_2/A_1=1$, $Z=0.45$; (c) $A_2/A_1=1$, $Z=0.30$.}
\label{Overlay_Comparisons}
\end{figure}

Using the theoretical framework introduced in Section~\ref{Theory}, the wave structure is analyzed by determining the angles of the shock interactions using the polar analysis. For sample Case A (Fig.~\ref{fig:mach_schlieren_A2A1_1}(a)(i)), the detonation--front curvature is modeled using an Eyring construction. Since $A_2/A_1 = 1$ for this case, the incident shock in the inert layer is taken to be straight, and the shock angles are obtained through a polar analysis, as illustrated in Fig.~\ref{fig:detonation_transition_inert}(a)(ii). For sample Case C (Fig.~\ref{fig:mach_schlieren_A2A1_1}(b)(i)), the orientation of the shock interaction at the triple point is determined from the polar analysis shown in Fig.~\ref{RR_RMR_Transition}(a)(i). These angles are then used in a set of geometric relations that provide the Mach-stem height, thereby introducing a length scale to the angles obtained from the polar analysis. The corresponding equations, along with the derivation of the precursor shock and Mach-stem curvature, are given in Appendix~\ref{Detonation_Geometric}. For sample Case D (Fig.~\ref{fig:mach_schlieren_A2A1_1}(b)(ii)), the inclination of the shock and reflected oblique detonation is determined from the intersection of the oblique detonation polar with the $\delta = 0$ axis, as shown in Fig.~\ref{RR_RMR_Transition}(a)(ii). In all cases, overlays of the analytical solutions are compared against numerical schlieren from the simulations, as presented in Fig.~\ref{Overlay_Comparisons}.

\begin{figure}[H]
\centering
\includegraphics[width=0.45\textwidth]{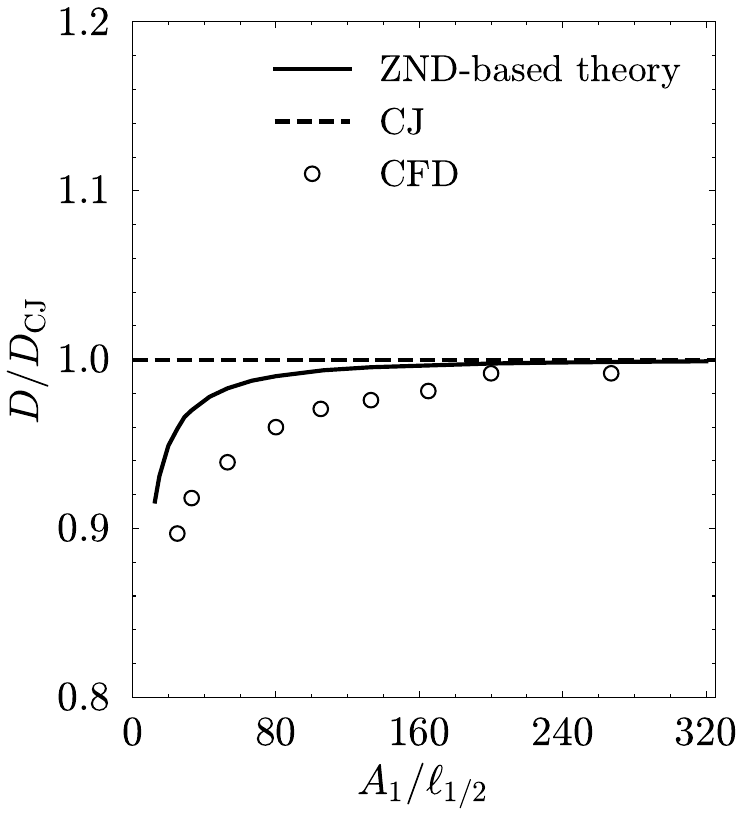}
\caption{ Comparison of velocity deficit predicted by the \textcolor{black}{curvature-based ZND formulation (Wood–Kirkwood closure applied to the Eyring-constructed front)} with velocity deficits obtained from CFD simulations (unfilled circles) for different reactive-layer thicknesses ($A_1 = 25, 33, 53, 80, 105, 133, 165, 200, 267$) at $Z = 0.70$.}
\label{ZND_CFD}
\end{figure}

With the detonation-front curvature constructed via Eyring’s model, the Wood–Kirkwood $D_{n}-\kappa$ relation  \cite{Wood_Kirkwood} yields the velocity deficit. \textcolor{black}{The Wood–Kirkwood formulation reduces the two-dimensional curved detonation to a quasi-one-dimensional reaction-zone problem by incorporating curvature-induced lateral divergence into the governing equations.}
Figure~\ref{ZND_CFD} compares the theoretical predictions with the CFD results.

\FloatBarrier

\begin{figure}[H]
\centering
\includegraphics[width=0.45\textwidth]{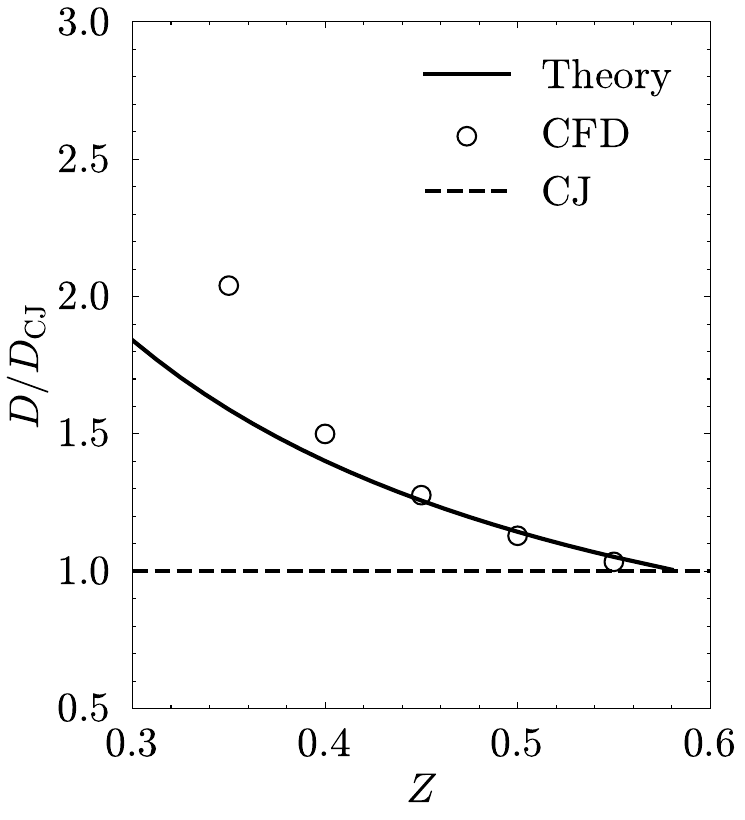}
\caption{Velocity increase predicted by Mitrofanov theory (solid line) compared with CFD simulation results (unfilled circles) for $A_2/A_1 = 1$. The comparison is shown for $Z$ values up to the point where the CJ velocity is attained.}
\label{Mitrofanov_CFD}
\end{figure}

For the overdriven cases ($Z < Z_\mathrm{Kant}$), the comparison between Mitrofanov's wave-velocity model against the numerical simulations are shown in Fig.~\ref{Mitrofanov_CFD} for $A_2/A_1 = 1$.

\begin{figure}[H]
\centering
\includegraphics[width=0.8\textwidth]{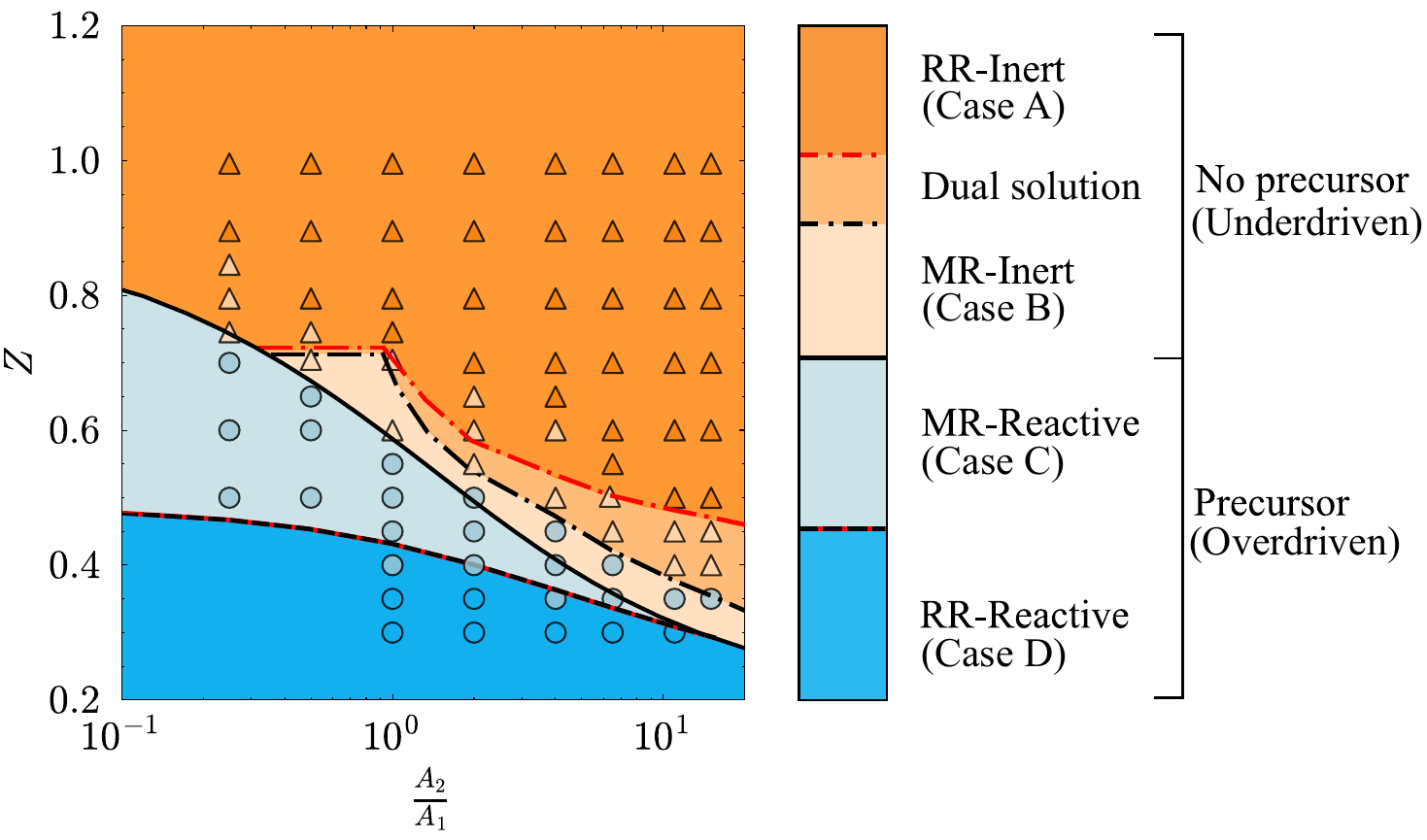}
\caption{Phase map of the different wave structures observed. CFD results are shown as symbols, with triangles ($\triangle$) denoting no precursor and circles ($\circ$) denoting precursor. Symbol shading denotes Mach or
regular reflection. Theoretical/analytical predictions are denoted with shaded regions. Solid black line is Kantrowitz
criterion. Broken lines are regular--Mach reflection transition, with red and black used for sonic and detachment criterion, respectively.}
\label{phase_map}
\end{figure}

\begin{figure}[H]
\centering
\includegraphics[width=1.0\textwidth]{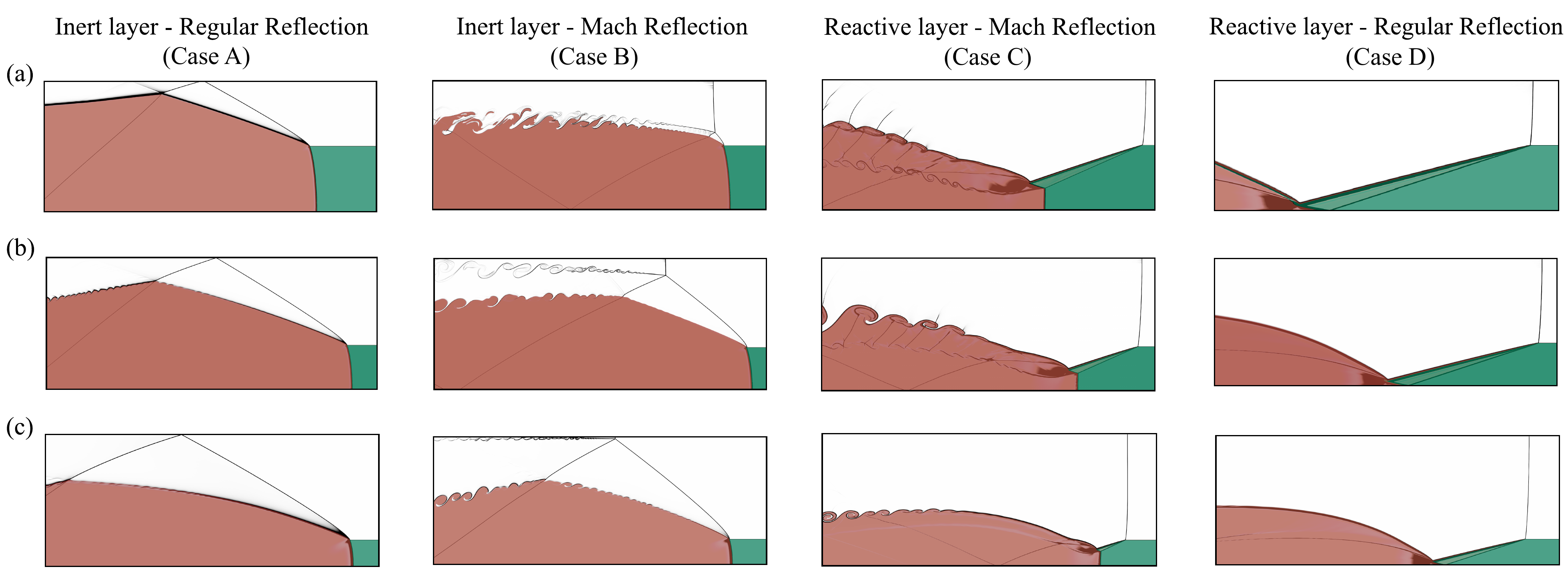}
\caption{Sample observed wave structures for various $A_2/A_1$ and $Z$ values. Rows (a), (b), and (c) correspond to $A_2/A_1 = 1, 2,$ and $4$, respectively. The columns correspond to $Z = 0.80$, $0.60$, $0.40$, and $0.30$, illustrating cases of regular reflection (inert), Mach reflection (inert), Mach reflection (reactive), and regular reflection (reactive), respectively. The figures represent overlays of $\lambda$ with schlieren with green, red, and white representing the unburnt reactive gas, detonation products, and inert gas, respectively.}
\label{mosaic}
\end{figure}

A phase map of the $Z$ and $A_2/A_1$ parameter space summarizes the wave-structure outcomes, with analytically predicted regions shaded and CFD results shown as symbols (Fig.~\ref{phase_map}). Both the sonic and detachment transition criteria are indicated. Between these two boundaries, \textcolor{black}{a \emph{dual-solution} region exists in which both regular and Mach reflection configurations are theoretically admissible, depending on whether the transition is governed by the sonic criterion or the detachment criterion. In this regime, the sonic condition predicts transition earlier than the detachment condition.} A visual catalog of sample structures across the parameter space is provided in Fig.~\ref{mosaic}.
\vspace{-20pt}

\section{Interpretation and Discussion}
\label{Discussion_Interpretattion}

\subsection{Overview of results}

The results show that confinement impedance and relative layer thicknesses are key in determining whether a precursor shock forms ahead of the detonation or an attached shock persists within the inert layer. When the acoustic impedance of the inert layer is sufficiently low, a precursor shock forms and propagates within the inert gas, with thinner inert layers favoring its formation. In contrast, a high-impedance inert layer with greater height tends to support a shock wave that remains attached to the detonation, which then interacts with the upper boundary, leading to regular or Mach reflection. 

Analytical comparison with CFD was facilitated by adopting a linear reaction rate, thereby isolating the effects of complex cellular dynamics from $Z$ and $A_2/A_1$. Variables were nondimensionalized by the half-reaction-zone length of the CJ detonation, so the cases correspond to reactive layers that are thick in comparison to the reaction-zone length. In this regime, the detonation behaves as a hydrodynamic wave, and the finite thickness of the reaction zone influences the outcome only for the very thinnest reactive layers. The flow was modeled using quasi-one-dimensional assumptions and shock-polar analysis, which showed qualitative agreement with CFD. The linear reaction rate removes the strong temperature sensitivity of Arrhenius kinetics, yielding a laminar detonation structure and eliminating quenching behavior.

While it is well established from experiments \cite{RADULESCU200229, KNYSTAUTAS198263, annurev:/content/journals/10.1146/annurev.fl.16.010184.001523} and numerical simulations \cite{HAN2018295, HOUIM2017185, Reynaud2020-rx} that detonations subject to loss mechanisms exhibit \emph{detonability limits}, a consequence of the reaction model in this study was that no detonation failures were observed, even for the thinnest layers, as shown in Fig.~\ref{phase_map}. For Arrhenius-governed kinetics, confinement-induced curvature weakens the leading shock and generates nonuniform post-shock temperatures, potentially resulting in shock–reaction decoupling and eventual failure. In ZND theory for curved detonations, this appears as a turning point in the $D_n-\kappa$ relation, indicating a critical curvature and corresponding layer thickness below which sustained propagation is not possible. In this study, the $D_n-\kappa$ relation yields no such turning point. This is consistent with CFD results that show sustained propagation even at large $A_2/A_1$, corresponding to very thin layers. A comparison of $D_n-\kappa$ relations for a linear and Arrhenius rate is provided in Fig.~\ref{Criticality} of Appendix~\ref{Eyring_Appendix}. This absence of critical behavior extends beyond the present linear-rate formulation: Mi et al. \cite{mi_effect_2018} demonstrated that even relatively low activation energies ($E_a/RT_1 = 10$) in two-dimensional simulations of detonations under yielding confinement still failed to produce a detectable critical layer thickness.

While kinetic effects influence overall detonation stability, the emergence of a precursor or attached shock is governed primarily by the confinement properties. In all cases where a precursor was obtained, the sound speed in the inert layer remains lower than the detonation speed, so the precursor propagates as a true shock rather than the shockless (acoustic) precursor wave that forms in other confinement configurations \cite{Sharpe2006-xc, JACKSON20112219}. 

The location of the sonic locus also distinguishes whether the inert shock is a precursor or a detached shock. In detached shocks, such as those ahead of a blunt body, the sonic line originates at the shock and separates subsonic and supersonic regions. In contrast, for a precursor shock, the sonic locus is spatially separated from the shock. The present results show that only configurations exhibiting a precursor shock lead to overdriven detonation behavior. Examples of detached shocks observed in this study are shown in Appendix~\ref{Detached_Precursor_Comparison} and similar detached-shock structures also appear in the results of Li et al. \cite{LiMiHiggins} for a case with an inert confiner of lower density than the reactive mixture.

The flowfield associated with a precursor shock is shown in Fig.~\ref{Overlay_Comparisons}(b). The formation of the precursor generates a transmitted oblique shock that precompresses the unreacted gas ahead of the detonation front. The simulations were designed such that this transmitted shock did not prematurely ignite the mixture, preserving behavior reported in prior experimental \cite{CHEEVERS20233095} and numerical \cite{HOUIM2017185, Reynaud2020-rx} studies. The resulting structure produces a triple-point interaction between the detonation, the transmitted shock, and an oblique detonation wave, causing the detonation front to curve toward the reactants. This curvature influences the sonic locus, which originates at the triple point and bends to form a throat-like structure behind the detonation, as shown in Fig.~\ref{fig:detonation_structures}(b). The region between the sonic locus and the front behaves as an effective contraction, enabling a quasi-one-dimensional interpretation of the flow and its relation to the overdriven speed \cite{Li_Yang_Wang_2025}.

The phase map, Fig.~\ref{phase_map}, also shows that, in the reactive layer, the detachment and sonic criteria are practically indistinguishable, whereas in the inert layer they differ appreciably. This reflects the structure of the corresponding polars: on the oblique-detonation polar (Fig.~\ref{RR_RMR_Transition}), the maximum-deflection and sonic points are nearly coincident, with $Z_\mathrm{detach(react)}-Z_\mathrm{sonic(react)} \approx 0.0001$, while on the reflected-shock polar (Fig.~\ref{fig:detonation_transition_inert}) they are more clearly separated, yielding $Z_\mathrm{sonic(inert)}-Z_\mathrm{detach(inert)} \approx 0.02$, leading to a more pronounced dual‑solution region between Mach and regular reflection in the inert layer.

Finally, while precursor shock formation is discussed here in a reactive context, similar behavior arises in inert multiphase problems. An example is the \emph{triple point problem} \cite{BILLAUDFRIESS2014488, PAN2018870}, which is used to assess CFD capability in capturing shocks and interfaces. In this configuration, the computational domain is partitioned into three regions with distinct phases, typically featuring a low-density (lower-impedance) region adjacent to a denser gas. The interaction of a propagating shock with this stratification produces a precursor shock in the low-density region, similar to that observed in Case C, which subsequently drives a Mach reflection in the denser gas. This comparison highlights that the observed precursor phenomena are not unique to reactive detonation systems but arise from general compressible flow physics whenever a strong acoustic impedance mismatch exists.

\subsection{Limitations of modeling}
Simple analytic or semi-analytic methods cannot be expected to fully reproduce the flowfield observed in computational simulations.  The models developed in this paper rely on assumptions whose validity varies across the parameter space, and the main sources of discrepancy are discussed below.

The most consequential assumption is that of a planar, normal CJ detonation in the polar analysis of the underdriven regime (Sections~\ref{Straight_Incident_Shock} and~\ref{Decaying_Incident_Shock}) and in the prediction of precursor onset. In practice, confinement-induced divergence within the reaction zone causes the detonation front to curve, and this curvature is neglected by the planar assumption. Accordingly, the Kantrowitz criterion performs best for $A_2/A_1 < 4$, as shown in Fig.~\ref{phase_map}, where thicker reactive layers yield smaller velocity deficits relative to the CJ speed and thus make the planar CJ assumption more accurate. At larger values of $A_2/A_1$, the increasing velocity deficit creates greater departures from the CJ state, and the criterion fails to capture several cases in the phase map at $A_2/A_1 > 4$ in which a precursor was nevertheless observed.

A further limitation of the present analysis arises from the thickening of the slipstream generated by the detonation--inert gas interaction. Downstream of the shock--detonation complex, this slipstream is amplified by Kelvin--Helmholtz instabilities, leading to the formation of vortical shear-layer structures. As these instabilities grow, the flow can no longer be described adequately by a simple quasi-one-dimensional model. This is particularly relevant to the velocity-based prediction of the overdriven detonation using the Mitrofanov model, which does not account for the downstream growth of these instabilities. In addition, within an Eulerian calculation, the evolution of the vortical shear-layer structures is not analytically tractable owing to the inherently grid-dependent numerical diffusion \cite{RaviPullin}. As a result, the velocity deviations observed in Fig.~\ref{Mitrofanov_CFD} at low $Z$ values, where the detonation is strongly overdriven, may be attributed to instability growth and mixing effects that are difficult to capture within the present framework.

\subsection{Implication for rotating detonation engines (RDEs)}

\noindent For the problem considered in this study, the precursor-shock state is driven by choking of the inert layer made possible by a confining upper wall. In an RDE, the upper boundary is an exhaust rather than a wall, and the lower boundary is the injector supplying the fuel–oxidizer mixture; the injection imparts an axial velocity component that tilts the detonation front. Without an upper wall, the inert-layer choking mechanism that sustains a precursor-driven overdrive is absent, so the overdriven behavior reported here is unlikely to be encountered in an RDE. Detached shocks similar to Case E may still occur; however, as discussed in Appendix~\ref{Detached_Precursor_Comparison}, such configurations are consistent with an underdriven detonation.

\section{Conclusion}
\label{conclusion}

A detonation weakly confined in a two-layer system of a reactive gas bound by an inert gas can either exhibit underdriven or overdriven behavior, depending on the acoustic impedance ratio $Z$ and the area ratio $A_2/A_1$ between both layers. This paper proposes theoretical approaches to comprehensively link the roles that $Z$ and $A_2/A_1$ play in determining which type of detonation behavior is observed. A set of analytical tools, primarily grounded in traditional compressible flow theory, is used to predict the wave speed, front curvature, and resulting wave structure in each regime. The predictions are evaluated through \emph{a posteriori} comparisons of analytic predictions with numerical simulations, and a phase map delineating four distinct wave-structure regions in the ($A_2/A_1$, $Z$) parameter space is constructed to elucidate the scope and limitations of the analytical models.

A Kantrowitz-type onset criterion, $Z_\mathrm{Kant}$, was formulated to predict whether or not a precursor shock forms in the inert layer as a function of the area ratio $A_2/A_1$. This criterion accounts for the expansion of the detonation products of a CJ detonation into the inert layer, which reduces the available flow area until the inert layer chokes and a shock is forced ahead of the detonation front. Cases with $Z > Z_\mathrm{Kant}$ remain underdriven (no precursor) and exhibit positive curvature, whereas cases with $Z < Z_\mathrm{Kant}$ yield overdriven detonations with negative curvature.

In the underdriven regime, the detonation velocity was theoretically predicted using the Wood-Kirkwood $D_n-\kappa$ relation combined with an Eyring geometric method to construct the curved detonation front. The front curvature, and therefore the velocity deficit, was found to depend almost entirely on $A_2/A_1$ and to be essentially independent of $Z$, a consequence of the sonic locus attaching directly to the detonation shock front. The trailing shock in the inert layer was classified through shock-polar analysis: for $A_2/A_1 \leq 1$, the oblique shock was treated as straight, while for $A_2/A_1 > 1$, its decay was modeled using a characteristic-based shock-expansion method that provides the shock angle at the upper wall, from which the type of reflection, regular or Mach, is determined.

In the overdriven regime, Mitrofanov's two-layer conservation model supplies the wave speed, and a polar analysis at the triple point is used to classify the reactive-layer interaction as either a Mach reflection with a detonative Mach stem or a regular reflection in which an oblique shock reflects off the lower wall as an oblique detonation. Geometric parameters derived from the polar solution yield the Mach-stem height, enabling a complete analytical reconstruction of the wave structure that closely reproduces the CFD results.

  The resulting phase map shows qualitative agreement with the CFD results across the parameter space. Where deviations arise, they are primarily attributable to features, such as slipstream instabilities and strong two-dimensional effects, that break the quasi-one-dimensional assumptions underlying the models.

\section{Acknowledgments}
This work was supported by Fonds de recherche du Québec – Nature et technologies (No. 328741). The authors gratefully acknowledge Digital Alliance Canada for providing computing resources.

\newcounter{figsave}
\setcounter{figsave}{\value{figure}}

\newcounter{tabsave}
\setcounter{tabsave}{\value{table}}

\newcounter{eqnsave}
\setcounter{eqnsave}{\value{equation}}

\begin{appendices}

\renewcommand{\thesection}{\arabic{section}}
\setcounter{section}{0} 

\setcounter{figure}{\value{figsave}}
\setcounter{table}{\value{tabsave}}
\setcounter{equation}{\value{eqnsave}}

\renewcommand{\thefigure}{\arabic{figure}}
\renewcommand{\thetable}{\arabic{table}}
\renewcommand{\theequation}{\arabic{equation}}

\section{Verification of results reaching steady~state}
\label{Appendi_steady_state}

\textcolor{black}{This appendix provides a verification that the wave structures and propagation velocities reported in Section~\ref{results} correspond to terminal steady-state solutions rather than transient behavior.} To confirm that the wave structures and propagation velocities reported in Section~\ref{results} have reached a terminal steady state, a shock-tracking method is implemented along the top and bottom boundaries. This approach allows for the calculation of the instantaneous velocities of the shock (top boundary) and detonation (bottom boundary), which are compared over time. A steady state is deemed achieved once both velocities converge and remain constant for the remainder of the simulation; the average propagation speed is then computed as \(D_{\mathrm{avg}} = \frac{x_\mathrm{final} - x_{0}}{t_{\mathrm{final}} - t_{0}}\), where \(x_{0}\) denotes the position at which the detonation has advanced sufficiently far from the initiation transient (here taken as \(10{,}000\)~$l_\mathrm{1/2}$), while 
\(x_{\mathrm{final}}\) is the final position through which the average velocity is 
taken (here taken as \(20{,}000\)~$l_\mathrm{1/2}$). The times \(t_{0}\) and \(t_{\mathrm{final}}\) correspond to the instants at which the detonation reaches \(x_{0}\) and \(x_{\mathrm{final}}\), respectively, and thus vary from case to case.

The shock-tracking method is based on detecting pressure jumps relative to an undisturbed reference pressure in each layer. For example, letting $p_1$ denote the undisturbed pressure in the bottom (reactive) layer at $j = 0$, and $p_2$ represent the undisturbed pressure in the top (inert) layer at $j = j_{\text{max}}$, where typically $p_1 = p_2$. A shock is identified when, for a given horizontal index $i$, the pressure satisfies
\[
p(i, j_{\text{max}}) > p_2 + \varepsilon \quad \text{and} \quad p(i, 0) > p_1 + \varepsilon,
\]
with $\varepsilon = 10^{-3}$ acting as the pressure threshold for detecting discontinuities. Tracking the spatial index $i$ where these conditions are first satisfied allows for measuring the propagation speed of both the shock and the detonation front. 

\begin{figure}[H]
\centering
\includegraphics[width=1\textwidth]{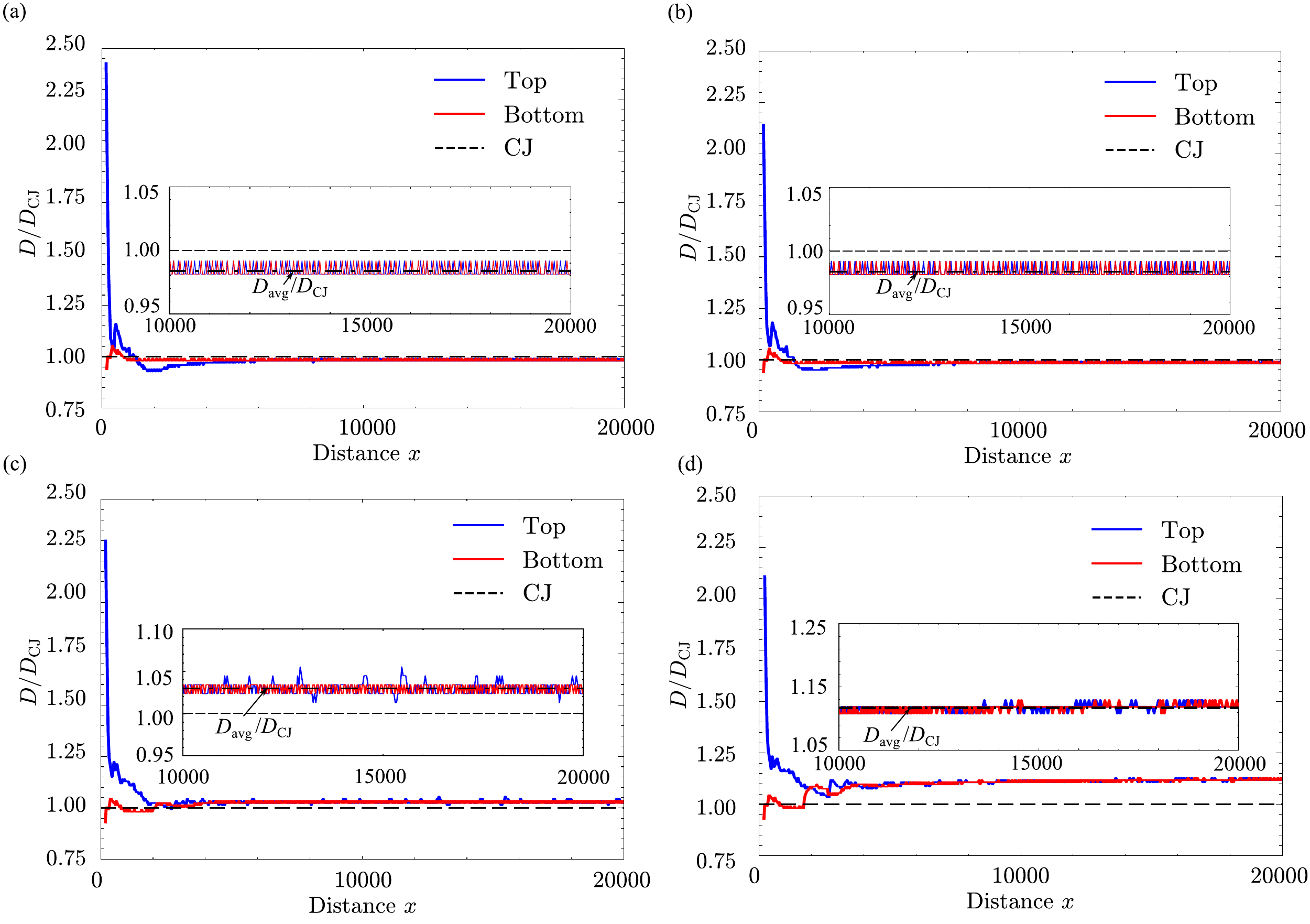}
\vspace{-10pt}
\caption{Velocity histories tracked along the top and bottom boundaries for 
$A_2/A_1 = 1$ and (a) $Z = 0.75$, (b) $Z = 0.70$, (c) $Z = 0.55$, and (d) $Z = 0.50$. 
(a) and (b) correspond to underdriven detonations, where an attached shock trails 
behind the detonation front, while (c) and (d) correspond to overdriven detonations, characterized by the formation of a precursor shock. The average velocity is indicated in the inset, taken at a sufficient distance downstream of the initiation region.}

\label{velocity_histories}
\end{figure}

Figure~\ref{velocity_histories} presents the velocity histories for four different cases with an area ratio of $\frac{A_2}{A_1} = 1$ in the lab-fixed reference frame. \textcolor{black}{Velocity histories throughout the simulation are provided for cases with different detonation front curvatures, $\kappa$, with $\kappa > 0$ corresponding to the attached shock scenario presented in the main text while $\kappa < 0$ corresponding to the precursor-shock driven detonations.} Figures~\ref{velocity_histories}(a) and~\ref{velocity_histories}(b) correspond to cases where $\kappa > 0$, indicating underdriven detonations. In contrast, Figs.~\ref{velocity_histories}(c) and~\ref{velocity_histories}(d) show cases with $\kappa < 0$, corresponding to overdriven detonations.

\vspace{-20pt}
The findings in this appendix verify that the wave structures and velocities reported in the main text achieved their terminal steady-state configurations.
\newpage
\section{Influence of initial conditions on results}
\label{Initial_conditions_test}
This appendix examines whether the method used to initiate the simulation affects the final wave configuration.
\textcolor{black}{Two different initialization methods were employed to address limiting cases in which a precursor shock is either present or absent. Both approaches demonstrate that the resulting detonation front–shock wave structure is independent of the initialization method, yielding identical results.}
Phenomena such as supersonic-inlet starting and the transition to Mach reflection in steady flows are known to exhibit \emph{hysteresis}, where identical geometries and inflow conditions yield different flowfields depending on the initialization \cite{BENDOR2002347}. To assess whether a similar sensitivity exists in the present problem, two dissimilar methods of initializing the density profile in the inert layer were employed. These approaches explore whether carefully controlled acoustic impedance gradients can suppress or promote precursor shock formation.

Specifically, for cases where precursor formation was previously observed (e.g., at low $Z$), the modified setup introduces a high acoustic impedance region directly above the ignition kernel to suppress this behavior. Conversely, for cases where no precursor was present (e.g., at high $Z$), a low impedance layer is imposed above the ignition kernel to encourage precursor development. The gradual transition in geometry was chosen to minimize artificial disturbances and deliberately invert the previously observed behavior, thereby revealing any hysteresis in the system's response to different initializations.

\begin{figure}[H]
\centering
\includegraphics[width=0.75\textwidth]{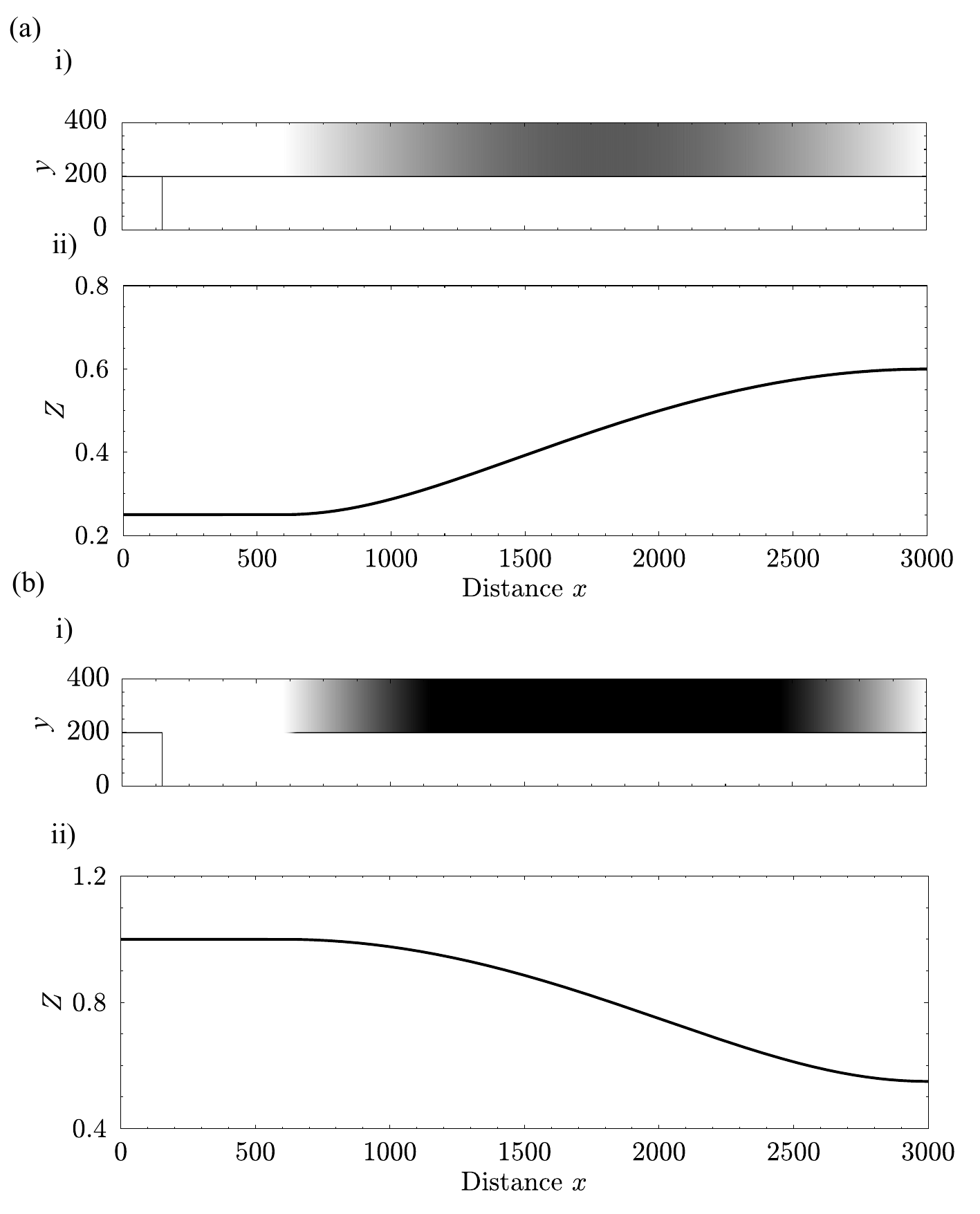}
\caption{Comparison of gradual initiation approaches for two impedance ratios: (a) $Z = 0.60$, where a low-impedance region is introduced above the ignition kernel to promote precursor formation; and (b) $Z = 0.55$, where a high-impedance region is imposed to attempt precursor suppression. For each case, (i) the schlieren visualization of the initiation setup and (ii) the corresponding acoustic impedance variation in the inert region are shown.}
\label{initiation_setup}
\end{figure}

As shown in Fig.~\ref{initiation_setup}, case (a) corresponds to a configuration with $Z = 0.60$, in which no precursor shock forms under the original conditions. In the adjusted setup (a)(i), a low-impedance region is initialized above the ignition kernel to test whether precursor formation can be induced. Conversely, case (b) tests a $Z = 0.55$ configuration where a precursor naturally forms; in response, (b)(i) modifies the setup by initializing a high-impedance inert layer above the kernel to attempt suppression of precursor formation.

This control over the inert region is implemented via a piecewise density distribution. A constant high or low-density region is first assigned over a uniform area of width 600$l_{1/2}$, beyond which a smooth transition to the target density is done through a cosine bell taper. The cosine profile ensures a smooth gradient in acoustic impedance and also has a zero gradient at both ends. The cosine taper is defined over the interval $x \in [x_0, x_1]$ as:

\begin{equation}
\rho(x) =
\begin{cases}
\rho_1 + \frac{1}{2}(\rho_0 - \rho_1)\left[1 + \cos\left( \pi \frac{x - x_0}{x_1 - x_0} \right)\right], & x_0 \leq x \leq x_1 \\
\rho_0, & x < x_0 \\
\rho_1, & x > x_1
\end{cases}
\end{equation}

\noindent
Here, $\rho_0$ is the density in the uniform region above the ignition zone (e.g., 0.0625), and $\rho_1$ is the final density corresponding to the desired impedance ratio. The taper smoothly varies the density from $\rho_0$ to $\rho_1$ over the region $[x_0, x_1]$, which spans from the end of the uniform high-impedance region to the end of the inert layer.

\vspace{-20pt}

\begin{figure}[H]
\centering
\includegraphics[width=1\textwidth]{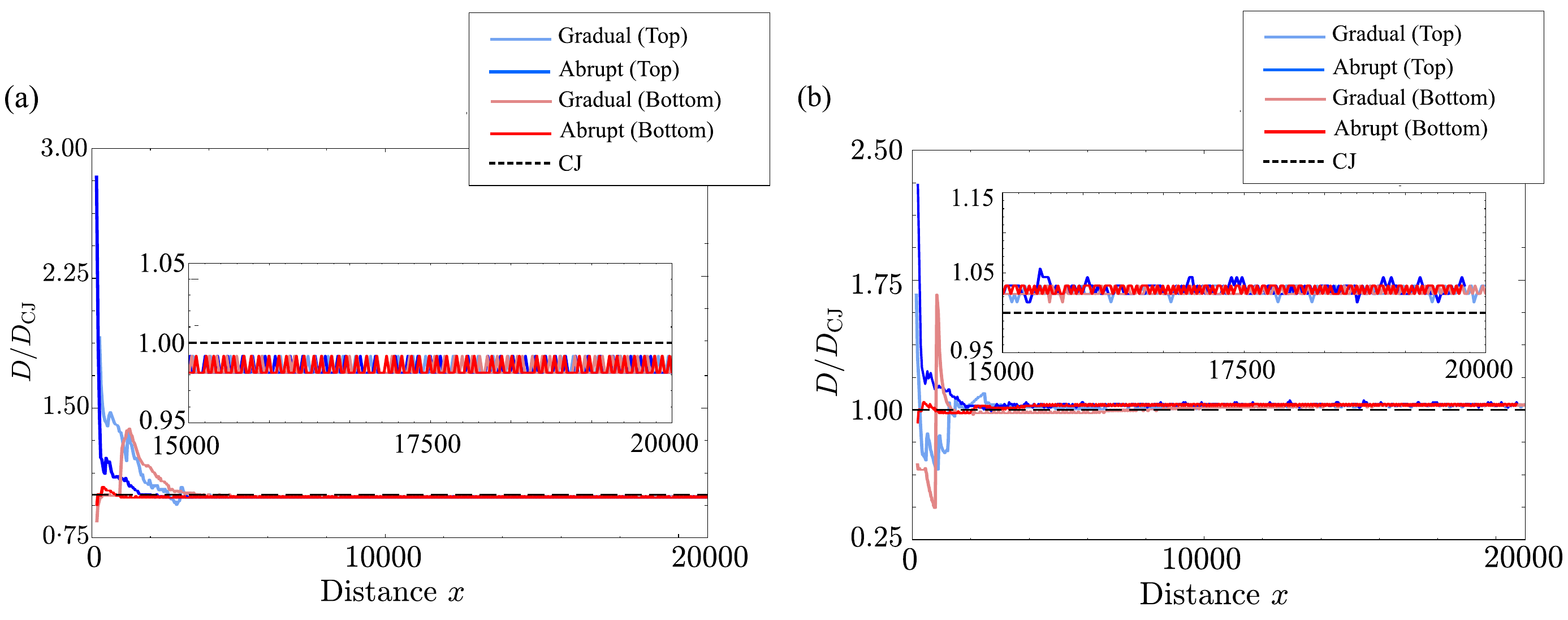}
\caption{Velocity histories for $A_2/A_1 = 1$ test cases. The top boundary is shown in blue and the bottom boundary in red. Gradual transition cases are plotted with slight transparency, while abrupt initiation cases are shown without transparency: (a) $Z = 0.60$, (b) $Z = 0.55$.}
\label{velocity_cosine}
\end{figure}

Figure~\ref{velocity_cosine} shows the evolution of the velocity recorded using the top and bottom shock trackers for (a) $Z$ = 0.60 and (b) $Z$ = 0.55. Thus, the different initiation mechanisms lead to different transient behavior early in the simulation but eventually relax to the same terminal velocity regardless of whether the transition to the confining wall is gradual or abrupt.

\begin{figure}[H]
\centering
\includegraphics[width=1\textwidth]{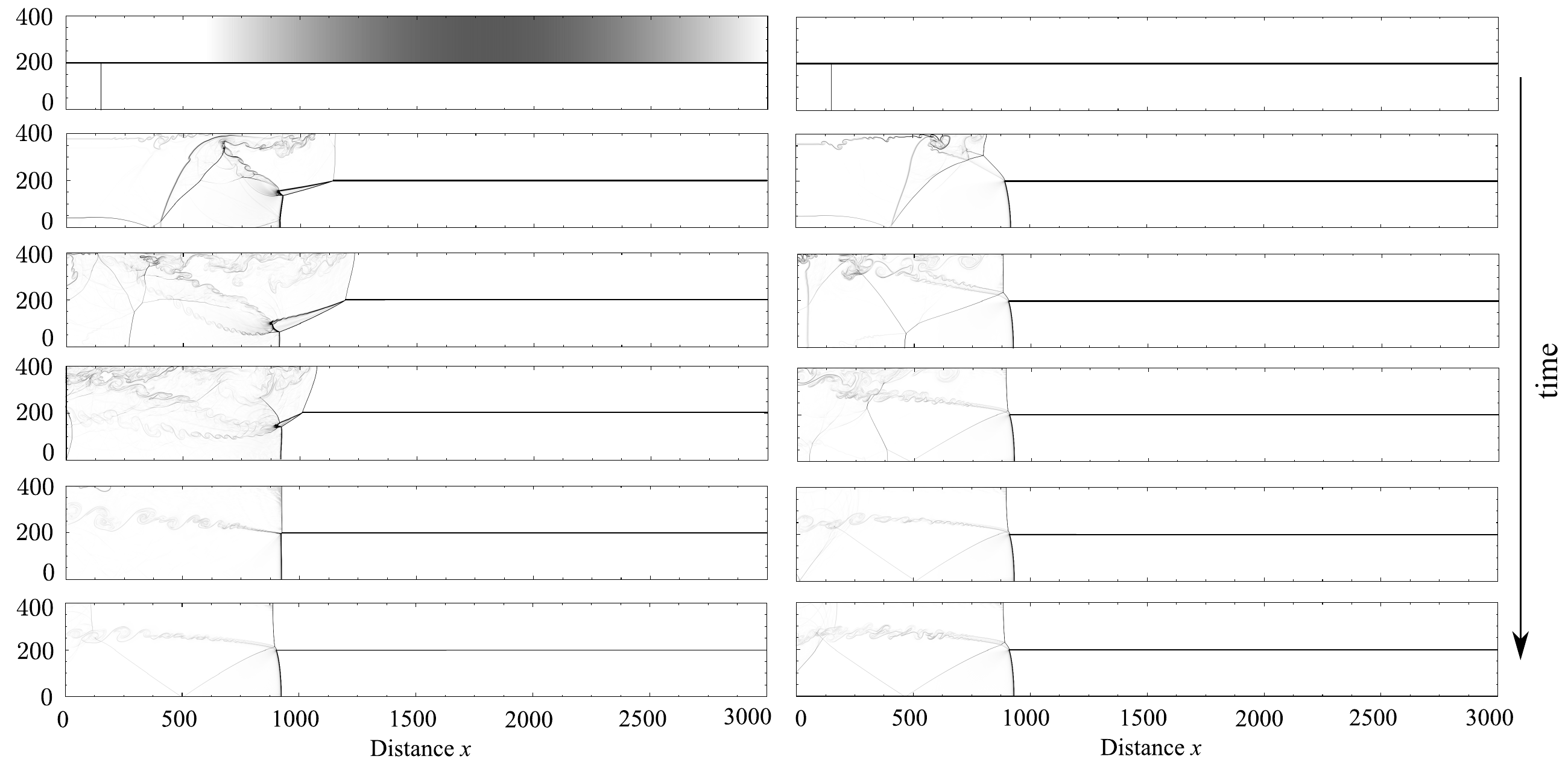}
\caption{Evolution of wave-structure relaxation for $Z = 0.60$. The left column shows gradual transition to the confining layer, while the right column shows abrupt transition to the confining layer. Rows correspond to identical numerical timeframes, with time progressing from top to bottom as indicated by the arrow.}
\label{Z0.60_Initiation_Test}
\end{figure}

\begin{figure}[H]
\centering
\includegraphics[width=1\textwidth]{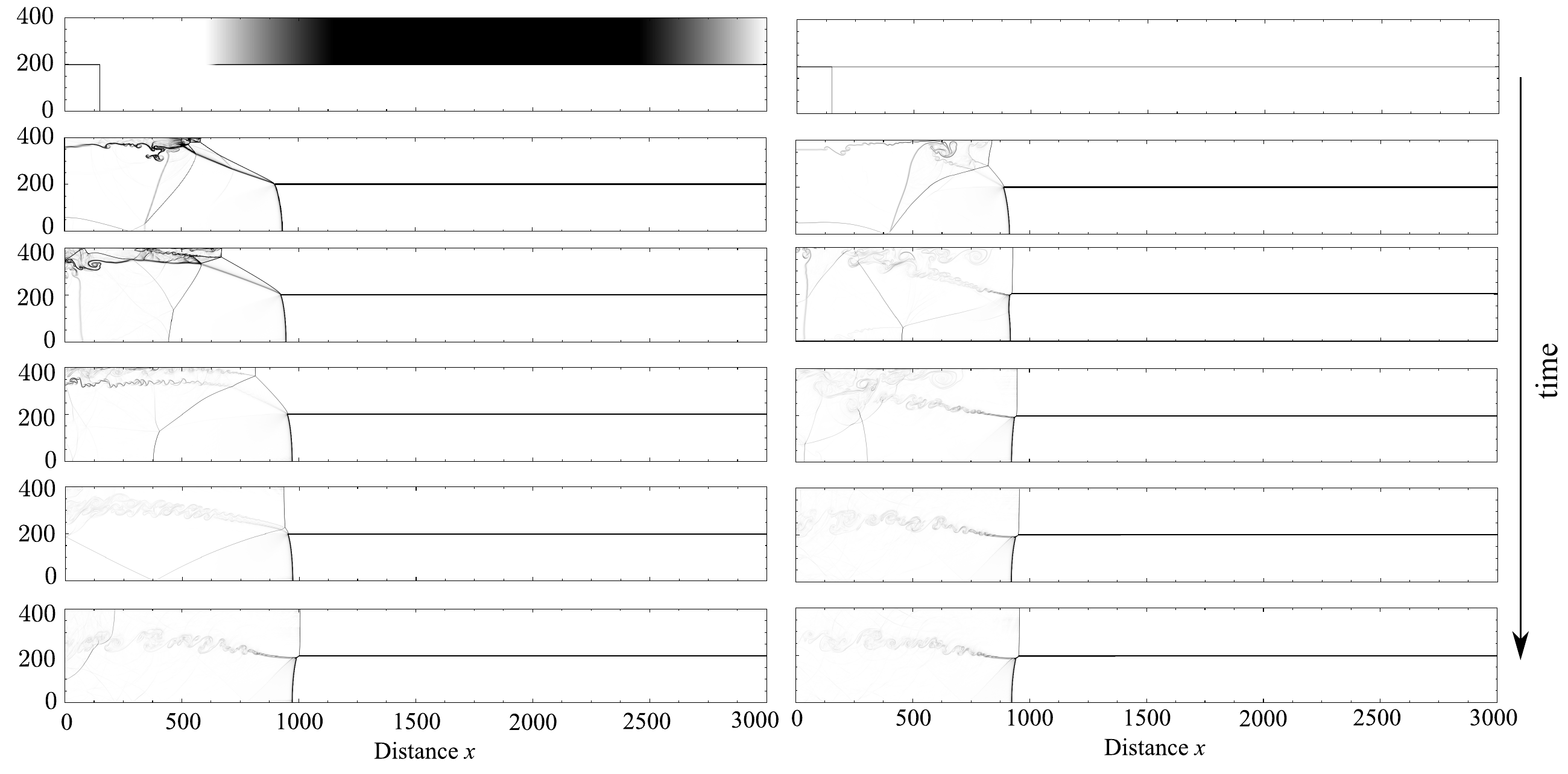}
\caption{Evolution of wave-structure relaxation for $Z = 0.55$. The left column shows gradual transition to the confining layer, while the right column shows abrupt transition to the confining layer. Rows correspond to identical numerical timeframes, with time progressing from top to bottom as indicated by the arrow.}
\label{Z0.55_Initiation_Test}
\end{figure}

Figures~\ref{Z0.60_Initiation_Test} and~\ref{Z0.55_Initiation_Test} compare the schlieren outputs obtained from the gradual initiation tests with those from the abrupt initiation mechanism used in the main body of the study, for $Z = 0.60$ and $Z = 0.55$, respectively. In each figure, rows correspond to identical numerical timeframes. Although the simulations are conducted in the laboratory reference frame, the schlieren images are translated so that the detonation structures at different times are aligned vertically for ease of comparison.  

For the $Z = 0.60$ case (Fig.~\ref{Z0.60_Initiation_Test}), the abrupt initiation produces no precursor at any stage, while the gradual initiation initially induces a precursor that later relaxes to the same terminal wave structure as the abrupt case. Conversely, for $Z = 0.55$ (Fig.~\ref{Z0.55_Initiation_Test}), the abrupt initiation shows precursor formation early in the simulation, whereas the gradual initiation initially suppresses precursor formation before eventually relaxing to the same wave configuration. These results indicated that the terminal flowfield configuration observed is independent of the details of initiation.
A precursor wave will or will not form independent of whether the transition to the confining layer is abrupt or
gradual.

\section{Discussion on the distinction between a detached and precursor shock}
\label{Detached_Precursor_Comparison}
In some cases with \(\kappa>0\), the transmitted shock in the inert layer detached ahead of the detonation front. Despite this detachment, the detonation remained underdriven, and a subsonic pocket formed immediately downstream of the detached shock, analogous to the flow behind a bow shock over an inclined wedge that exceeds the maximum deflection angle for an oblique shock. These cases belonged to a family of solutions that were introduced as Case E in Section~\ref{sample_results}. In shock--polar terms, detachment occurs when the inert-layer shock polar fails to intersect the Prandtl–Meyer expansion fan polar; i.e., the required flow deflection exceeds the oblique-shock limit $\delta_\mathrm{max}$, leaving a bow-shock–like structure as the sole admissible solution. This behavior is distinct from the precursor--shock cases discussed in this paper, where a shock precedes the detonation and drives it to overdriven speeds.

\begin{figure}[H]
\centering
\includegraphics[width=0.96\textwidth]{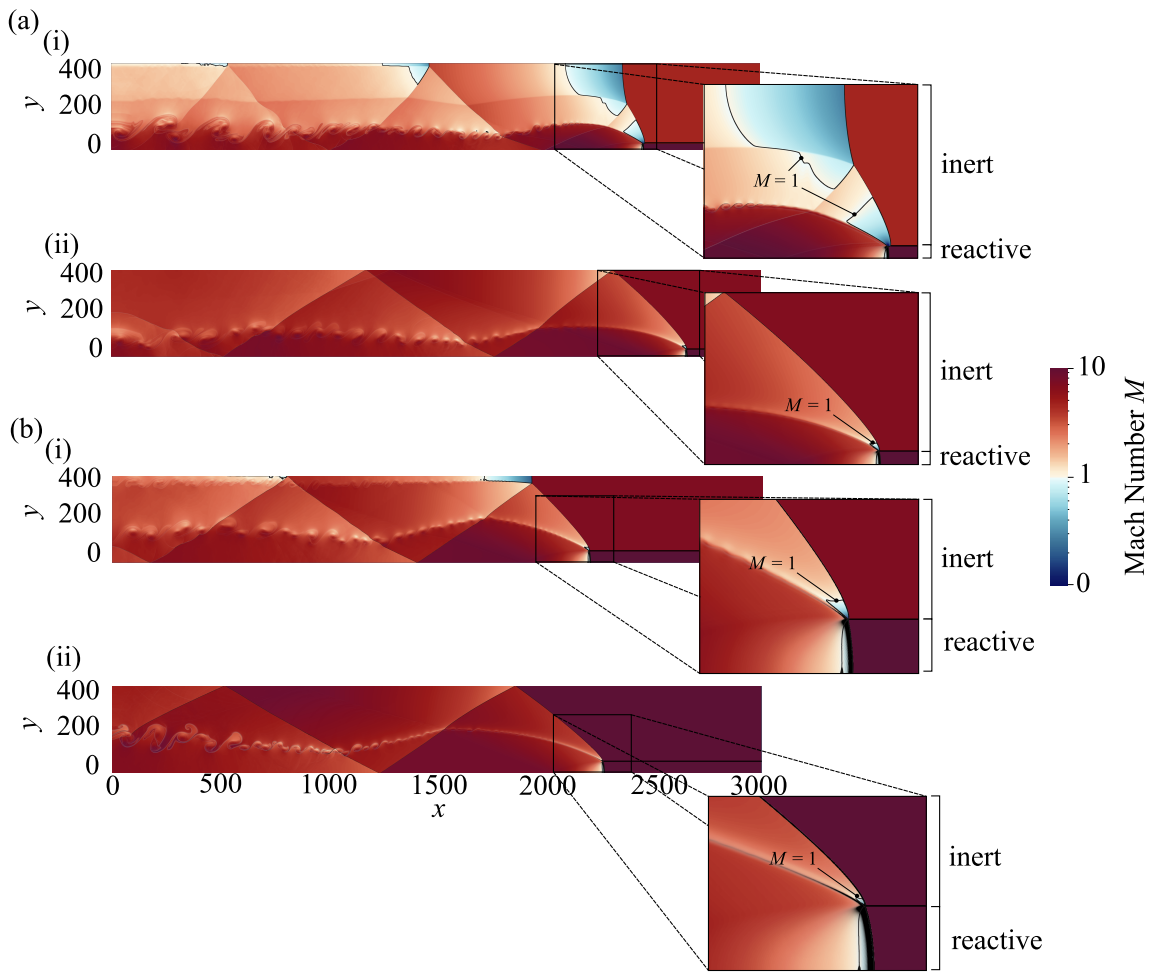}
\caption{Wave-fixed Mach number fields with numerical schlieren for detached-shock cases. The $M = 1$ contour is outlined in black. (a) $A_2/A_1 = 11$: (i) $Z = 0.40$ with a Mach reflection in the inert layer, (ii) $Z = 0.50$ with regular reflection in the inert layer. (b) $A_2/A_1 = 6.5$: (i) $Z = 0.50$ with a Mach reflection in the inert layer, (ii) $Z = 0.60$ with a regular reflection in the inert layer. Subsonic regions appear in blue and supersonic regions in red. A magnified view of the detonation structure is provided for clarity.}
\label{fig:detached_shock}
\end{figure}

Figure~\ref{fig:detached_shock} shows the flowfields for the detached-shock cases, defined by the appearance of a subsonic patch behind the shock in the inert layer near the interface. The sonic loci are overlaid to clearly highlight this region. The occurrence of a detached shock was independent of whether a Mach stem formed in the inert layer, as illustrated in Fig.~\ref{fig:detached_shock}(a) and (b). Subfigures (i) and (ii) correspond to cases where the detached shock appeared with Mach reflection and regular reflection, respectively. Figure~\ref{fig:detached_shock}(a) corresponds to $A_2/A_1 = 11$, while Fig.~\ref{fig:detached_shock}(b) corresponds to $A_2/A_1 = 6.5$.

\begin{figure}[H]
\centering
\includegraphics[width=1\textwidth]{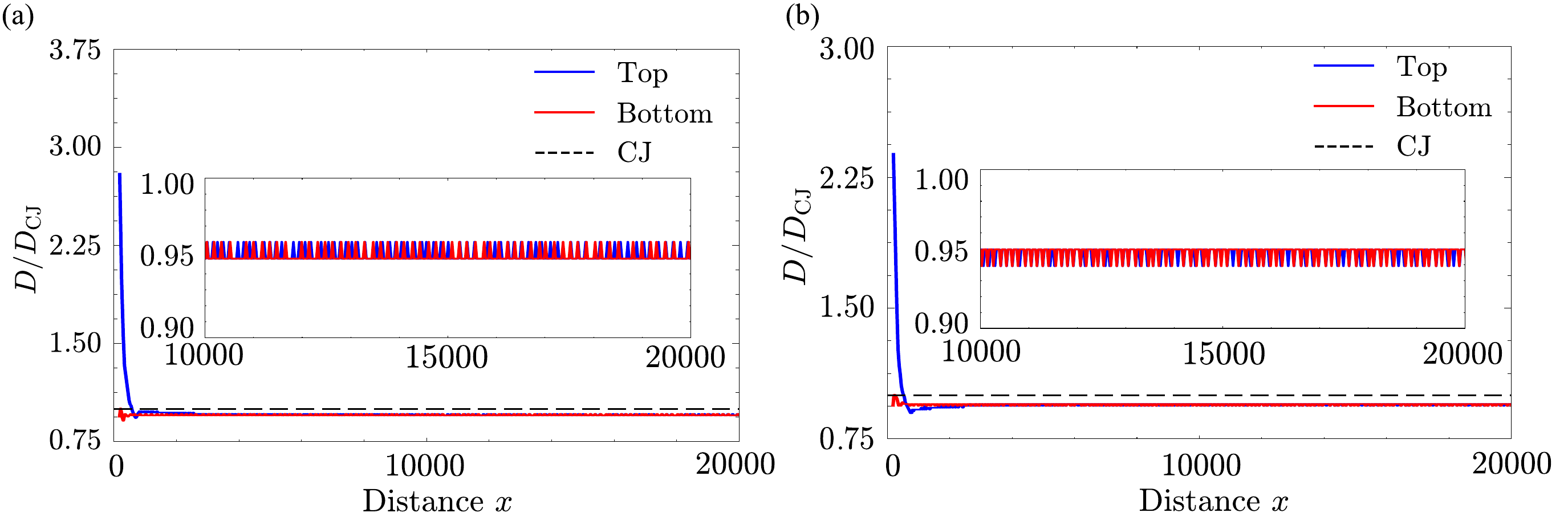}
\caption{Velocity deficit for representative detached-shock cases: (a) $A_2/A_1 = 11$, $Z = 0.50$; (b) $A_2/A_1 = 6.5$, $Z = 0.50$. The detached-shock velocity is tracked along the top boundary, while the detonation velocity is tracked along the bottom boundary.}
\label{fig:detached_velocity}
\end{figure}

\begin{figure}[H]
\centering
\includegraphics[width=0.8\textwidth]{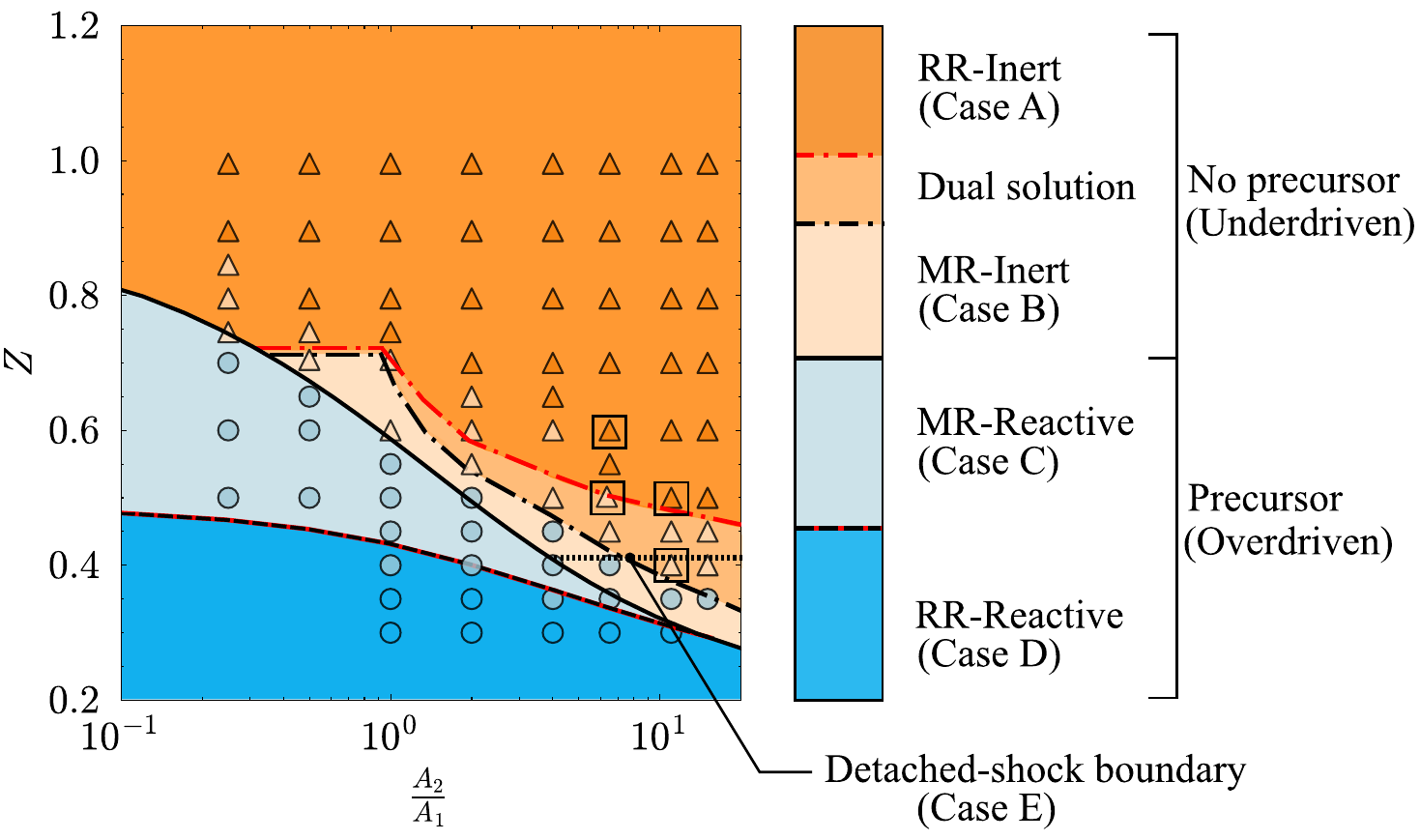}
\caption{Phase map showing predicted analytical regions with CFD simulation results indicated by symbols. Red and black lines mark the sonic and detachment criterion for the RR–MR transition in both the inert and reactive layers, respectively. The detached-shock boundary criterion is labeled, and the detached-shock cases from Fig.~\ref{fig:detached_shock} are denoted by \textcolor{black}{open square markers}}
\label{fig:detachment_phasemap}
\end{figure}

The velocities of the representative Mach reflection flowfields shown in Fig.~\ref{fig:detached_shock} are plotted in Fig.~\ref{fig:detached_velocity} with (a) representing $A_2/A_1 = 11$ and (b) $A_2/A_1 = 6.5$. Both results confirm that the detonation remained underdriven with a detached shock appearing. 
\vspace{-20pt}
The representative polar solution is constructed by assuming a CJ detonation, drawing an expansion fan polar from the $\delta = 0$ axis at $p_\mathrm{CJ}$, and plotting the inert shock polar with $M_2 = Z \cdot M_\mathrm{CJ}$. The value of $Z$ is then adjusted until the expansion polar intersects the maximum deflection point on the inert shock polar, which defines the detachment condition.  
\begin{figure}[H]
\centering
\includegraphics[width=0.4\textwidth]{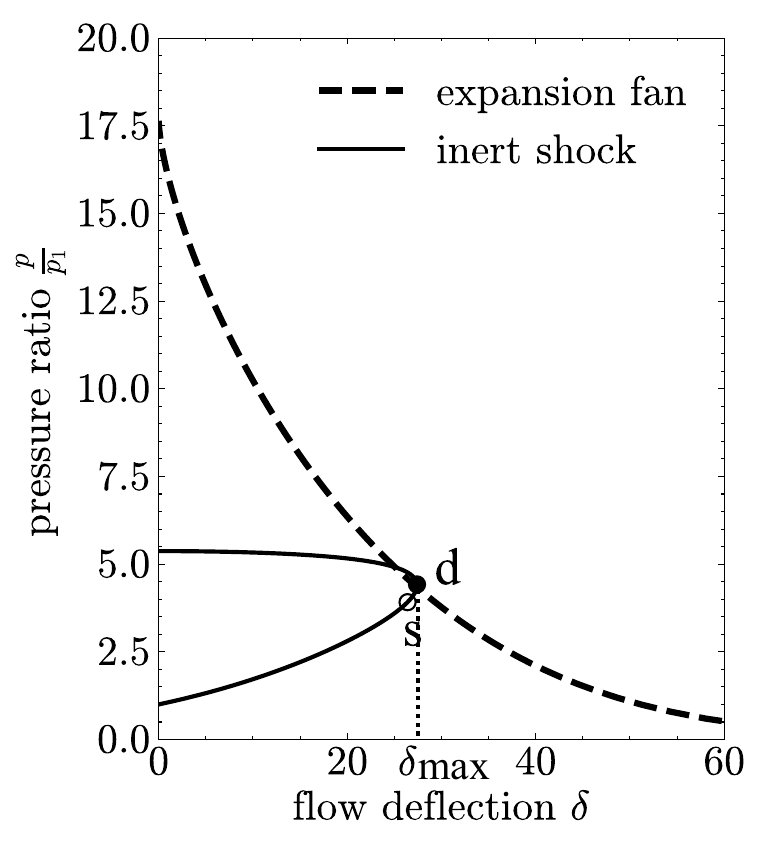}
\caption{Shock–polar analysis for a CJ detonation. The inert-layer shock polar is constructed with \(M_2 = Z \cdot M_\mathrm{CJ}\). Detachment occurs at \(Z = 0.4015\), where the expansion-fan polar meets the maximum-deflection point. For \(Z < 0.4015\), the solution is detached, as illustrated in Fig.~\ref{fig:detached_shock} (\(A_2/A_1 = 11\), \(Z = 0.40\)).}
\label{fig:detachment_polar}
\end{figure}
The accuracy of this detachment criterion can be compared with the results from the simulations shown in Fig.~\ref{fig:detached_shock}. The phase map introduced earlier has been updated to include the analytical detached-shock boundary. \textcolor{black}{CFD cases in which a detached inert-layer shock was observed are indicated by open square markers in Fig.~\ref{fig:detachment_phasemap}, which can enclose either regular reflection (inert) or Mach reflection (inert) cases.} The analytical detachment criterion does not accurately predict the onset of detached shocks observed in the simulations.

\section{Detonation curvature effects: eigenvalue and Eyring analyses ($\kappa > 0$)}
\label{Eyring_Appendix}
In order to investigate the velocity deficits induced by the curvature effect in the detonation, the $D_{n}-\kappa$ relations are formulated by solving a coupled set of ODEs, associated with the conservation of mass, momentum, and energy, in addition to a reaction rate, $\dot{\lambda}$. An eigenvalue solution is iterated upon and found once the detonation products reach and surpass the sonic speed in the reaction zone.

\begin{equation}
\mathrm{d}(\rho u A) = 0
\label{eq:conservation_mass}
\end{equation}
\begin{equation}
\mathrm{d}(p + \rho u^2) = -\rho u^2 \, \mathrm{d}A
\label{eq:conservation_momentum}
\end{equation}
\begin{equation}
\mathrm{d}(h + \frac{1}{2}u^2) = 0
\label{eq:conservation_energy}
\end{equation}
\begin{equation}
\dot{\lambda} = k (1 - \lambda) 
\label{eq:reaction_rate}
\end{equation}

\

\noindent
We can rewrite $\mathrm{d}u$ in terms of the area variation within the reaction zone \cite{Higgins2012}
\begin{equation}
\frac{\mathrm{d}u}{\mathrm{d}x} = \frac{\dfrac{\Delta q \dot{\lambda}}{c_p T} - u \left(\dfrac{1}{A} \dfrac{\mathrm{d}A}{\mathrm{d}x} \right)}{1 - M^2}.
\label{eq:du}
\end{equation}
If the area-change term of the numerator in Eq.~(\ref{eq:du}) is 0, the ideal Chapman--Jouguet equilibrium solution is recovered. Several closures may be used to specify the area-variation term; a common choice relates it to the local shock curvature, which is the basis of Detonation Shock Dynamics (DSD) theory \cite{Bdzil_1981}.

The area variation term can be expressed in terms of the gradient in the wave-fixed transverse velocity
\begin{equation}
\frac{1}{A} \frac{\mathrm{d}A}{\mathrm{d}x} = \frac{\alpha}{u} \left( \frac{\partial v}{\partial y} \right).
\end{equation}

\noindent
The transverse velocity gradient  in the reaction zone can be described in terms of the local radius of curvature, $R$, along the shock front, and the wave-fixed axial velocity at the von Neumann point, $u_\mathrm{vN}$
\begin{equation}
\frac{\partial v}{\partial y} = \frac{D - u_\mathrm{vN}}{R}.
\end{equation}

\noindent
Introducing curvature, $\kappa$, will simplify the analysis, and $\alpha$ is a geometric term which is equal to 1 since the problem investigated is analogous to a two-dimensional slab,
\begin{equation}
\kappa = \frac{\alpha}{R}, \quad \text{with} \quad \alpha = 1.
\end{equation}

\begin{figure}[H]
\centering
\includegraphics[width=0.8\textwidth]{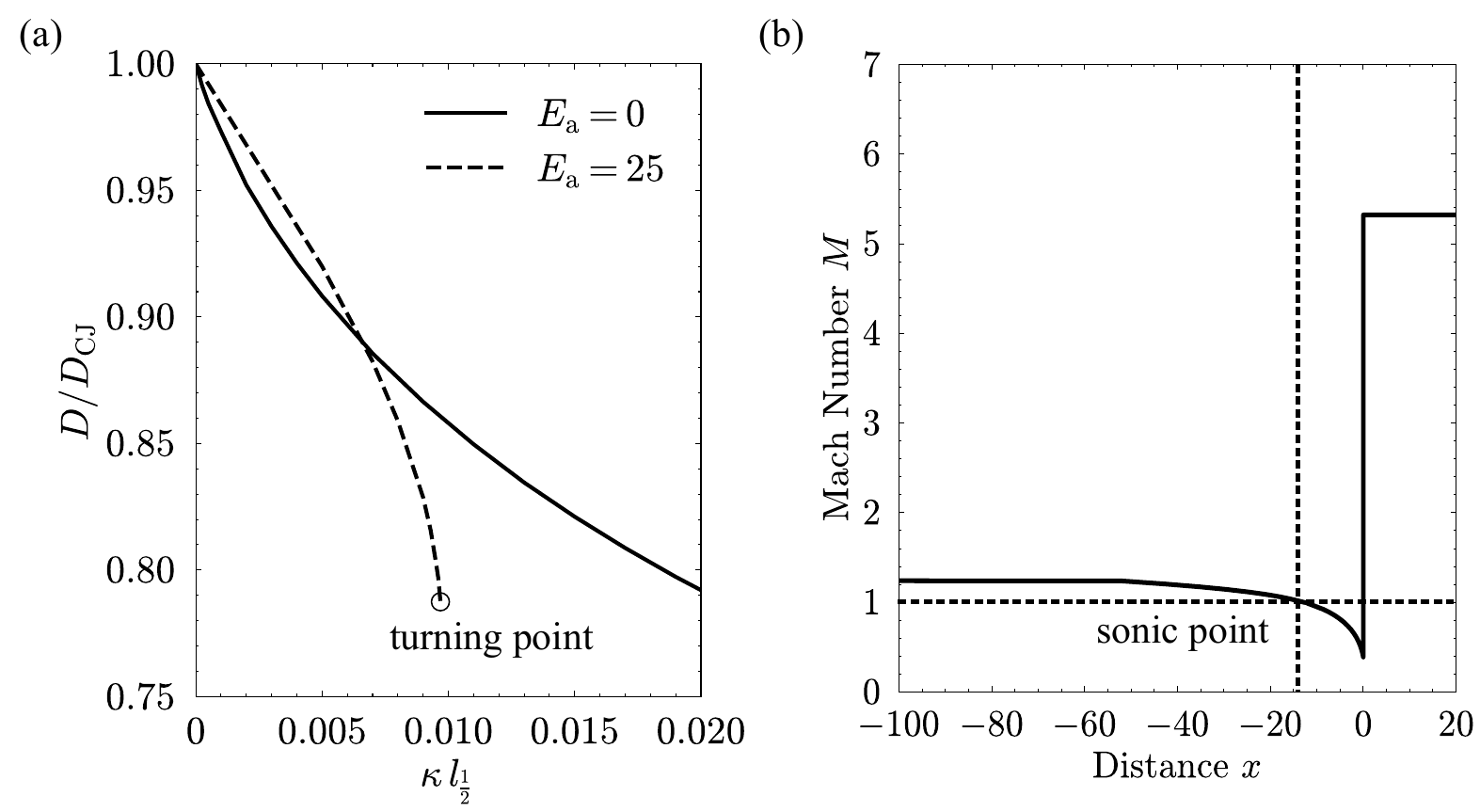}
\caption{(a) Normal detonation velocity–curvature ($D_{n}$–$\kappa$) relations from the eigenvalue solution, comparing $E_\mathrm{a} = 0$ (no turning point) with $E_\mathrm{a} = 25$ (showing criticality). (b) Wave-fixed Mach number profile in the reaction zone for a representative eigenvalue solution ($E_\mathrm{a} = 0, \kappa = 0.001/l_\mathrm{1/2}, D/D_\mathrm{CJ} = 0.9733$), integrated through the sonic point to the supersonic regime with the detonation shock located at $x = 0$.}
\label{Criticality}
\end{figure}

\noindent The curvature can therefore be expressed in terms of the steady-state detonation front profile, $x_\mathrm{s}(y)$, as
\begin{equation}
\kappa = \frac{\dfrac{\mathrm{d}^2 x_\mathrm{s}}{\mathrm{d}y^2}}{\left[1 + \left(\dfrac{\mathrm{d} x_\mathrm{s}}{\mathrm{d}y}\right)^2\right]^{3/2}}.
\end{equation}
Figure~\ref{Criticality} provides the results obtained from integrating the coupled ODE set for different activation energies. 

\vspace{-20pt}

In Fig.~\ref{Criticality}(a) a turning point and criticality is exhibited when Arrhenius kinetics are employed with $E_\mathrm{a} = 25$. By contrast, for the cases simulated in the paper, $E_\mathrm{a} = 0$, no turning point was exhibited and as such, detonation was initiated even at very thin reactive layer heights since the total area of the domain was kept constant and the area ratio was varied. Figure~\ref{Criticality}(b) provides the wave-fixed Mach number evolution for a sample iterated eigenvalue solution for $E_\mathrm{a} = 0$ in which the sonic point is reached within the reaction zone.

\begin{figure}[H]
\centering
\includegraphics[width=0.85\textwidth]{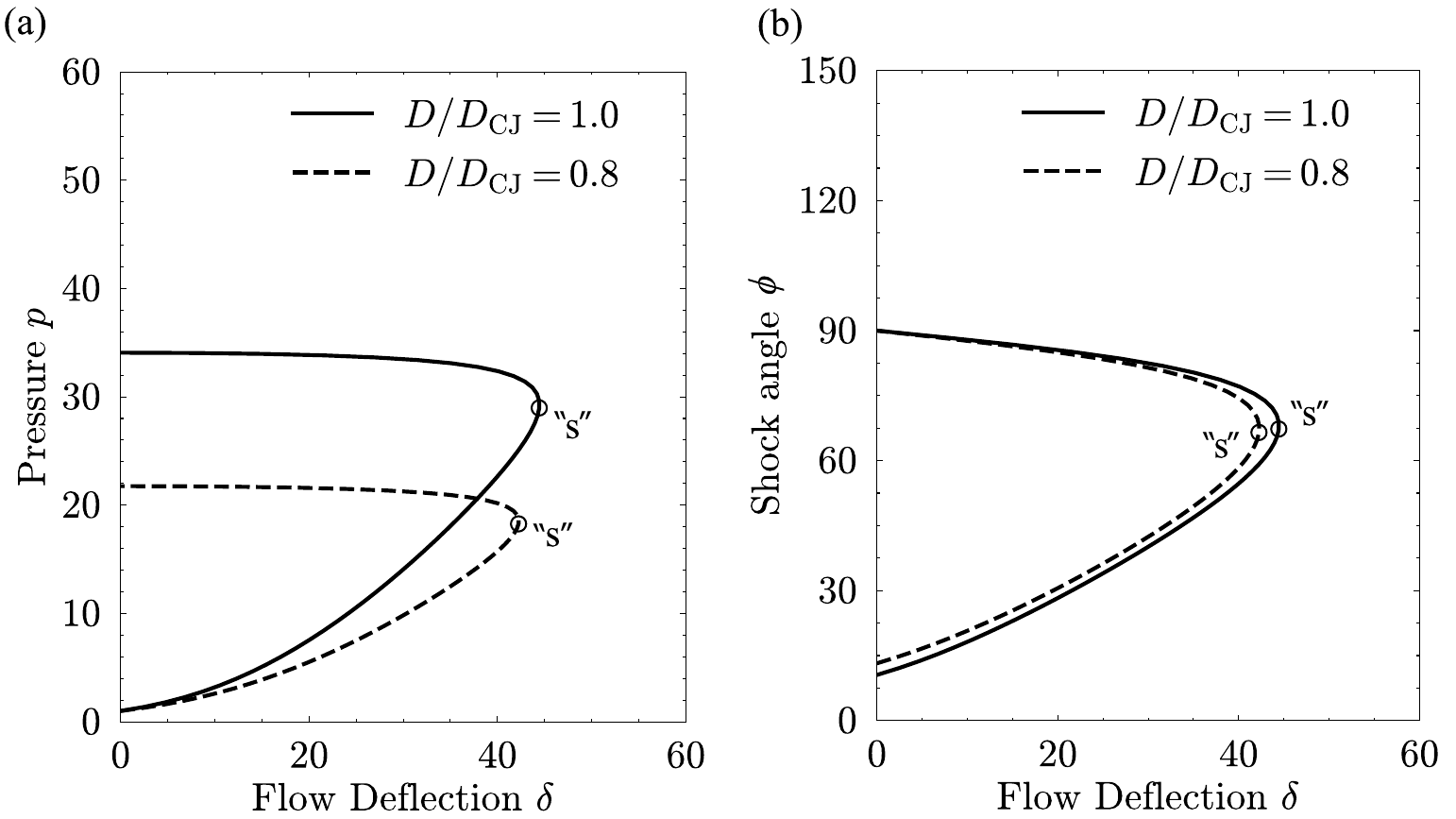}
\caption{Shock polars for $D/D_\mathrm{CJ} = 1.00$ and $0.80$, with the point ``s'' denoting the sonic condition. (a) $p$–$\delta$ polar; (b) $\phi$–$\delta$ polar. The shock angle $\phi$ associated with the sonic point ($\phi_\mathrm{s}$) determines the inclination of the detonation shock at the reactive--inert interface}

\label{polars}
\end{figure}

It was demonstrated in Section~\ref{sample_results} that for underdriven cases ($\kappa > 0$), a sonic locus extends from the reactive–inert interface behind the detonation front. Accordingly, Fig.~\ref{polars} shows the location of this sonic solution on the polar diagrams: (a) the $p$–$\delta$ polar and (b) the corresponding $\phi$–$\delta$ polar. The latter allows determination of the detonation front inclination by identifying the shock angle $\phi$ associated with the sonic point.

\begin{figure}[H]
\centering
\includegraphics[width=0.8\textwidth]{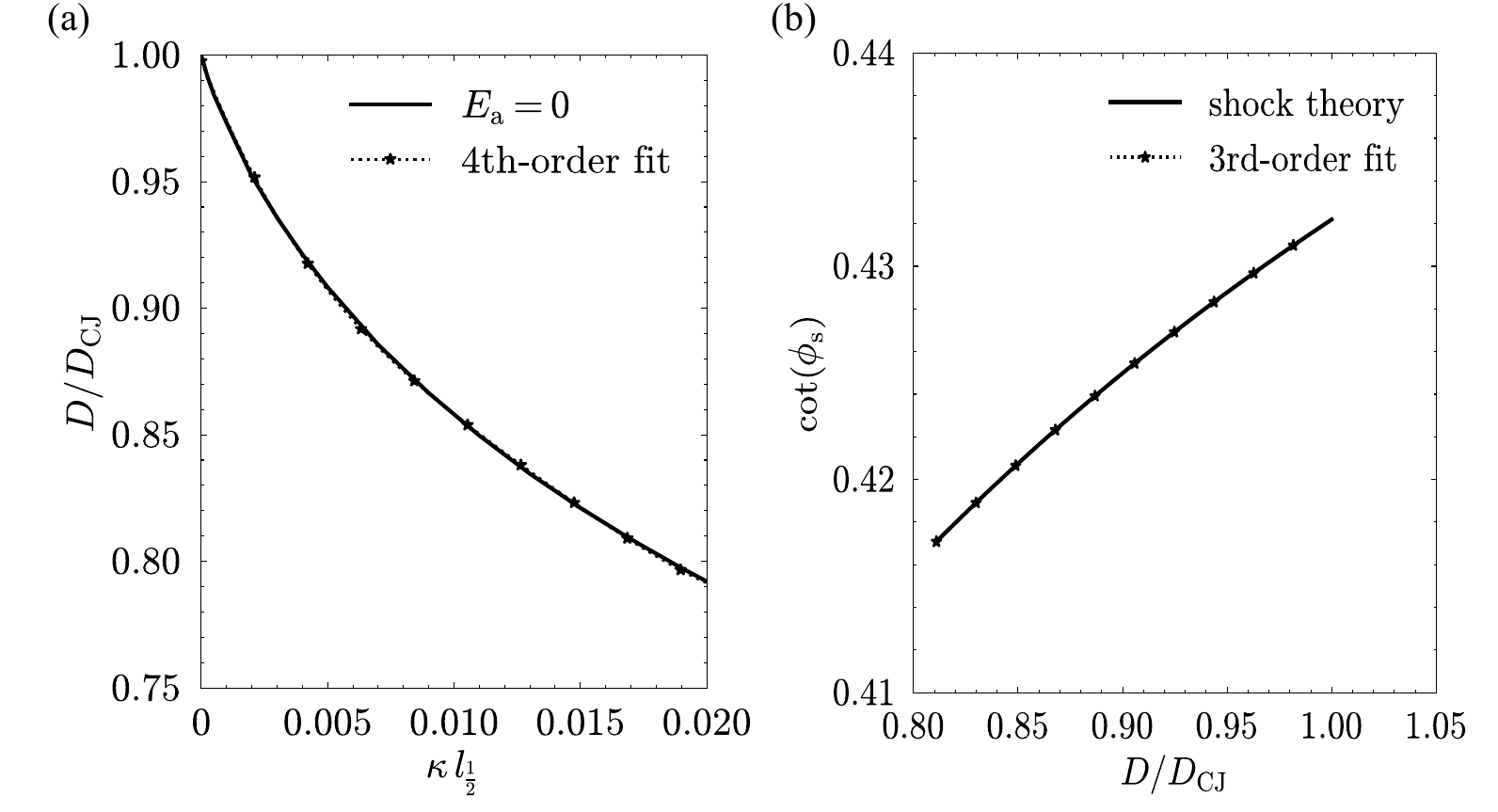}
\caption{(a) $D_{n}$–$\kappa$ relation shown by a solid line together with a fourth-order polynomial fit; (b) $\cot(\phi_\mathrm{s})$ determined from shock polars that were constructed for different $D/D_\mathrm{CJ}$ values shown with a third-order polynomial fit.}
\label{fit}
\end{figure}

\vspace{-20pt}
A fourth-order polynomial can be fitted across the $D_{n}-\kappa$ in order to interpolate for intermediate values not directly calculated from the coupled ODE set, as shown in Fig.~\ref{fit}(a). It is clear that at different Mach numbers, the sonic shock angle exhibits a slight difference depending on the Mach number.  As such, a calibration fit is obtained that can relate the sonic shock angle, $\phi_\mathrm{s}$, to the velocity deficit, $D/D_{\mathrm{CJ}}$, as shown in Fig.~\ref{fit}(b). A solution is obtained once the gradient obtained from the Eyring construction at the interface matches the associated $\mathrm{cot}(\phi_\mathrm{s})$. 

\vspace{-10pt}
The analysis presented in this appendix defines the quantitative dependence of the detonation velocity on curvature and connects the eigenvalue, Eyring, and shock-polar formulations within a unified framework to model the curvature and velocity deficits associated with the underdriven detonations in Cases A and B of the main text.
\newpage

\section{Eyring construction comparison with CFD simulations}
\label{Eyring_Construction_CFD}

\begin{figure}[H]
\centering
\includegraphics[width=1\textwidth]{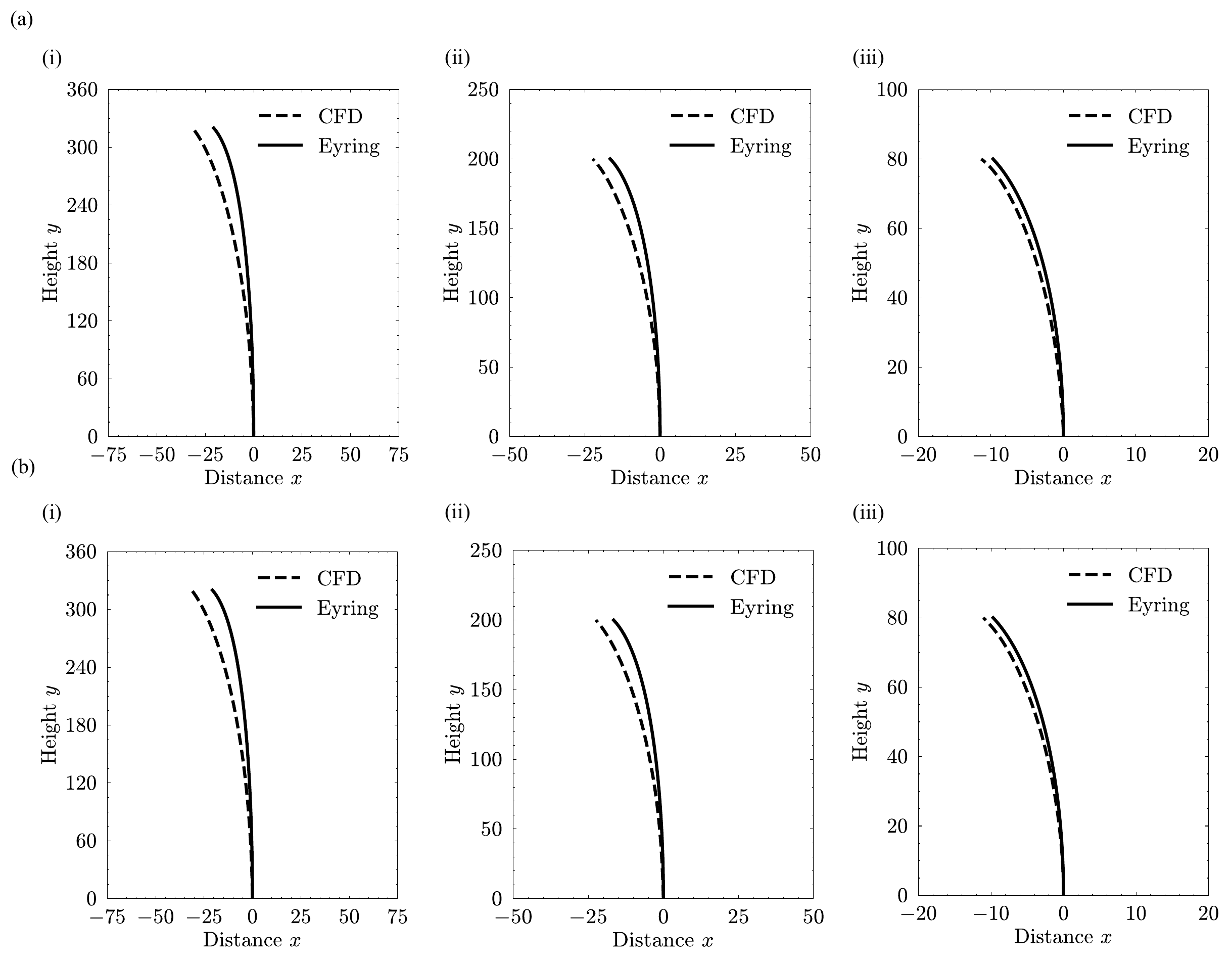}
\caption{Comparison of detonation front curvature obtained from CFD simulations with theoretical predictions from the Eyring construction method. Results are shown for (a) $\textit{Z} = 0.80$ and (b) $\textit{Z} = 0.90$, with subfigures corresponding to (i) $A_2/A_1 = 0.25$, (ii) $A_2/A_1 = 1$, and (iii) $A_2/A_1 = 4$.}

\label{Detonation_Front}
\end{figure}

To assess the accuracy of Eyring’s geometric construction, the detonation front from CFD simulations is extracted by perturbing the pressure contour by $\epsilon = 10^{-3}$ from the undisturbed value $p = 1$ ahead of the detonation. A Bézier curve is fitted through 20 evenly spaced control points along the extracted contour and compared with the theoretical prediction of Eyring’s model, as shown in Fig.~\ref{Detonation_Front}.

\begin{figure}[H]
\centering
\includegraphics[width=0.45\textwidth]{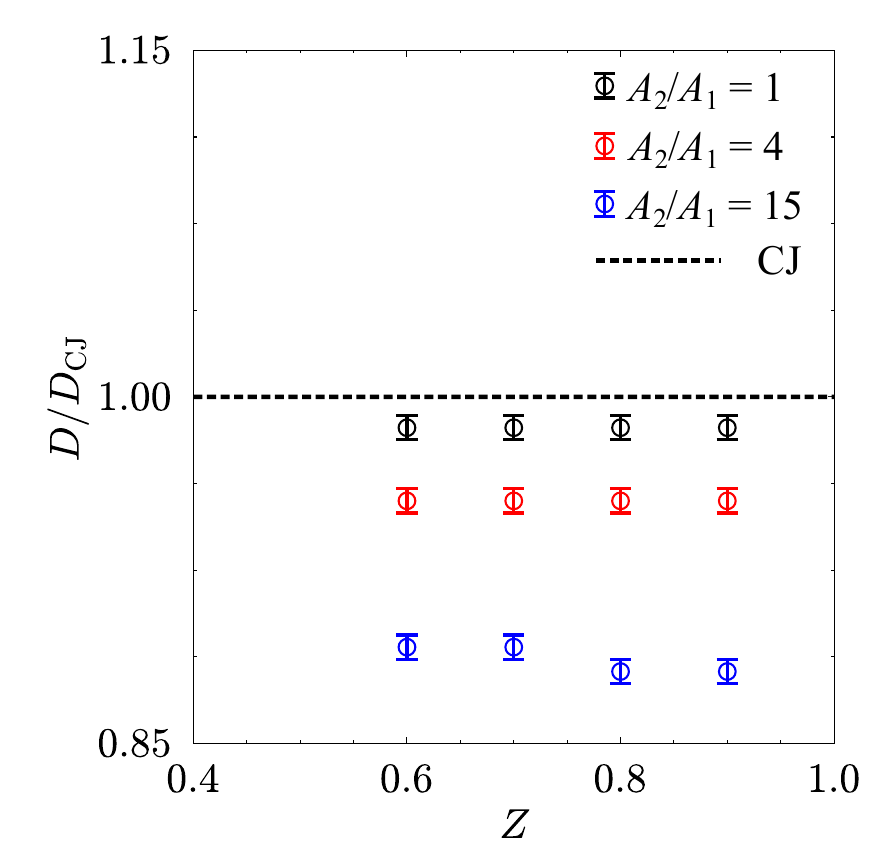}
\caption{Comparison of $D/D_\mathrm{CJ}$ for different $A_2/A_1$ values at $Z = 0.60$, $0.70$, $0.80$, and $0.90$. All cases correspond to underdriven detonations with $\kappa > 0$. Error bars indicate the range of velocities obtained in the simulations (upper bound = maximum, lower bound = minimum), with the unfilled circle denoting the arithmetic-averaged value.}

\label{Detonation_Velocity_Comparison}
\end{figure}

Figure~\ref{Detonation_Front}(a) compares results for different $A_2/A_1$ values at $Z = 0.80$, while Fig.~\ref{Detonation_Front}(b) shows the corresponding cases at $Z = 0.90$. The two sets are nearly indistinguishable, indicating that in weakly confined cases the detonation behavior—provided no precursor forms—is governed primarily by $A_2/A_1$ and is largely independent of $Z$. This observation is further supported by Fig.~\ref{Detonation_Velocity_Comparison}, where the velocity deficits $D/D_\mathrm{CJ}$ display a strong dependence on $A_2/A_1$ but only a weak sensitivity to $Z$. Physically, this regime corresponds to the limit of an \emph{infinitely weak confinement}, where the impedance of the surrounding layer becomes too small to influence the detonation. In this limit, the sonic surface intersects the leading shock, fully isolating the detonation from the external medium. Consequently, the wave structure and propagation velocity become identical to those that would exist even if the detonation were expanding into vacuum.

\section{Shock--expansion method}
\label{Shock_Expansion}
\begin{figure}[H]
\centering
\includegraphics[width=0.6\textwidth]{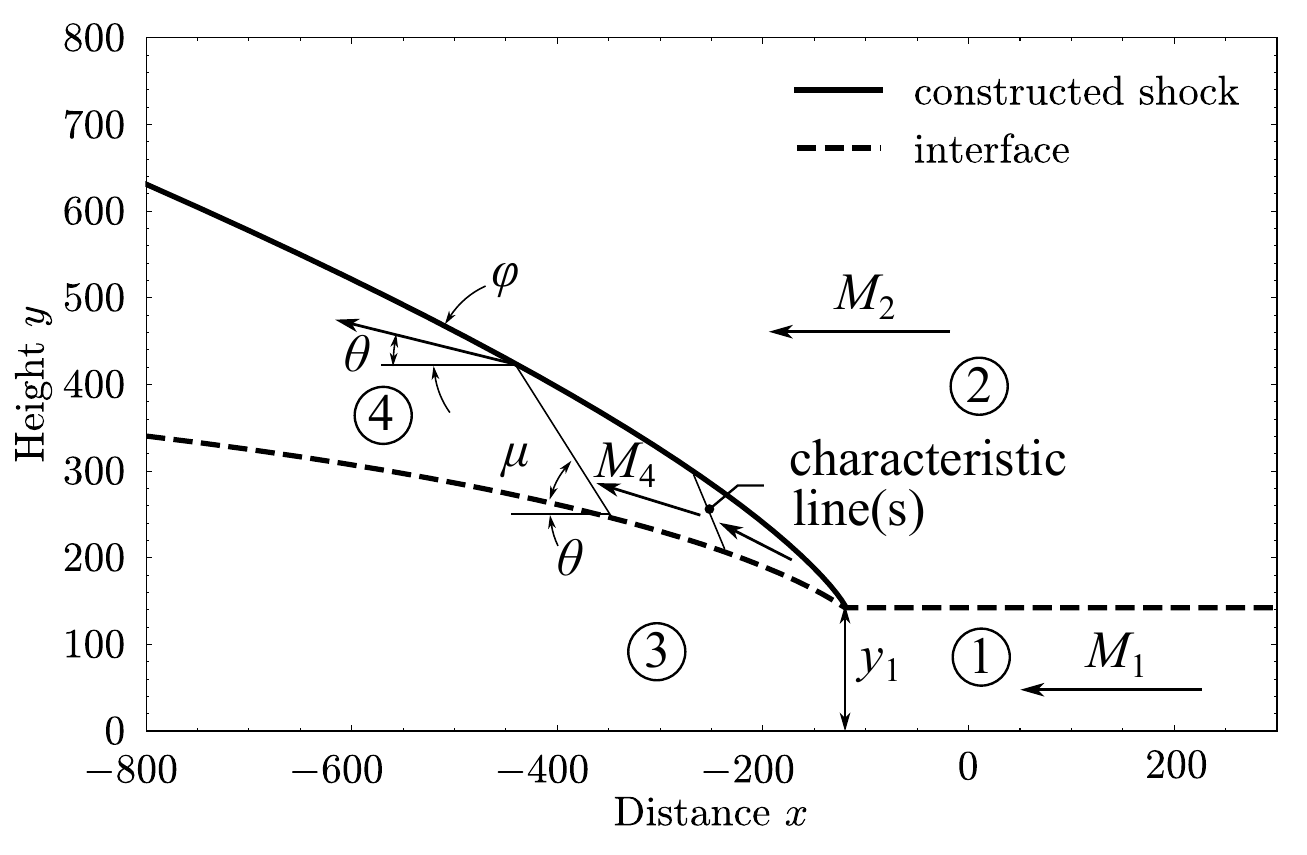}
\caption{Illustration of a decaying shock constructed using the shock–expansion method applied to the region of expanding detonation products. The parameter $y_1$ denotes the initial thickness of the reactive layer, and straight-line characteristics are drawn between the interface and the attached shock.}
\label{fig:shock_decay}
\end{figure}

A geometric construction of the decaying shock in the inert layer can be obtained by assuming its interaction with straight\footnote{While the assumption of a straight shock neglects the non-simple wave region behind the decaying shock, this
assumption is frequently employed in the hypersonics literature; see discussion of its validity in \cite{WALDMAN_PROBSTEIN} and \cite{Jones_1963}.} characteristics extending from the reactive--inert interface~\cite{WALDMAN_PROBSTEIN}. This proceeds by drawing characteristics whose inclination is set by the local Mach angle, \(\mu\), and by the deflection angle of the detonation products, \(\theta\). By locating their intersections with the shock, a decaying shock can be reconstructed, as illustrated in Fig.~\ref{fig:shock_decay}.

\vspace{-20pt}
\noindent
The deflection of the detonation products as a function of $x$, $\theta(x)$, was solved for in Section~\ref{Decaying_Incident_Shock} along with the variation along $x$ of the local Mach number between the interface and shock, $M_4(x)$ by solving the coupled ODE set. The shock angle \(\phi(x)\) is determined by enforcing that the inert layer turns by \(\theta(x)\) across the shock:

\begin{equation}
\label{tan_theta}
{\;\tan \theta \;=\;
\frac{\big(M_2^{2}\sin^{2}\phi - 1\big)\,\cot\phi}
     {\left(\frac{\gamma+1}{2}\right) M_2^{2} - M_2^{2}\sin^{2}\phi + 1}\;}\,
\end{equation}
This allows a simple construction of the straight-line characteristics transmitted from the interface with gradient:
\begin{equation}
{\;\left(\frac{\mathrm{d} y}{\mathrm{d} x}\right)_{C^{+}} = \tan\!\big(\theta(x) + \mu(x)\big)\;}
\end{equation}
The varying gradient of the shock can also be described as 
\begin{equation}
{\;\left(\frac{\mathrm{d} y}{\mathrm{d} x}\right)_{\text{shock}} = \tan \phi(x)\;}
\end{equation}
The equation for the $C^{+}$ characteristic is represented by
\begin{equation}
{\;f(x) = x\tan(\theta+\mu)\, + f_{0}\;}
\end{equation}
The intercept $f_{0}$ is obtained from the known point on the interface $(x_\mathrm{interface},y_\mathrm{interface})$ 
through which the characteristic emanates, such that
\begin{equation}
{\;f_{0} = y_\mathrm{interface} - \tan(\theta+\mu)\,x_\mathrm{interface }\;}
\end{equation}
The shock is written in the form
\begin{equation}
{\;g(x) = x\tan\phi(x)\ + g_{0}\;}
\end{equation}
At the origin, $g_{0}=0$ and for subsequent constructions, $g_{0}$ is determined by enforcing the intersection condition between $f(x)$ and $g(x)$.

\begin{figure}[H]
\centering
\includegraphics[width=0.76\textwidth]{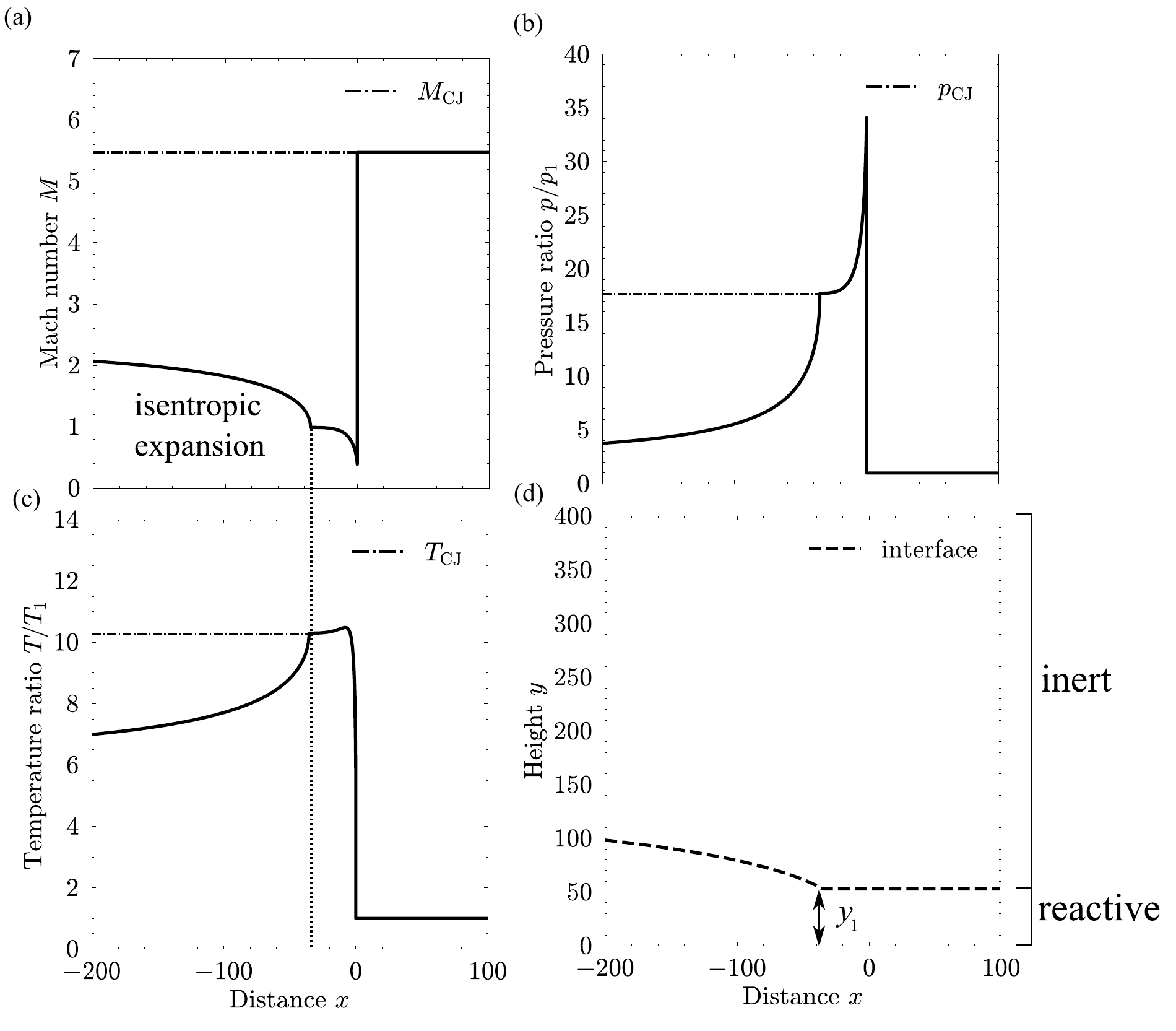}
\caption{Modified-ZND shock--expansion model results for a sample CJ detonation with $A_2/A_1 = 6.5$ and $Z = 0.60$: (a) wave-fixed Mach number, (b) pressure, (c) temperature, and (d) interface deflection profiles with the detonation shock located at $x = 0$}
\label{ZND_isentropic}
\vspace{-30pt}
\end{figure}

\begin{figure}[H]
\centering
\includegraphics[width=0.85\textwidth]{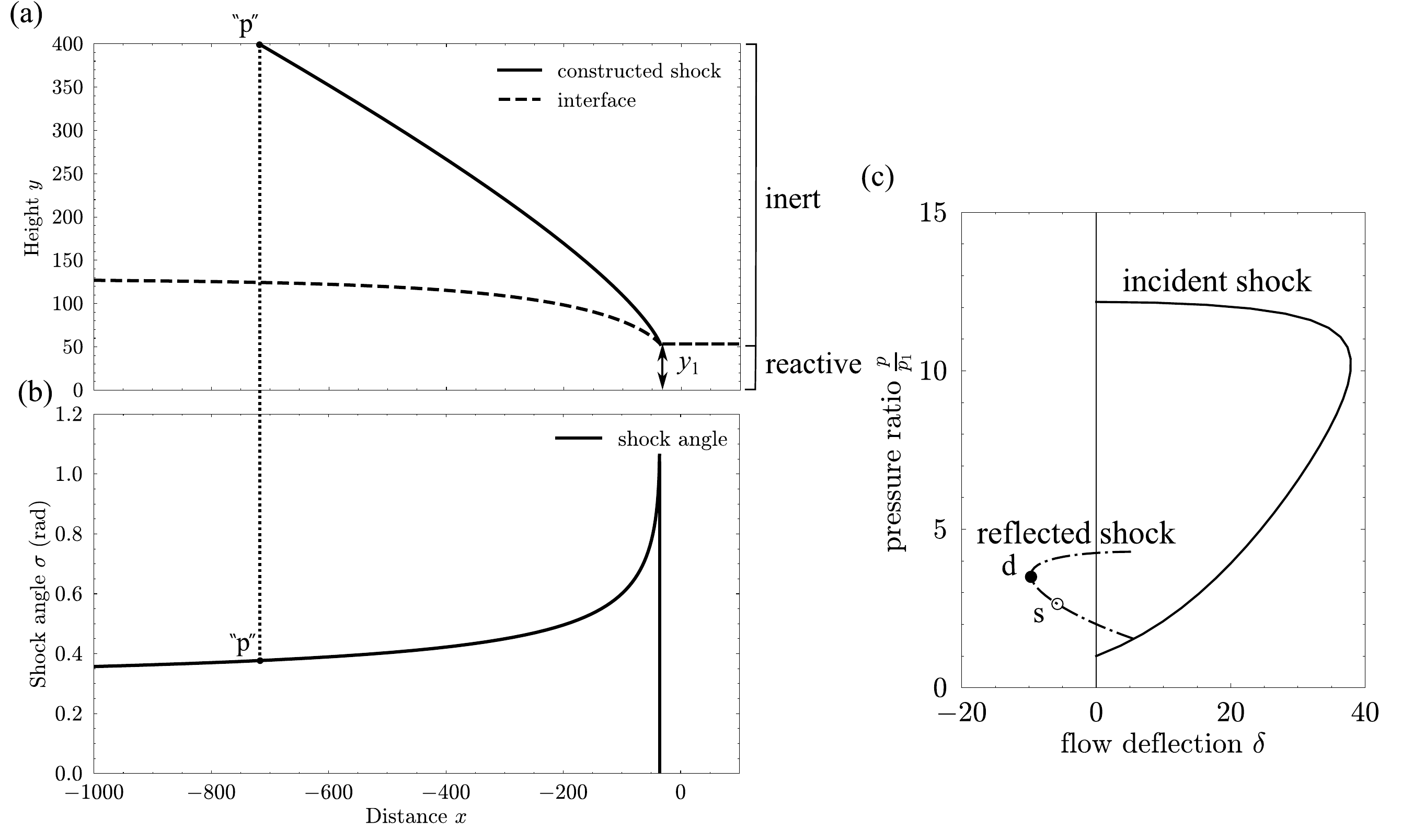}
\caption{Shock-construction method applied to a CJ detonation with $A_2/A_1 = 6.5$ and $Z = 0.60$, where point ``p'' marks the intersection between the decaying shock wave and the upper boundary: (a) shock profile in the inert layer, (b) shock angle variation, (c) shock-polar analysis performed at point ``p''.}
\label{constructed_shock}
\end{figure}

 A representative solution obtained from integrating Eqs.~(\ref{eq:dy_dx}--\ref{eq:du_dx}) for the expanding products is shown in Fig.~\ref{ZND_isentropic} for the case $Z = 0.60$ and $A_2/A_1 = 6.5$, where $y_1 = 53 l_{1/2}$ denotes the initial thickness of the undisturbed reactive layer.
Once the deflection of the detonation products into the inert layer is specified, the decaying shock within the inert gas can be constructed using the shock–expansion method, with details provided earlier in this section. The constructed shock wave is shown in Fig.~\ref{constructed_shock}(a). The variation of the shock angle can also be tracked, such that its value at the upper boundary ($y = 400~ l_{1/2}$) is known. Subsequent polar analysis of the shock interaction with the top boundary allows determination of whether a regular or
Mach reflection is established in the inert layer, depending on either the maximum deflection or sonic criteria; this forms
the boundary between Case A and Case B in the main text of the paper.

\section{Geometric construction of Case C: modeling of precursor shock and detonative Mach-stem curvature}
\label{Detonation_Geometric}

The case where a Mach stem emerges in the reactive layer and a triple point forms can be analyzed using a shock polar method to determine the orientations of the detonation and reflected shocks at the triple point. An analogy can be drawn to overexpanded nozzles, where the geometric interaction pattern (ADTG) resembles the shock interactions observed at nozzle exits. Such problems have been studied in the literature to predict Mach-stem heights. Equations~(\ref{first_geometric}–\ref{last_geometric}) give the geometric relations connecting the Mach-stem length and height to angles obtained from polar analysis. Equations~(\ref{first_isentropic}–\ref{last_isentropic}) enforce the requirement that the flow above the sonic throat in the reactive layer becomes parallel to the lower wall. These relations follow from the analysis of \cite{Paramanantham_Janakiram_Gopalapillai_2022}.

In order to construct the oblique detonation polar for DT, the angle $\phi_{1}$ must be initially evaluated. Its value is determined by equating the pressure behind the precursor shock BA with that behind the incident oblique shock AT, as given in Eqs.~(\ref{first_sinphi}–\ref{last_sinphi}). This relation simplifies to a dependence solely on the acoustic impedance ratio $Z$.

\begin{figure}[H]
\centering
\includegraphics[width=0.8\textwidth]{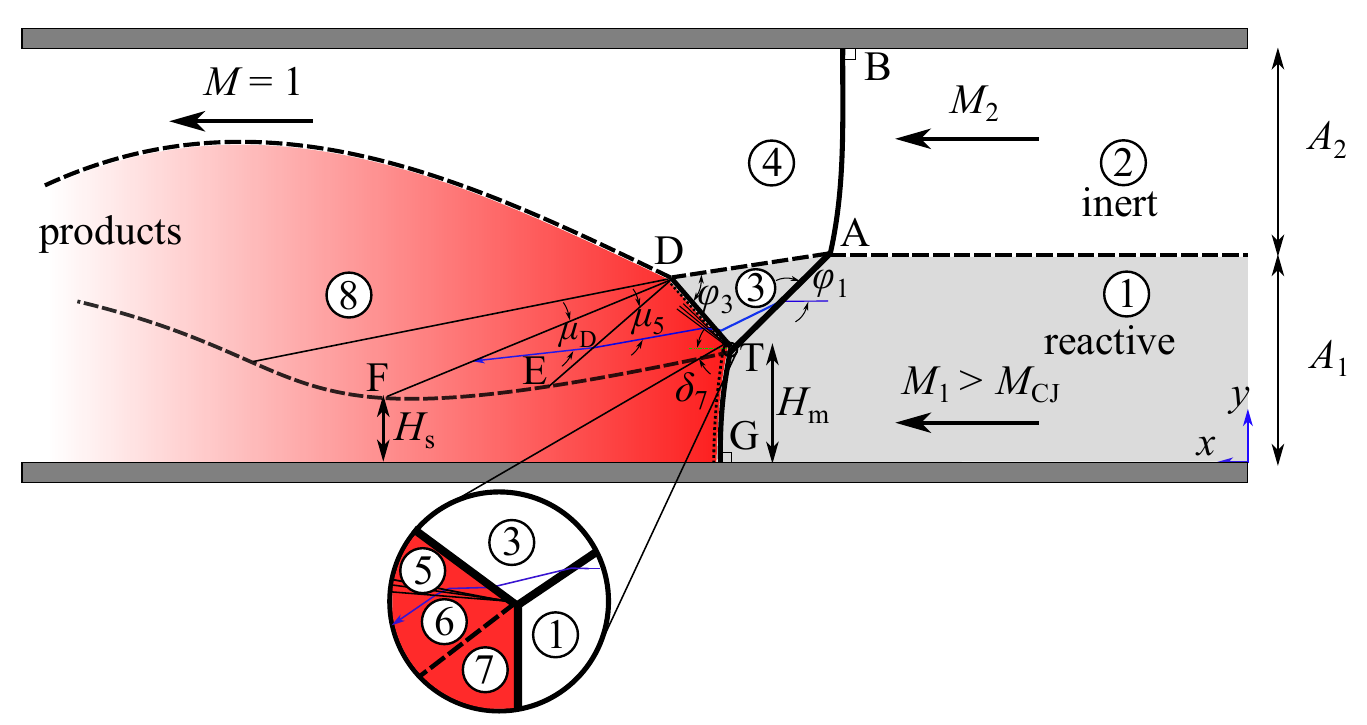}
\caption{Labeled schematic of Case C. Gray-shaded regions correspond to unburned reactive gas, while red-shaded regions denote detonation products. In the reactive layer, a nonreactive shock (AT), an oblique detonation (DT), and a detonative Mach stem (GT) intersect at the triple point T. The lines BA and TG are drawn with slight curvature to emphasize the nonplanar nature of the precursor shock and the detonative Mach stem, respectively.}
\end{figure}
\vspace{-40pt}
\begin{align}
    &X_\mathrm{T} \tan \phi_\mathrm{1} = H_1 - H_\mathrm{m},\label{first_geometric}  \\
    &X_\mathrm{D} \tan \delta_\mathrm{1} = H - Y_\mathrm{D}, \\
    &(X_\mathrm{D} - X_\mathrm{T}) \tan (\phi_\mathrm{3} - \delta_\mathrm{1}) = Y_\mathrm{D} - H_\mathrm{m}, \\
    &(X_\mathrm{E} - X_\mathrm{T}) \tan \delta_7 = H_\mathrm{m} - Y_\mathrm{E}, \\
    &(X_\mathrm{F} - X_\mathrm{D}) \tan \mu_\mathrm{D} = Y_\mathrm{D} - H_\mathrm{s}, \\
    &(X_\mathrm{E} - X_\mathrm{D}) \tan (\mu_\mathrm{5} + \delta_\mathrm{7}) = Y_\mathrm{E} - Y_\mathrm{D}, \\
    &(X_\mathrm{F} - X_\mathrm{E}) \tan \delta_\mathrm{7} = \left( 2 + \tan^2 \delta_\mathrm{7} \right) (Y_\mathrm{E} - H_\mathrm{s}).
    \label{last_geometric}
\end{align}

\begin{align}
    \nu(M_\mathrm{D}) &= \sqrt{\frac{\gamma + 1}{\gamma - 1}} \arctan \sqrt{\frac{\gamma - 1}{\gamma + 1} \left( M_\mathrm{D}^2 - 1 \right)} - \arctan \sqrt{\left(M_\mathrm{D}^2 - 1\right)}, \label{first_isentropic}\\
    \nu(M_\mathrm{D}) &= \nu(M_\mathrm{5}) + \delta_\mathrm{7}.
    \label{last_isentropic}
\end{align}

\begin{equation}
M_{\mathrm{1}} = \frac{D}{\sqrt{\gamma R T_{\mathrm{1}}}} \quad \text{and} \quad M_{\mathrm{2}} = \frac{D}{\sqrt{\gamma R T_{\mathrm{2}}}}
\label{first_sinphi}
\end{equation}

\begin{equation}
\frac{p_3}{p_1} = \frac{2\gamma M_{\mathrm{1}}^2 \sin^2 \phi_{\mathrm{1}} - (\gamma - 1)}{\gamma + 1}
\end{equation}

\begin{equation}
\frac{p_{\mathrm{4}}}{p_{\mathrm{2}}} = \frac{2\gamma M_{\mathrm{2}}^2 - (\gamma - 1)}{\gamma + 1}
\end{equation}

\begin{equation}
\frac{p_{\mathrm{3}}}{p_{\mathrm{1}}} = \frac{p_4}{p_2} ,\quad M_{\mathrm{2}}^2 = M_{\mathrm{1}}^2 \sin^2 \phi_{\mathrm{1}}
\end{equation}

\begin{equation}
\frac{D^2}{\gamma R T_{\mathrm{1}}} \sin^2 \phi_{\mathrm{1}} = \frac{D^2}{\gamma R T_{\mathrm{2}}}
\end{equation}

\begin{equation}
\sin^2 \phi_{\mathrm{1}} = \frac{T_{\mathrm{1}}}{T_{\mathrm{2}}} = Z^2 , \quad \sin \phi_{\mathrm{1}} = Z
\label{last_sinphi}
\end{equation}
\noindent
The Mach-stem height $H_\mathrm{m}$ is related to the sonic throat height in the reactive layer $H_\mathrm{s}$ through
\begin{equation}
\frac{H_\mathrm{m}}{H_\mathrm{s}} = \frac{1}{\overline{M}} \left[ \frac{2}{\gamma + 1} + \left(\frac{\gamma - 1}{\gamma + 1}\right) \overline{M}^2 \right]^{\frac{\gamma + 1}{2(\gamma - 1)}}
\end{equation}
\noindent
where $\overline{M}$ is the wave-fixed average Mach number behind the detonative Mach stem. This value is obtained from a first-order approximation by integrating the momentum flux across the Mach-stem height, following the method of \cite{LI_BEN-DOR_1997}, and can be expressed as

\begin{equation}
\label{Mbar}
\overline{M} = \frac{2 \left( \rho_7 u_7 \cos \delta_7 + \rho_\mathrm{G} u_\mathrm{G} \right)}{\left( \rho_7 + \rho_\mathrm{G} \right) \left( c_7 + c_\mathrm{G} \right)}
\end{equation}
\noindent
where point G denotes the state of the flow behind the nearly normal detonative Mach stem coinciding with the lower wall. Since the Mach stem is nearly normal in this region, a one-dimensional model is most appropriate. The state at G can therefore be determined from the ZND conservation equations by imposing the overdrive predicted by Mitrofanov (a factor of 1.25 in Case~C). While this can be obtained by numerically integrating the ZND equations, Appendix~\ref{Closed-Form Model} provides a closed-form model that explicitly calculates the state of the combustion products using only the thermodynamic variables and the degree of overdrive, without requiring a numerical solver. Point 7, by contrast, is located at the tip of the Mach stem near the triple point. Because a sonic locus extends from the triple point of the reactive Mach reflection, the state at point 7 is taken to coincide with the Chapman--Jouguet conditions.

\begin{figure}[H]
\centering
\includegraphics[width=0.4\textwidth]{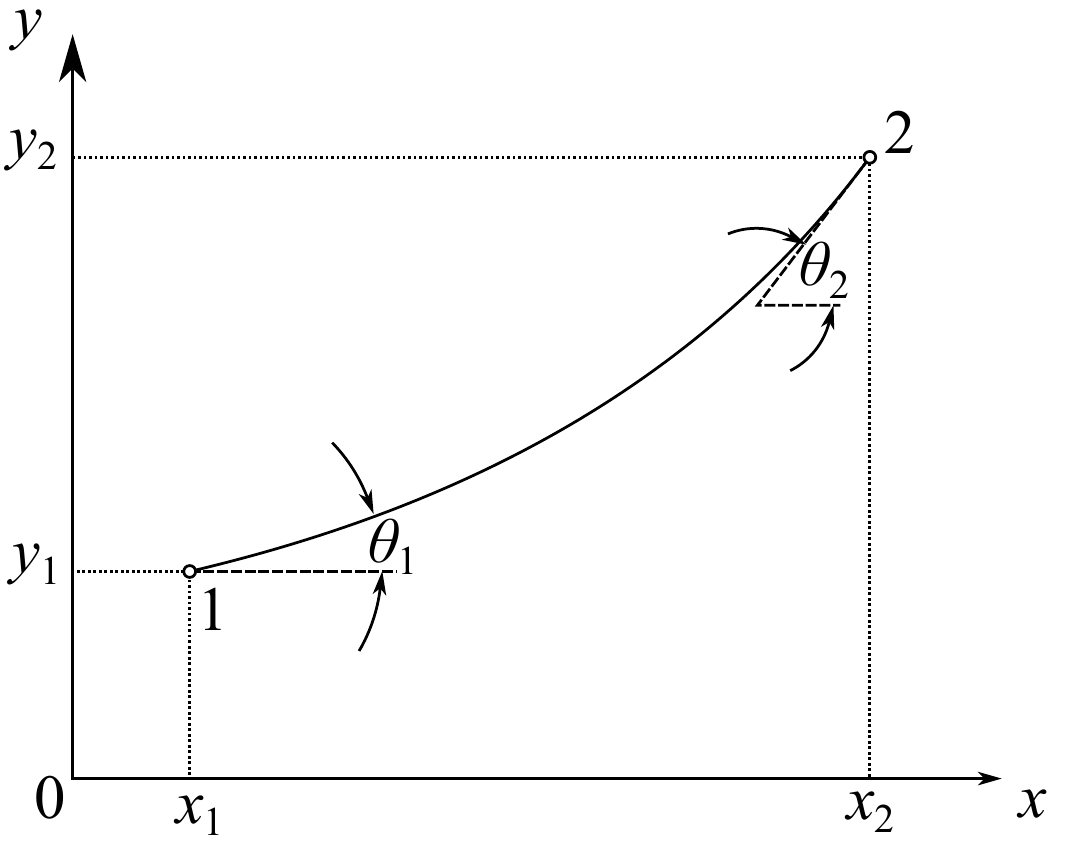}
\caption{Monotonic curve passing through points 1 and 2 and having a gradient of tan $\theta_1$ and tan $\theta_2$ at both points}
\label{Mononotonous}
\end{figure}

The precursor shock AB and the detonation Mach stem TG can be modeled as a
single monotonic curve passing through two points (Fig.~\ref{Mononotonous}).
Following the first-order, small-turning approximation used in the literature
(Ben-Dor~\cite{BenDor}) to represent weakly curved Mach stems, one obtains
an  implicit relation for the curve based on the coordinates and
slopes at its endpoints. In this framework, Eq.~(\ref{J_eq}) is derived in terms
of the positions of the two endpoints and their local gradients.
\vspace{-40pt}

\begin{align}
J(x, y, x_1, y_1, x_2, y_2, \theta_1, \theta_2) = & \left[ (y - y_1) \tan \theta_1 + (x - x_1) \right]^2 \tan(\theta_2 - \theta_1) \notag \\
& + 2 \left[ (x_2 - x_1) + (y_2 - y_1) \tan \theta_1 \right] \left[ (x - x_1) \tan \theta_1 - (y - y_1) \right] = 0
\label{J_eq}
\end{align}

\begin{figure}[H]
\centering
\includegraphics[width=0.4\textwidth]{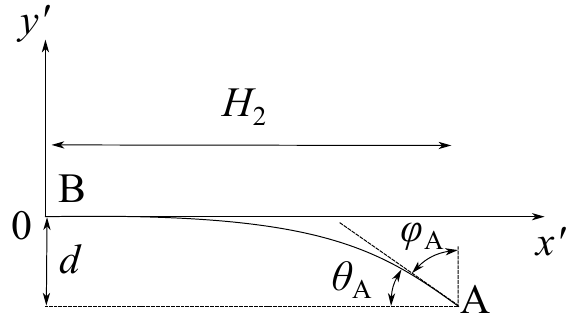}
\caption{Precursor–shock segment AB in the \(x'\)–\(y'\) plane. The local slope satisfies \(\left.\frac{dy'}{dx'}\right|_{x'=0}=0\) and \(\left.\frac{dy'}{dx'}\right|_{x'=H_2}=-\tan\theta_\mathrm{A}\). The offset of the shock tip at A from the \(x'\)-axis is denoted \(d\). The shock angle is \(90^{\circ}\) at B and \(\phi_\mathrm{A}\) at A.}
\label{curved_shock}
\end{figure}

A simplified translated view of the precursor shock AB is provided in Fig.~\ref{curved_shock} onto the $y'-x'$ plane. The shock is initially normal to the upper wall at point B ($\theta_\mathrm{B} = 0$) and curves to an angle of $\theta_\mathrm{A}$ at point A.
\noindent
$\theta_\mathrm{A}$ can be related to the shock angle $\phi_\mathrm{A}$ using the following relation
\begin{equation}
    \theta_\mathrm{A} = 90 - \phi_\mathrm{A}
\end{equation}
\noindent
where $\phi_\mathrm{A}$ is determined by matching the deflection induced by the oblique shock AT. As such, an equation for the curved shock can be obtained if the coordinates of points A and B are substituted into Eq.~(\ref{J_eq}) with values (0,0) and ($H_2, -d$) respectively as

\begin{equation}
    y' = -\frac{x'^2 \cot \phi_\mathrm{A}}{2H_\mathrm{2}}
    \label{curved}
\end{equation}
by setting $x'$ to $H_2$ in Eq.~(\ref{curved}), the magnitude of the value of the deviation, $d$, can be determined
\begin{equation}
    d = \frac{H_\mathrm{2} \cot \phi_\mathrm{A}}{2}
\end{equation}

These geometric relations, together with the closed-form model for predicting the exit state of an overdriven detonation, as presented in Appendix~\ref{Closed-Form Model}, provide a complete analytical framework for predicting the Mach-stem height, as well as curvature, of the detonative triple-point configuration corresponding to Case~C.

\section{Closed-form model for prediction of state at exit of an overdriven detonation}
\label{Closed-Form Model}
As discussed in Appendix~\ref{Detonation_Geometric}, in order to obtain the height of the Mach stem in the reactive layer, knowledge of $\bar{M}$ is required. As shown in Eq.~(\ref{Mbar}), this requires the sound speed, density, and velocity at the exit of an overdriven detonation. The state at the exit of the overdriven detonation near the lower boundary, i.e., state G, can be determined by imposing an overdrive factor on the conservation-based ZND equations and numerically integrating to determine the product state at the exit. While this procedure requires the use of a numerical solver, it is also possible to derive an analytical expression that explicitly gives the pressure, density, temperature, and Mach number of the products at the exit of the overdriven detonation. This is achieved by solving the conservation equations without imposing the Chapman--Jouguet sonic criterion. The derivation is given below, and the analysis is conducted in the wave-fixed reference frame, with state 1 representing the mixture ahead of the overdriven detonation and state 2 representing the combustion products at the exit of the reaction zone.

For convenience, the pressure, density, and temperature ratios are expressed as
\begin{equation}
\alpha = \frac{p_2}{p_1}, \qquad 
\zeta = \frac{\rho_2}{\rho_1}, \qquad 
\eta = \frac{T_2}{T_1},
\end{equation}

\noindent the continuity equation dictates that

\begin{equation}
\rho_1 u_1 = \rho_2 u_2
\end{equation}
where $u_1 = D$. The conservation equations for momentum and energy can therefore be expressed as

\begin{equation}
\label{Momentum}
p_1 + \rho_1D^2 = p_2 + \rho_2u_2^2 
\end{equation}

\begin{equation}
\label{energy}
h_2 +\frac{1}{2} u_2^2 = h_1 + \frac{1}{2} D^2 + \Delta q
\end{equation}

\noindent $u_2$ can be expressed as
\begin{equation}
\label{u2}
u_2 = \frac{u_1}{\zeta} = \frac{D}{\zeta}
\end{equation}
\noindent Eq.~(\ref{Momentum}) can be rearranged as
\begin{equation}
p_2 - p_1 = \rho_1 D^2 (1 - 1/\zeta)
\end{equation}

\begin{equation}
\frac{p_2}{p_1} - 1 = \frac{\rho_1D^2}{\rho_1RT_1}\left(1-\frac{1}{\zeta}\right)
\end{equation}

\noindent which allows $\alpha$ to be expressed explicitly in terms of $\zeta$

\begin{equation}
\label{alpha_eq}
\alpha = 1 + \gamma M_1^2 \left(1 - \frac{1}{\zeta}\right)
\end{equation}

\begin{equation}
M_1 = \frac{u_1}{c_1}
\end{equation}

\begin{equation}
c_1^2 = \frac{\gamma p_1}{\rho_1}
\end{equation}

\noindent Using the fact that $h = c_pT$ for a calorically perfect gas, and given Eq.~(\ref{u2}), Eq.~(\ref{energy}) can be re-expressed as

\begin{equation}
c_\mathrm{p}T_2 + \frac{D^2}{2\zeta^2} = c_\mathrm{p}T_1 + \frac{D^2}{2} + \Delta q
\end{equation}

\noindent Normalizing by $c_{p}T_1$
\begin{equation}
\label{fraction_T2_T1}
\frac{T_2}{T_1} + \frac{D^2}{2\zeta^2 c_pT_1} = 1 + \frac{D^2}{2c_pT_1} + \frac{\Delta q}{c_pT_1}
\end{equation}
the term $\frac{D^2}{c_pT_1}$ can be expressed in terms of $M_1$ as follows

\begin{equation}
\label{Dsquared}
\frac{D^2}{c_pT_1} =\frac{D^2\left(\gamma - 1\right)}{\gamma RT_1} = M_1^2\left(\gamma - 1\right) 
\end{equation}
\noindent Substituting Eq.~(\ref{Dsquared}) into Eq.~(\ref{fraction_T2_T1}), $\eta$ can be expressed explicitly in terms of $M_1$
\begin{equation}
\label{eta_long}
\eta = 1 + \frac{\Delta q}{c_pT_1} + \frac{\left(\gamma - 1\right)}{2} M_1^2 \left(1 - \frac{1}{\zeta^2}\right)
\end{equation}
The ideal gas equation of state relates the variables, $\eta$, $\zeta$, and $\alpha$ through
\begin{equation}
\label{eta_eq}
\eta = \frac{\alpha}{\zeta}
\end{equation}
\noindent substituting Eqs.~(\ref{alpha_eq}) and~(\ref{eta_eq}) into Eq.~(\ref{eta_long}), a quadratic equation of $\zeta$ with coefficients involving $M_1$ can be derived
\begin{equation}
\label{unsimplified_zeta}
\frac{1 + \gamma M_1 ^2 \left(1 - \frac{1}{\zeta}\right)}{\zeta} = 1 + \frac{\Delta q}{c_p{T_1}} + \frac{\left(\gamma - 1\right)}{2}M_1 ^2 \left(1-\frac{1}{\zeta^2}\right)
\end{equation}

\noindent multiplying both sides of Eq.~(\ref{unsimplified_zeta}) by $\zeta^2$ leaves

\begin{equation}
\zeta^2\left[\left(1+ \frac{\Delta q}{c_pT_1}\right) + \frac{\left(\gamma - 1\right)}{2}M_1^2\right] - \zeta\left[1 + \gamma M_1^2\right] + \frac{\left(\gamma + 1\right)}{2}M_1^2 = 0
\end{equation}
which can be expressed as 

\begin{equation}
a\zeta^2 + b\zeta + c = 0
\end{equation}
with $\zeta$ calculated as
\begin{equation}
\zeta = \frac{-b + \sqrt{b^2 - 4ac}}{2a}
\end{equation}
Once $\zeta$ is obtained, $\eta$ and $\alpha$ follow from Eqs.~(\ref{eta_eq}) and (\ref{alpha_eq}), respectively. The Mach number of the products at the exit is then
\begin{equation}
M_2  = \frac{M_1}{\zeta \sqrt{\eta}}   
\end{equation}

\begin{figure}[H]
\label{overdriven_ZND}
\centering
\includegraphics[width=0.77\textwidth]{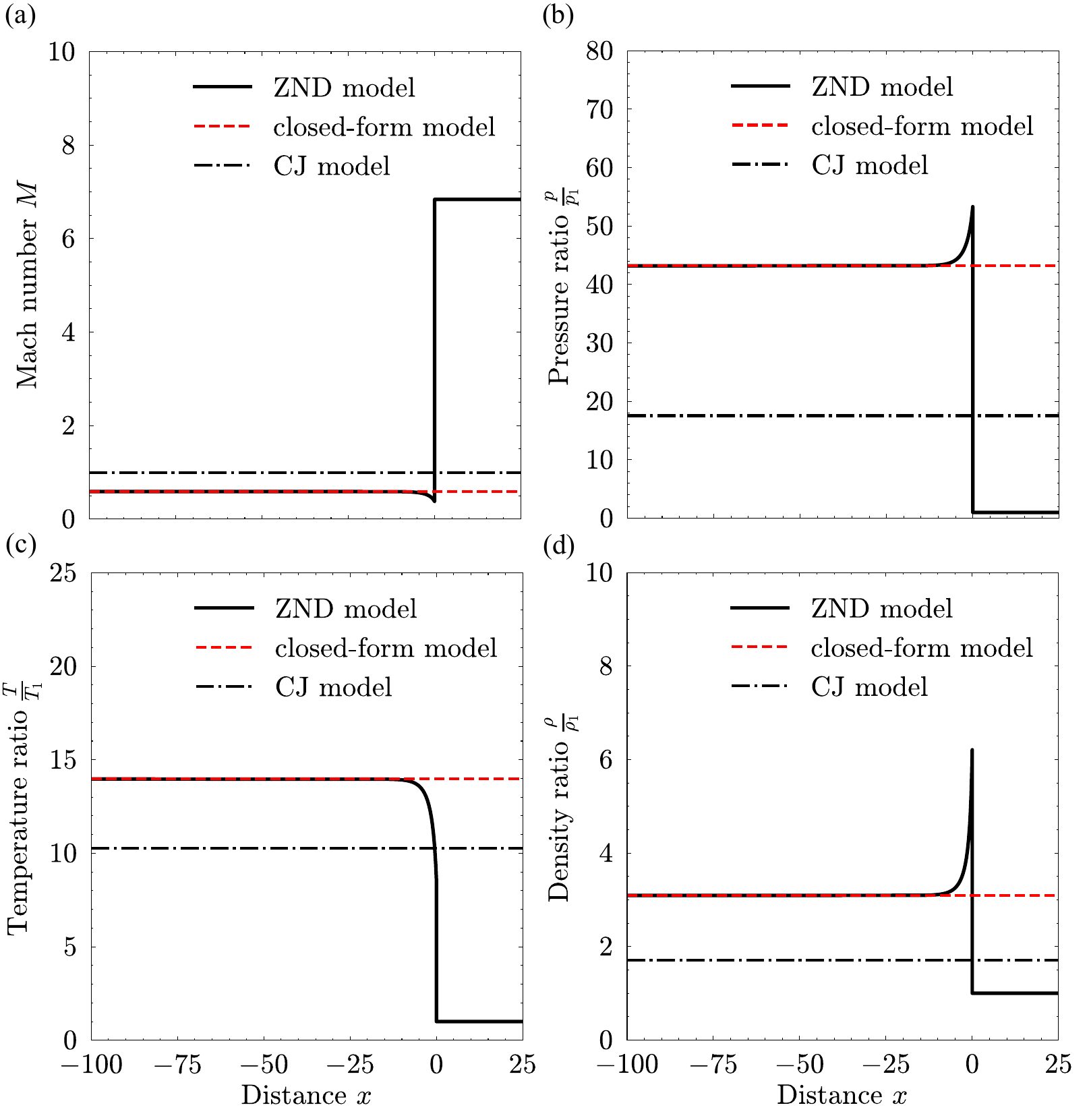}
\caption{Comparison of the closed-form model with ZND integration for an overdriven detonation with $M_1 = 1.25M_\mathrm{CJ}$. The ZND profiles of (a) Mach number, (b) pressure, (c) temperature, and (d) density are shown together with predictions from the closed-form model. Results are presented in the wave-fixed reference frame with the detonation located at $x = 0$. The corresponding CJ properties are provided for reference.}
\label{overdriven}
\end{figure}

Figure~\ref{overdriven} compares a representative solution from the present closed-form model with results obtained from numerical integration of the ZND equations. The comparison corresponds to an overdrive factor of 1.25 relative to the CJ speed, i.e., $M_1 = 1.25M_\mathrm{CJ}$, obtained using the Mitrofanov-based approach for predicting the detonation speed. The case shown corresponds to $A_2/A_1 = 1$ and $Z = 0.45$.

\vspace{-20pt}
The closed-form model presented in this appendix can therefore be used, together with the overdrive ratio determined from the Mitrofanov model, to predict the exit state of an overdriven detonation. This exit state provides the necessary conditions to construct and determine the Mach-stem height of a Mach reflection in the reactive layer, corresponding to Case~C discussed in the main text.
\end{appendices}
\clearpage
\bibliography{sn-bibliography}
\end{document}